\documentclass[preprint,aps,nofootinbib,tightenlines]{revtex4}

\usepackage{graphicx}
\usepackage{amsmath}
\usepackage[hypertex]{hyperref}

\newcommand{\eq}[1]{Eq.~\eqref{#1}}
\newcommand{\eqs}[2]{Eqs.~\eqref{#1} and \eqref{#2}}
\renewcommand{\sec}[1]{Sec.~\ref{#1}}

\newcommand{\fig}[1]{Fig.~\ref{#1}}

\newcommand{\ie}{\emph{i.e.}}

\newcommand{\geneva}{\texttt{GenEvA}}
\newcommand{\GenEvA}{\texttt{GenEvA}}
\newcommand{\madgraph}{\texttt{Madgraph}}

\newcommand{\nn}{\nonumber}

\newcommand{\cM}{\mathcal{M}}
\newcommand{\ord}{\mathcal{O}}
\newcommand{\df}{\mathrm{d}}
\newcommand{\MC}{\mathrm{MC}}
\newcommand{\LO}{\mathrm{LO}}
\newcommand{\NLO}{\mathrm{NLO}}
\renewcommand{\max}{\mathrm{max}}
\newcommand{\start}{\mathrm{start}}
\newcommand{\cut}{\mathrm{cut}}
\renewcommand{\a}{\alpha}
\newcommand{\GeV}{\:\mathrm{GeV}}

\begin{document}

%%%%%%%%%%%%%%%%%%%%%%%%%%%%%%%%%%%%%%%%%%%%%%%%%%%%%%%%%%%%%%%%%%%%%%%%%%%%%%%%
% Title page
%%%%%%%%%%%%%%%%%%%%%%%%%%%%%%%%%%%%%%%%%%%%%%%%%%%%%%%%%%%%%%%%%%%%%%%%%%%%%%%%

\title{GenEvA (I): A new framework for event generation}

\author{Christian W.~Bauer\footnote{Electronic address: cwbauer@lbl.gov}}

\author{Frank J.~Tackmann\footnote{Electronic address: ftackmann@lbl.gov}}

\author{Jesse Thaler\footnote{Electronic address: jthaler@jthaler.net}}

\affiliation{Ernest Orlando Lawrence Berkeley National Laboratory,
University of California, Berkeley, CA 94720
\vspace{2ex}}

\begin{abstract}
We show how many contemporary issues in event generation can be recast in terms of partonic calculations with a matching scale. This framework is called \GenEvA, and a key ingredient is a new notion of phase space which avoids the problem of phase space double-counting by construction and includes a built-in definition of a matching scale. This matching scale can be used to smoothly merge any partonic calculation with a parton shower. The best partonic calculation for a given region of phase space can be determined through physics considerations alone, independent of the algorithmic details of the merging. As an explicit example, we construct a positive-weight partonic calculation for $e^+ e^- \to n \text{ jets}$ at next-to-leading order (NLO) with leading-logarithmic (LL) resummation. We improve on the NLO/LL result by adding additional higher-multiplicity tree-level (LO) calculations to obtain a merged NLO/LO/LL result. These results are implemented using a new phase space generator introduced in a companion paper~\cite{genevatechnique}.
\end{abstract}

\maketitle

\tableofcontents

%%%%%%%%%%%%%%%%%%%%%%%%%%%%%%%%%%%%%%%%%%%%%%%%%%%%%%%%%%%%%%%%%%%%%%%%%%%%%%%%
% Main body of the paper
\newpage

%%%%%%%%%%%%%%%%%%%%%%%%%%%%%%%%%%%%%%%%%%%%%%%%%%%%%%%%%%%%%%%%%%%%%%%%%%%%%%%%
\section{Introduction}
%%%%%%%%%%%%%%%%%%%%%%%%%%%%%%%%%%%%%%%%%%%%%%%%%%%%%%%%%%%%%%%%%%%%%%%%%%%%%%%%

Despite facing a complicated detector environment, top quark~\cite{Brubaker:2006qt} and $W$ boson~\cite{:2007ypa,Aaltonen:2007ps} studies at the Tevatron have proved that it is nevertheless possible to make precision measurements at hadron colliders. With the upcoming Large Hadron Collider (LHC), a complete understanding of Standard Model backgrounds will be essential for discovering new physics at the energy frontier \cite{:1999fq,:1999fr,Ball:2007zza}.
Therefore, precision theoretical calculations are needed to complement the increasingly sophisticated experimental techniques available at hadron colliders. At lepton colliders, data is often compared to theoretical predictions for inclusive quantities, but at hadron colliders, it is more typical for data to be compared to theoretical predictions for exclusive quantities in order to more readily apply experimental cuts that may not be well defined in an inclusive theoretical framework.

Monte Carlo programs have proved indispensable for making exclusive theoretical predictions. Together with parton distribution functions, hadronization models, and underlying event models, traditional event generators~\cite{Lonnblad:1992tz, Sjostrand:2000wi, Sjostrand:2006za, Corcella:2000bw, Mangano:2002ea, Maltoni:2002qb, Gieseke:2003hm, Gieseke:2006ga, Gleisberg:2003xi, Paige:2003mg, Boos:2004kh, Kilian:2007gr, Cafarella:2007pc} agree remarkably well with Tevatron data over a wide variety of experimental observables~\cite{Aaltonen:2007dg, :2007kp}. However, in anticipation of further theoretical progress in refining Standard Model (and Beyond the Standard Model) predictions, it is worthwhile to consider possible improvements to the traditional Monte Carlo approach. There has been much work in recent years on merging fixed-order matrix element calculations with parton showers~\cite{Catani:2001cc, Lonnblad:2001iq, Krauss:2002up, MLM, Mrenna:2003if, Schalicke:2005nv, Lavesson:2005xu, Hoche:2006ph, Alwall:2007fs, Giele:2007di, Lavesson:2007uu, Nagy:2007ty, Nagy:2008ns}. Furthermore, there are several programs which implement higher-order calculations to produce inclusive cross sections~\cite{MCFM,NLOJET,PHOX,VBFNLO}, though these next-to-leading (NLO) order programs are not cast in the form of a Monte Carlo program that can generate fully-hadronized exclusive events. While several ideas exist in the literature of how to implement a combination of NLO results and parton showers~\cite{Collins:2000gd,Collins:2000qd,Potter:2001ej, Dobbs:2001dq, Frixione:2002ik, Frixione:2003ei, Nason:2004rx, Nason:2006hfa, LatundeDada:2006gx, Frixione:2007vw, Kramer:2003jk, Soper:2003ya, Nagy:2005aa, Kramer:2005hw}, only a few publicly available programs currently exist~\cite{Frixione:2006gn,Frixione:2007nu} that implement this properly.

At the level of partons, current event generators are based on two independent frameworks: fixed-order matrix element calculations and parton showers. While fixed-order calculations include quantum interference and can be systematically improved through perturbative loop calculations, parton showers are always necessary to generate additional QCD radiation in a Monte Carlo framework. This is because fixed-order calculations cannot handle the large number of final states typically present in high-energy collisions. Schematically, the way the fully differential hadronic cross section $\df \sigma$ is achieved in a traditional approach is through
%%%
\begin{equation}
\label{eq:mastertrad}
\df\sigma^{\rm trad.} = \MC\left( \lvert\cM\rvert^2 \, \df\Phi \right)
\,,\end{equation}
%%%
where $\lvert\cM\rvert^2$ represents a fixed-order QCD calculation, $\df\Phi$ represents a fixed-multiplicity phase space algorithm, and $\MC$ represents the action of a showering/hadronization scheme. The main challenge of \eq{eq:mastertrad} is that the actual partonic four-momenta are generated both through the action of $\df\Phi$ and $\MC$, and in order to avoid phase space double-counting, either $\lvert\cM\rvert^2$ has to be modified from the value calculated in QCD, or the action of $\MC$ has to modified to accommodate the fixed-order calculation, or both.

In this paper, we present a new Monte Carlo framework \GenEvA---for \textbf{Gen}erate \textbf{Ev}ents \textbf{A}nalytically---that allows almost any parton-level calculation to be translated into hadron-level events in a generic way with only a minimal modification of field theoretic methods.\footnote{This framework is unrelated to the ``Geneva'' jet algorithm \cite{Bethke:1991wk}.} The key idea is to separate the physics considerations, which determine the appropriate distribution to use, from the algorithmic details, which define how phase space is generated and what models are used to perform additional showering and hadronization. As we will see, it is possible to separate the physics and algorithmic issues using a matching scale $\mu$. Schematically, the fully differential hadronic cross section $\df \sigma$ in \GenEvA\ is given by
%%%
\begin{equation}
\label{eq:mastergeneva}
\df\sigma^{\tt GenEvA} = \lvert\cM(\mu)\rvert^2\, \df\MC(\mu)
\,.\end{equation}
%%%
The quantity $\df\MC(\mu)$ represents both a phase space algorithm \emph{and} a showering/hadronization scheme that starts generating radiation at the scale $\mu$. The only information the ``partonic calculation'' $|\cM(\mu)|^2$ needs to know about the specific implementation of phase space generation and showering is this matching scale $\mu$. From the point of view of the partonic calculation, $\mu$ is an infrared scale and can be thought of as the scale at which the partonic calculation is interfaced with a parton shower. We will see that the details of this matching are independent of any specific parton shower algorithm. While the partonic calculation $|\cM(\mu)|^2$ does not necessarily correspond to a traditional QCD amplitude,\footnote{In the context of soft-collinear effective theory (SCET) \cite{Bauer:2000ew,Bauer:2000yr,Bauer:2001ct,Bauer:2001yt}, $\lvert\cM(\mu)\rvert^2$ corresponds to the square of a Wilson coefficient and $\df\MC(\mu)$ corresponds to the matrix element of an SCET operator~\cite{Bauer:2006qp,Bauer:2006mk}. This explains why there should be a cancellation of the $\mu$ dependence in $\lvert\cM(\mu)\rvert^2 \df\MC(\mu)$. While SCET offers formal definitions for what $|\cM(\mu)|^2$, $\df\MC(\mu)$, and $\mu$ are, these formal definitions are not necessarily required.} we will show various ways that $\lvert\cM(\mu)\rvert^2$ can be determined solely in the context of perturbative QCD, independent of the algorithmic details of $\df\MC(\mu)$.

In this way, \GenEvA\ allows the user to focus on determining the most accurate differential cross section in a given region of phase space, instead of trying to figure out an algorithm which not only covers phase space with the right distributions, but also interfaces cleanly with a showering/hadronization scheme. As a concrete example of the power of the \GenEvA\ framework, we will present a conceptually simple implementation of two known methods to improve the accuracy of Monte Carlo: parton shower/matrix element merging (PS/ME)~\cite{Catani:2001cc, Lonnblad:2001iq, Krauss:2002up, MLM, Mrenna:2003if, Schalicke:2005nv, Lavesson:2005xu, Hoche:2006ph, Alwall:2007fs, Giele:2007di, Lavesson:2007uu, Nagy:2007ty, Nagy:2008ns} and Monte Carlo at next-to-leading order (PS/NLO)~\cite{Collins:2000gd,Collins:2000qd,Potter:2001ej, Dobbs:2001dq, Frixione:2002ik, Frixione:2003ei, Nason:2004rx, Nason:2006hfa, LatundeDada:2006gx, Frixione:2007vw, Kramer:2003jk, Soper:2003ya, Nagy:2005aa, Kramer:2005hw}. In previous approaches, PS/ME merging and PS/NLO merging relied on different Monte Carlo algorithms, but in the context of \GenEvA, they correspond just to different choices for $|\cM(\mu)|^2$ and share the same algorithmic underpinnings.

In fact, \GenEvA\ allows PS/ME merging and PS/NLO merging to be \emph{combined} to create a positive-weight Monte Carlo sample that merges NLO information with higher-order tree-level (LO) matrix elements and leading-logarithmic (LL) resummation. The simplicity with which we achieve an NLO/LO/LL merged sample through a special choice for $|\cM(\mu)|^2$ suggests obvious generalizations to including next-to-next-to-leading order (NNLO) or next-to-leading logarithmic (NLL) information in event generators, though we will not pursue those directions in the present work.

Our focus will be on getting the most accurate distribution possible using available theoretical tools, and not necessarily on generating these distributions efficiently. In a companion paper \cite{genevatechnique}, we describe a $\df\MC(\mu)$ generator based on parton shower reweighting \cite{Bauer:2007ad}, and we will see that certain gains in efficiency come as a bonus for using the parton shower as a phase space generator. However, the \GenEvA\ framework is more general than the specific algorithmic implementation in Ref.~\cite{genevatechnique}. Though we anticipate extending \GenEvA\ to handle the full range of Standard Model processes at hadron colliders, we will focus here on the process $e^+ e^- \to n$ jets, which is sufficiently complicated to highlight all of the novel techniques introduced by \GenEvA. In the companion paper \cite{genevatechnique}, we comment on additional technical issues that arise in trying to understand Tevatron or LHC physics.

In the next section, we review the challenges faced when constructing Monte Carlo tools and the insights offered by \GenEvA. In \sec{sec:toy}, we discuss the \geneva\ framework in a simple toy model. We then transition to the more complicated case of QCD, where we discuss the first emission in detail in \sec{sec:QCDfirst} and extend the results to multiple emissions in \sec{sec:NLO}. After showing results from the \geneva\ program in \sec{sec:results}, we conclude in \sec{sec:conclusions}.

%%%%%%%%%%%%%%%%%%%%%%%%%%%%%%%%%%%%%%%%%%%%%%%%%%%%%%%%%%%%%%%%%%%%%%%%%%%%%%%%
\section{Overview of the GenEvA Framework}
\label{sec:overview}
%%%%%%%%%%%%%%%%%%%%%%%%%%%%%%%%%%%%%%%%%%%%%%%%%%%%%%%%%%%%%%%%%%%%%%%%%%%%%%%%

%===============================================================================
\subsection{The Challenge of QCD}
%===============================================================================

To build a perfect Monte Carlo program, one would need a complete description of the Standard Model that is valid over all of phase space. Unfortunately, no such description exists, mainly because we do not have a complete description of QCD that is valid for every energy scale and every kinematic configuration. Instead, we know various limits of QCD, various perturbative expansions of QCD, and various phenomenological models based on QCD. The best Standard Model Monte Carlo program we can hope to build is one that coherently combines as many descriptions of QCD as possible.

\begin{figure}
\includegraphics[scale=0.7]{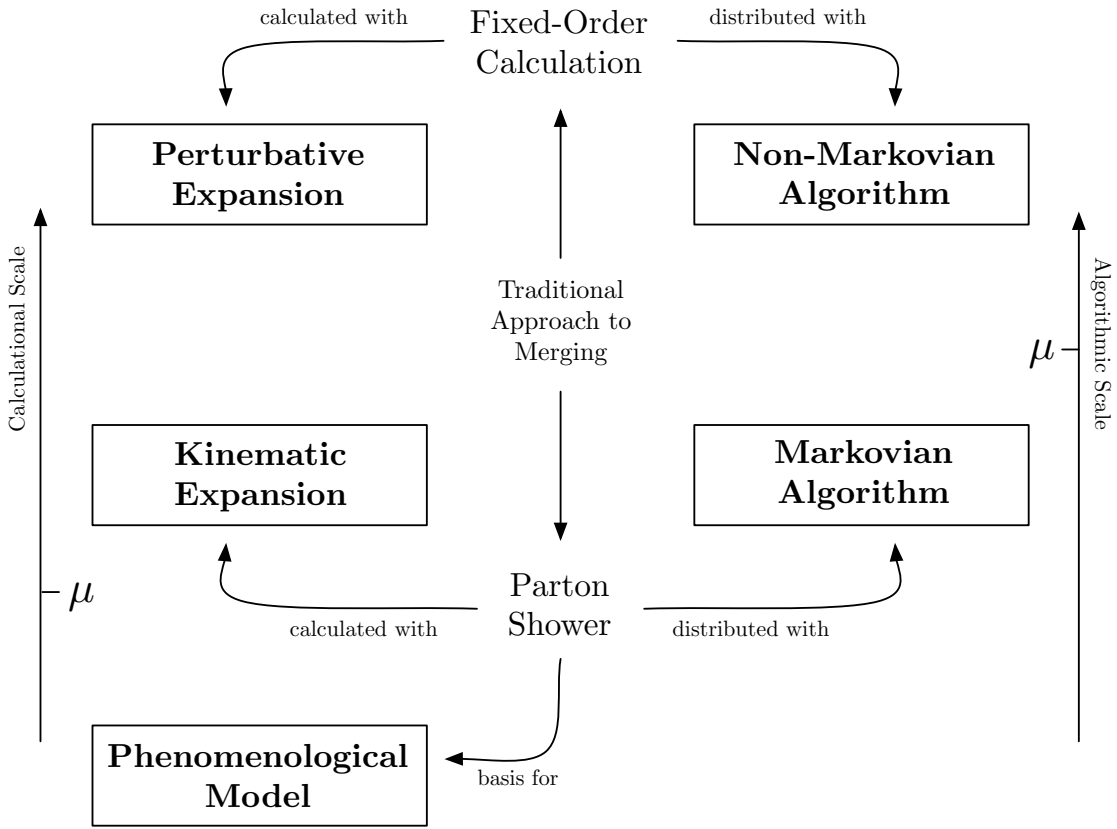}
\caption{The traditional approach to merging fixed-order calculations with parton showers. Traditionally, fixed-order calculations tie a perturbative expansion to a non-Markovian fixed-multiplicity phase space algorithm. The parton shower ties a kinematic expansion to a Markovian variable-multiplicity phase space algorithm. Hence, in trying to merge fixed-order calculations with parton showers, one is led to simultaneously having to merge two different QCD expansions and two different algorithmic methods to generate phase space. On the algorithmic side, imposing a phase space cut $\mu$ to separate the two different algorithms would results in residual $\mu$ dependence and uncanceled infrared divergences as explained in the text. \GenEvA\ is based on imposing a calculational $\mu$ cut that separates QCD calculations from QCD phenomenological models. This requires splitting the parton shower into a formal component and a phenomenological component.}
\label{fig:traditional}
\end{figure}

One crucial combination that has been the subject of many recent advances in Monte Carlo is the merging of fixed-order calculations with parton showers. Fixed-order calculations are reliable when pairs of partons are well separated in phase space, while parton showers are reliable in the soft-collinear limit. For this reason, a Monte Carlo program that combines both descriptions has a better chance to correctly describe experimental data over a broad range of observables and energy scales. The problem with the language of ``fixed-order calculation'' and ``parton shower'' is that these terms convolute calculational definitions with algorithmic ones, and the goal of the \GenEvA\ framework is to isolate the calculational from the algorithmic challenges involved in combining different descriptions of QCD.

As shown in \fig{fig:traditional}, a fixed-order calculation is based on a perturbative expansion of QCD to fixed order in $\alpha_s$. To distribute events according to a fixed-order calculation, one usually uses a non-Markovian algorithm (such as an adaptive grid) to generate points in fixed-multiplicity $n$-body phase space $\df\Phi_n$. The parton shower, on the other hand, is defined in the soft-collinear limit of QCD, and the splitting functions and Sudakov factors are calculated in this limit. Parton showers are usually used in the context of a Markovian phase space algorithm that recursively generates all of variable-multiplicity phase space through a probabilistic mapping of $\df\Phi_n \to \df\Phi_{n+1}$.\footnote{More general parton showers such as in Refs.~\cite{Nagy:2007ty, Nagy:2008ns} can also have non-Markovian aspects.} Therefore, in trying to combine a fixed-order calculation with a parton shower, one is simultaneously trying to merge two different expansions of QCD and two different phase space algorithms. Indeed, currently there does not exist a solution for how to merge \emph{generic} fixed-order calculations with parton showers, mainly because separate algorithmic merging procedures are currently necessary depending on whether one is considering tree-level calculations or one-loop calculations.

Naively, one might hope to define an algorithmic phase space cut $\mu$ to separate fixed-order calculations from parton showers as on the right side of \fig{fig:traditional}. One could imagine using a fixed-order calculation when the invariant mass between two partons is greater than $\mu$ and using a parton shower when the invariant mass is less than $\mu$. This \emph{algorithmic} $\mu$ would immediately solve the problem of double-counting that arises when the parton shower acting on $n$-body phase space covers exactly the same phase space regions as an $(m>n)$-body generator. However, this approach is suspect for two reasons. First, the parton shower resums the leading Sudakov logarithms~\cite{Sudakov:1954sw} in the problem, so if the parton shower is only used below $\mu$, it will exhibit double-logarithmic $\alpha_s \log^2 \mu$ sensitivity to this arbitrary, unphysical $\mu$ scale. Second, infrared divergences in virtual diagrams contributing to the fixed-order results are canceled by collinear and/or soft real-emission diagrams, which are by definition contributing below the scale $\mu$ and are therefore contained in the parton shower picture. Thus, one has to somehow link the fixed-order calculation to the parton shower in order to properly cancel infrared divergences.

We argue that a better definition of $\mu$ is as a \emph{calculational} scale that separates calculations performed in QCD from phenomenological models based on QCD as on the left side of \fig{fig:traditional}. This requires cleanly separating two different uses of the parton shower. Above the scale $\mu$, the parton shower corresponds directly to quantities that can be defined formally in the soft-collinear limit. In particular, the splitting functions~\cite{Gribov:1972ri, Altarelli:1977zs, Dokshitzer:1977sg} and Sudakov factors~\cite{Sudakov:1954sw} can be systematically derived~\cite{Bauer:2006qp, Bauer:2006mk} using soft-collinear effective theory (SCET)~\cite{Bauer:2000ew,Bauer:2000yr,Bauer:2001ct,Bauer:2001yt}. Below the scale $\mu$, the parton shower should be regarded as simply a QCD-inspired phenomenological model. For example, a phenomenological shower can be used to extrapolate into regions of phase space away from the soft-collinear limit if high-multiplicity fixed-order matrix elements are unavailable, creating partonic final states that include the correct singularity and symmetry structure of QCD, but lack the full quantum interference. Similarly, a phenomenological shower can serve as the entry point to nonperturbative hadronization in a general fragmentation scheme, and in that context, the shower can and should be tuned to data to reproduce the measured fragmentation properties seen in experiments.

It is the fact that the parton shower has meaning above and below $\mu$ that can be used to mitigate the unphysical $\mu$ dependence, since it implies that the double-logarithmic dependence of the parton shower is identical to the double-logarithmic dependence of QCD. This guarantees that partonic calculations with leading-logarithmic improvements will have no leading $\mu$ dependence when interfaced with a parton shower at the scale $\mu$. However, the challenge of defining $\mu$ in this way is that now it is not clear whether the issue of phase space double-counting will be solved, nor is it obvious that such a definition of $\mu$ will provide a method to combine generic fixed-order calculations with a parton shower, especially in the presence of infrared divergences. At this point, it is thus worthwhile to reconsider the various components of \fig{fig:traditional} to distill the essential challenges in combining fixed-order calculations with parton showers.

%===============================================================================
\subsection{Combining Different QCD Descriptions}
%===============================================================================

\begin{figure}
\includegraphics[scale=0.7]{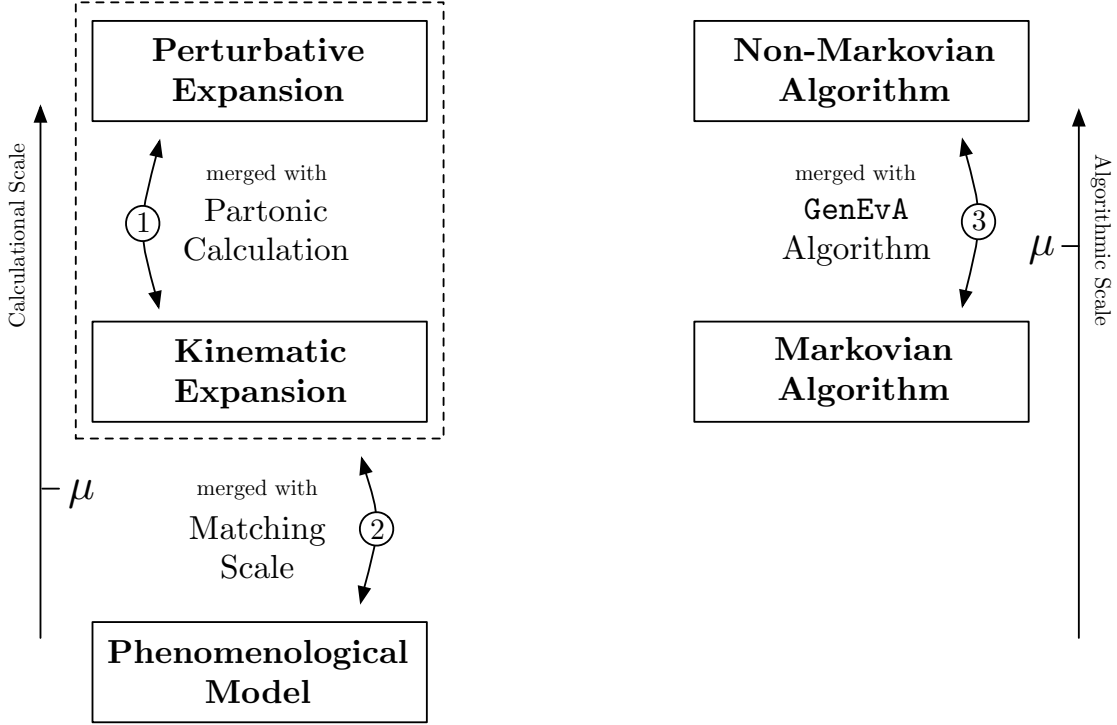}
\caption{The three different meanings of ``combine'':
\text{\large \textcircled{\normalsize 1}} Combining different expansions of QCD in a partonic calculation;
\text{\large \textcircled{\normalsize 2}} combining a partonic calculation with a phenomenological model;
and \text{\large \textcircled{\normalsize 3}} combining a non-Markovian phase space generator with a Markovian generator. By isolating the issues involved in merging fixed-order calculations with parton showers, \GenEvA\ offers a simple and generic method to improve Monte Carlo. Note that all these meanings of ``combine'' respect the division between calculations on one side and algorithms on the other, and the \GenEvA\ framework is devised to cleanly separate the calculational and algorithmic issues.}
\label{fig:boxes}
\end{figure}

In trying to combine different descriptions of QCD into a coherent Monte Carlo framework, there are really three definitions of what is meant by ``combine'' as shown in \fig{fig:boxes}: combining different formal expansions, combining formal calculations with phenomenological models, and combining different phase space algorithms. These definitions of ``combine'' are in one-to-one correspondence with three technical problems that arise when trying to merge fixed-order calculations with parton showers: regulating infrared divergences, canceling unphysical $\mu$ dependence, and eliminating double-counting. The power and simplicity of the \GenEvA\ framework comes from cleanly separating these three issues.

The first meaning of ``combine'' is to merge different formal expansions of QCD, in our case a perturbative expansion in $\alpha_s$, with a kinematic expansion that allows resumming logarithms of some ratio $r$ of energy scales. A successful combination of this form will result in a partonic calculation $\lvert\cM(\mu)\rvert^2$ that is simultaneously correct to the calculated order in $\alpha_s$ while also including the desired level of logarithmic resummation. While there are formal procedures to calculate $\lvert\cM(\mu)\rvert^2$ by matching QCD onto SCET~\cite{Bauer:2006mk,Bauer:2006qp}, there are numerous correct choices for $\lvert\cM(\mu)\rvert^2$ that satisfy the desired properties at the leading-logarithmic level. We will also see that the issue of infrared divergences is solved entirely in the context of a partonic calculation, because one can think of the cancellation of divergences between loop diagrams and tree diagrams as just another type of formal combination.

We have included an explicit $\mu$ dependence in the partonic calculation $\lvert\cM(\mu)\rvert^2$ in anticipation of the second meaning of ``combine'', which is to combine a formal QCD calculation with a phenomenological model based on QCD. Below some scale $\mu$, the phenomenological parton shower will be used to fill out phase space and interface with a nonperturbative hadronization model, but we already saw that this will introduce unphysical $\alpha_s \log^2 \mu$ dependence in the final results. However, if we include leading-logarithmic resummation in the partonic calculation $\lvert\cM(\mu)\rvert^2$, then this $\mu$ dependence will cancel by construction. At first glance, it appears that the partonic calculation now needs to know the details of the phenomenological shower in order to engineer the $\alpha_s \log^2 \mu$ cancellation. Remember though, that this double-logarithmic $\mu$ dependence is a property of QCD and not of a specific parton shower, so the correct $\mu$ dependence can be included in $\lvert\cM(\mu)\rvert^2$ independently of the details of the shower. The only requirement is that the same formal definition of $\mu$ is used in both the partonic calculation and the phenomenological shower. In practice, the choice of $\mu$ corresponds to the choice of shower evolution variable.

Finally, the third meaning of ``combine'' has to do with merging different algorithmic techniques. We saw that phase space can either be covered by a fixed-multiplicity non-Markovian algorithm or with a variable-multiplicity Markovian algorithm, and for computational reasons we may wish to use different algorithms in different regions of phase space. Because of the problem of double-counting, one needs to merge these two algorithms in such a way that a given point in phase space is covered once and only once. One option is to define a phase space cut $\mu$, as mentioned already, and use a non-Markovian algorithm above that scale $\mu$ and a Markovian algorithm below $\mu$. We denote such phase space with a matching scale as $\df\MC(\mu)$, and an example of an efficient $\df\MC(\mu)$ generator is presented in the companion paper~\cite{genevatechnique}.

Let us return to the traditional approach in \eq{eq:mastertrad} for creating a fully differential hadronic cross section, now with the final state multiplicity $n$ made manifest:
%%%
\begin{equation}
\df\sigma^{\rm trad.} = \sum_n \MC_{\mu_n} \left( \lvert\cM_n\rvert^2 \, \df\Phi_n \right)
\,.\end{equation}
%%%
Here, $\lvert\cM_n \rvert^2$ is a fixed-order calculation with $n$ final state partons, $\df\Phi_n$ is $n$-body phase space, and $\MC_{\mu_n}$ represents a phenomenological parton shower that starts at the scale $\mu_n$. As mentioned already, current approaches to merging fixed-order calculations with parton showers require either a modification to $\lvert\cM_n\rvert^2$, to $\MC_{\mu_n}$, or to both, and \GenEvA\ is no exception. The novelty of \GenEvA\ is that these modifications can be made completely generic by applying the logic of \fig{fig:boxes}.

First, a fixed-order calculation lacks the correct behavior for small $r$, so we have to substitute a partonic calculation that includes a proper logarithmic resummation,
%%%
\begin{equation}
\text{\large \textcircled{\normalsize 1}} \qquad
\lvert\cM_n\rvert^2 \to \lvert\cM_n (\tilde{\mu}_n)\rvert^2
\nn\,,\end{equation}
%%%
where at this point $\tilde{\mu}_n$ is an unspecified infrared scale. Second, to avoid large $\alpha_s \log^2 \mu_n$ dependence when the phenomenological parton shower is applied, the infrared scale in the partonic calculation must be the same as the starting scale of the shower,
%%%
\begin{equation}
\text{\large \textcircled{\normalsize 2}} \qquad \tilde{\mu}_n \to \mu_n
\nn\,.\end{equation}
%%%
Finally, the traditional approach to phase space involves two different phase space algorithms, a non-Markovian $\df\Phi_n$ generator and a Markovian $\MC_{\mu_n}$ generator. For any specific phase space point, the partonic calculation is simply a number that is unchanged by the action of $\MC_{\mu_n}$, so $\lvert\cM_n (\mu_n)\rvert^2$ can be factored out separately for each $n$. The quantity $\sum_n \MC_{\mu_n} (\df\Phi_n)$ involves multiple covering of phase space, since the action of the phenomenological model starting from a phase space point in $\df \Phi_n$ will populate regions of phase space $\df \Phi_{m>n}$ by the splitting of particles. To solve the problem of double-counting we introduce a $\mu$-aware phase space generator
%%%
\begin{equation}
\text{\large \textcircled{\normalsize 3}}
\qquad \MC_{\mu_n} (\df\Phi_n) \to \df\MC_n(\mu_n)
\nn\,,\end{equation}
%%%
which cleanly separates phase space populated by the $\df\Phi_n$ generator and the subsequent evolution using the phenomenological model.
These three steps lead to the \GenEvA\ master formula from \eq{eq:mastergeneva}
%%%
\begin{equation}
\label{eq:mastergeneva2}
\df\sigma^{\tt GenEvA} = \sum_n \lvert\cM_n (\mu_n)\rvert^2 \df\MC_n(\mu_n)
\,.\end{equation}
%%%

\begin{figure}
\includegraphics[scale=0.5]{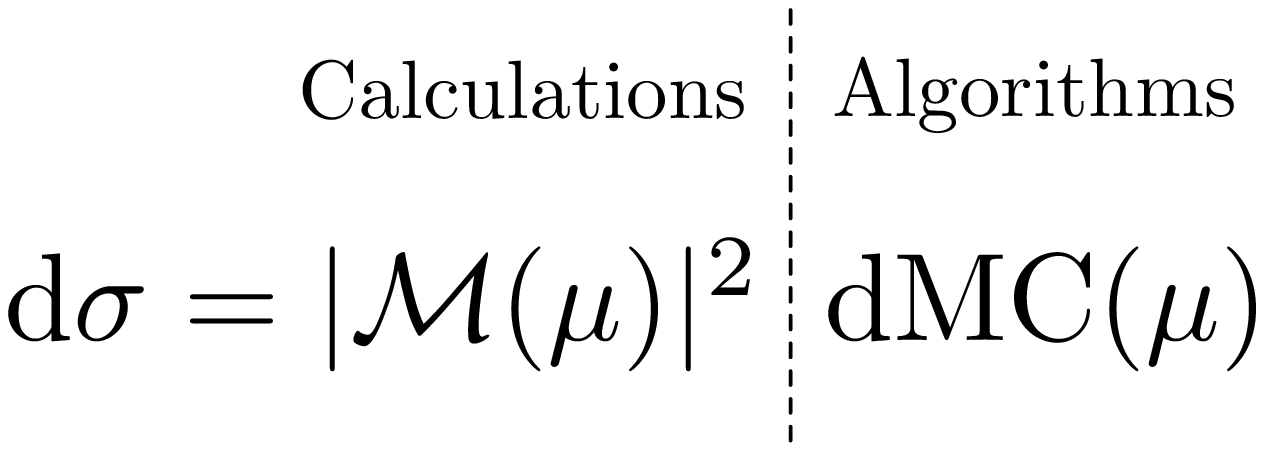}
\caption{A schematic summary of the \GenEvA\ master formula. The fully exclusive differential cross section comes from interfacing a partonic calculation $\lvert\cM(\mu)\rvert^2$ with a phenomenological model through the use of Monte Carlo space $\df\MC(\mu)$. The matching scale $\mu$ allows this interface to be smooth and defines a clean separation between calculational and algorithmic issues.}
\label{fig:masterseparation}
\end{figure}

There are two interesting features of \eq{eq:mastergeneva2}. First, as summarized in \fig{fig:masterseparation}, this formula can be interpreted as an explicit separation of calculational issues from algorithmic issues. $\lvert\cM_n (\mu_n)\rvert^2$ certainly encodes the challenges of merging together different QCD expansion schemes into a single partonic cross section. $\df\MC_n(\mu_n)$ is slightly more complicated because it encodes both algorithmic information as well as the physics of phenomenological models. However, to the extent to which we (unfairly) regard phenomenological models as QCD-inspired numerical algorithms, $\df\MC_n(\mu_n)$ encodes the algorithmic challenges of distributing QCD-like events. Because calculational issues are now separated from algorithmic ones, the partonic calculations $\lvert\cM_n (\mu_n)\rvert^2$ can be determined through physics consideration alone, independently of the details of $\df\MC_n(\mu_n)$.

\begin{figure}
\includegraphics[scale=0.7]{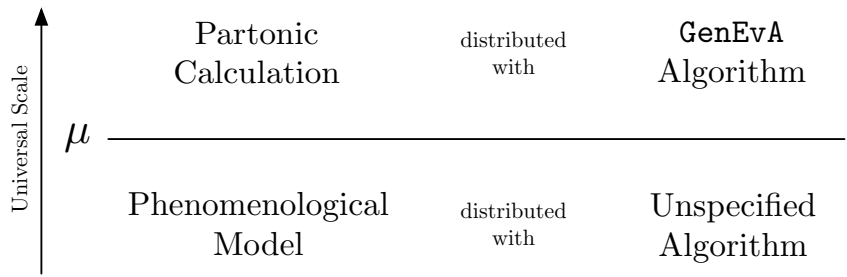}
\caption{The user's view of the \GenEvA\ framework. Partonic calculations as defined by $\lvert\cM(\mu)\rvert^2$ are used above a universal scale $\mu$ and can be distributed with the \geneva\ algorithm \cite{genevatechnique}. A phenomenological model is used below the scale $\mu$ and is implemented using an unspecified algorithm, usually a parton shower interfaced with a hadronization scheme. Because the same scale $\mu$ defines both the calculational separation and the algorithmic separation, the user is free to choose the best partonic calculations from physics considerations alone, without having to worry about the details of the algorithmic implementation.}
\label{fig:scales}
\end{figure}

Second, as summarized in \fig{fig:scales}, the same scale $\mu_n$ appears both in $\lvert\cM_n (\mu_n)\rvert^2$ and in $\df\MC_n(\mu_n)$. A potential mismatch as in \fig{fig:traditional} between the calculational scale $\mu$ and the algorithmic scale $\mu$ is avoided because these two scales are now forced to be the same to cancel the $\alpha_s \log^2 \mu$ dependence. In this way, all the user needs to know is that full QCD information as encoded in $\lvert\cM(\mu)\rvert^2$ will be used above $\mu$, and phenomenological models of QCD (which the user need not specify) will be used below $\mu$. Of course, the precise definition of $\mu$ must always be specified, and in this paper, we will use the invariant mass between particles to define $\mu$, leaving a discussion of generalizations for the companion paper \cite{genevatechnique}.

In the remainder of this section, we will consider the various elements in \eq{eq:mastergeneva2} in more detail, starting with a more precise definition of the phase space $\df\MC(\mu)$, explaining how to include additional flexibility in defining more scales like $\mu$, and then discussing methods to determine the partonic calculation $\lvert\cM(\mu)\rvert^2$.

%===============================================================================
\subsection{A New Approach to Phase Space}
\label{subsec:newphasespace}
%===============================================================================

\begin{figure}
\includegraphics[scale=0.7]{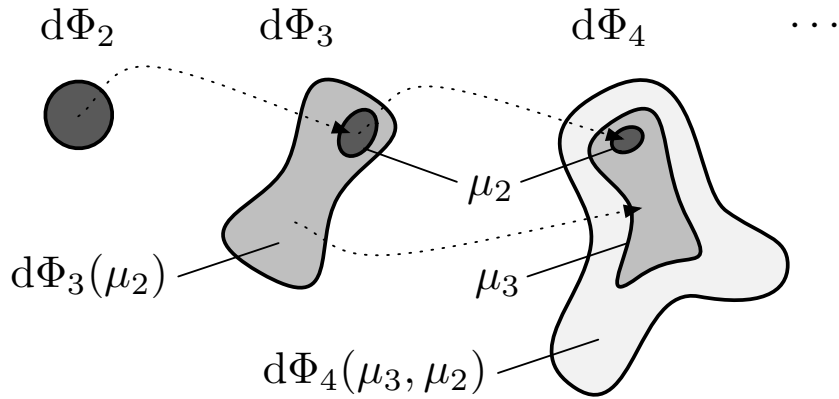}
\caption{The scale-dependent phase space $\df\Phi_n(\{\mu_i\}_{i<n})$. The regions of $\df\Phi_{3,4}$ labeled $\mu_2$ are those reached by the phenomenological model acting on $\df\Phi_2$ with starting scale $\mu_2$. The region of $\df\Phi_4$ labeled $\mu_3$ is reached by the phenomenological model acting on $\df\Phi_3$ when started at the scale $\mu_3$. Excluding these regions from $\df\Phi_{3,4}$ defines $\df\Phi_3(\mu_2)$ and $\df\Phi_4(\mu_2, \mu_3)$. When interfaced with a phenomenological model, the set of all $\df\Phi_n(\{\mu_i\}_{i<n})$ regions gives a complete covering of phase space with no double-counting by construction.}
\label{fig:phasecartoona}
\end{figure}

\begin{figure}
\includegraphics[scale=0.7]{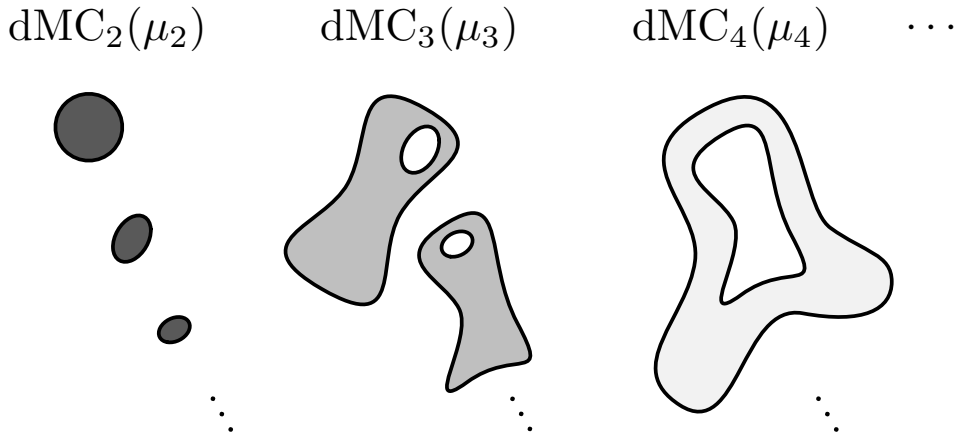}
\caption{Definition of ``Monte Carlo space'' $\df\MC_n(\mu_n)$ as all regions of scale-dependent phase space $\df\Phi_n(\{\mu_i\}_{i<n})$ that can be reached by the phenomenological model starting from $\mu_n$. $\df\MC_2(\mu_2)$ consists of $\df\Phi_2$ together with the part of $\df\Phi_{3,4,\ldots}$ labeled by $\mu_2$ in \fig{fig:phasecartoona}, $\df\MC_3(\mu_3)$ consists of $\df\Phi_3(\mu_2)$ together with the part of $\df\Phi_{4,\ldots}$ labeled by $\mu_3$ (but excluding that labeled by $\mu_2$), and so on. This way of organizing phase space emphasizes that while partonic calculations are defined on scale-dependent phase space $\df\Phi(\mu)$, they affect all regions of Monte Carlo space $\df\MC(\mu)$ through the phenomenological model.}
\label{fig:phasecartoonb}
\end{figure}

As discussed in the previous section, the \geneva\ framework requires a phase space generator $\df\MC(\mu)$ that, by construction, does not introduce double-counting when interfaced with a phenomenological model for filling out phase space. To be specific, we take the phenomenological model to be a parton shower that takes an $n$-body configuration $\Phi_n$ in $\df\Phi_n$ and starts showering at some scale $\mu_n$ to produce additional partons. In other words, the parton shower produces $m$-body configurations with $m \geq n$, where the precise regions of higher-dimensional phase space that are populated are determined by the choice of starting scale $\mu_n$ and the details of the parton shower. Calling these regions $\MC_n(\mu_n)$, we can think of the parton shower as defining a map
%%%
\begin{equation}
\label{eq:MC_map}
\Phi_n \mathop{\longrightarrow}^{\rm ~pheno.\,model~} \MC_n(\mu_n)
\,.\end{equation}
%%%

To eliminate double-counting, we simply have to exclude those regions from $\df\Phi_n$ that can be reached by acting the shower on $\df\Phi_i$ starting from the scale $\mu_i$ for any $i<n$. As illustrated in \fig{fig:phasecartoona}, the remaining parts of $\df\Phi_n$ define phase space with a matching scale $\df\Phi_n(\{\mu_i\}_{i<n})$.%
\footnote{%
In \fig{fig:phasecartoona}, $\df\Phi_4(\mu_2, \mu_3)$ strictly speaking only depends on $\mu_3$ since the region of $\df\Phi_4$ mapped out by $\mu_2$ lies completely inside $\mu_3$. In general, this need not be the case, see for example \fig{fig:toyphasespacematch}.} Note that $\df\Phi_n(\{\mu_i\}_{i<n})$ is a function of all the lower-dimensional matching scales $\{\mu_i\}_{i<n}$ and is independent of the matching scale $\mu_n$.

Having solved the issue of double-counting, it is now more convenient to directly talk about ``Monte Carlo space'' $\df\MC_n(\{\mu_i\}_{i\leq n})$, which is defined as $\df\Phi_n(\{\mu_i\}_{i<n})$ plus the collection of all relevant regions excluded from $\df\Phi_{m>n}$, as illustrated in \fig{fig:phasecartoonb}. More precisely, $\df\MC_n(\{\mu_i\}_{i\leq n})$ is the image of $\df\Phi_n(\{\mu_i\}_{i<n})$ under the parton shower map in \eq{eq:MC_map}. To simplify our notation, we will mostly suppress the implicit dependence on the scales $\mu_{i<n}$ and only write $\df\MC_n(\mu_n)$.

By construction, a covering of phase space with no double-counting is given by the map
%%%
\begin{equation}
\label{eq:MC_generator}
\sum_{n=2}^{n_\max} \df\Phi_n(\{\mu_i\}_{i<n}) \mathop{\longrightarrow}^{\rm ~pheno.\,model~} \sum_{n=2}^{n_\max} \df\MC_n(\mu_n)
\,,\end{equation}
%%%
where $n_\max$ will be determined by the maximal available number of external particles in the partonic calculation. As long as the scales $\mu_n$ for $n\leq n_\max$ and the parton shower satisfy certain mild constraints to guarantee no dead zones, \eq{eq:MC_generator} also provides a complete covering of phase space which is one-to-one and onto, \emph{i.e.}\ every region of phase space is covered exactly once:
%%%
\begin{equation}
\label{eq:MC_completemap}
\sum_{n=2}^{n_\max} \df\MC_n(\mu_n) \to \sum_{n=2}^{\infty} \df\Phi_n
\,.\end{equation}
%%%

Making the dependence on $n_\max$ explicit, the \geneva\ master formula is
%%%
\begin{equation}
\label{eq:master_generic}
\df\sigma = \sum_{n=2}^{n_\max} \lvert\cM_n(\mu_n)\rvert^2 \, \df\MC_n(\mu_n)
\,.\end{equation}
%%%
A crucial point is that an $n$-body partonic calculation $\lvert\cM_n(\mu_n)\rvert^2$ is defined just as a function of $\Phi_n$, but the extrapolation to all of $\df\MC_n(\mu_n)$ is provided by the phenomenological model. In principle, all \eq{eq:master_generic} requires is a phase space generator that can generate scale-dependent phase space $\df\Phi(\mu)$. Such a generator could be built out of existing Monte Carlo tools by including a phase space veto. Because $\df\Phi(\mu)$ implicitly depends on the precise choice of phenomenological model, doing so might be nontrivial, but there is no conceptual difficulty. Once such a generator exists, we can think entirely in the language of partonic calculations, but still generate fully hadronized events with the aid of QCD phenomenological models.

%===============================================================================
\subsection{Variable Resolution Scale}
\label{subsec:PSwithmatching}
%===============================================================================

\begin{figure}
\includegraphics[scale=0.7]{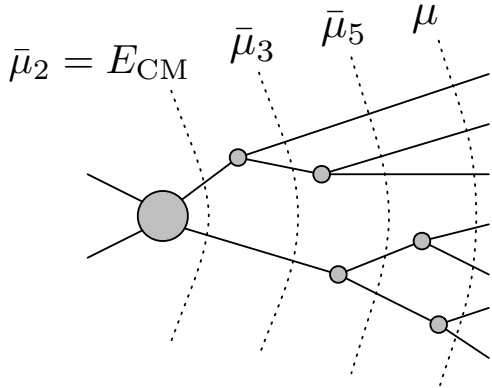}
\caption{Phase space with a variable resolution scale. The shaded blobs represent one possible phase space projection map as in \eq{eq:PSprojection} based on successive $2 \to 1$ recombinations. At the matching scale $\mu$, the event has 7 final state partons. Going from the matching scale $\mu$ to the resolution scale $\bar{\mu}_5$ maps a $7$-body configuration to a $5$-body configuration. Further increasing the scale to $\bar{\mu}_3$ and $\bar{\mu}_2 = E_\mathrm{CM}$ yields a $3$-body and eventually a $2$-body configuration. The ability to project a given point of phase space to a lower-dimensional phase spaces allows for partonic calculations $\lvert \cM(\mu,\bar{\mu}, \ldots) \rvert^2$ that depend on multiple scales $\{\mu\}$.}
\label{fig:truncation}
\end{figure}

When discussing strategies for calculating the partonic result $\lvert\cM(\mu)\rvert^2$, it will prove very useful to define a projection map from $n$-body phase space to $m$-body phase space
%%%
\begin{equation}
\label{eq:PSprojection}
\Phi_n \to \Phi_{m} \qquad \text{with} \qquad m < n
\,.\end{equation}
%%%
This phase space projection map will be the key to creating our NLO/LO/LL merged sample. Such a map can be achieved by defining a resolution scale $\bar{\mu}$, which would resolve more and more partons as the scale is lowered. Note that this scale does not have to coincide with the scale $\mu$ we use to divide up phase space.

As we raise the scale $\bar{\mu}$, the projection map will cluster together two partons once their ``distance'' becomes smaller than $\bar{\mu}$. The phase space map given in \eq{eq:PSprojection} will give $\Phi_n \to \Phi_{m}$ at the resolution scale $\bar{\mu}_m$. As an example, consider the event shown in Fig.~\ref{fig:truncation}, where at the matching scale $\mu$ there are 7 resolved partons. At the scale $\bar{\mu}_5$, two pairs of particles have been clustered together such that the event has 5 resolved partons. Similarly, at the scales $\bar{\mu}_3$ and $\bar{\mu}_2 = E_{\rm CM}$, we have 3 and 2 resolved partons, respectively.

There are several such maps available in the literature including the $k_T$ algorithm~\cite{Catani:1991hj, Ellis:1993tq} and the momentum mappings often found in dipole showers~\cite{Gustafson:1987rq, Lonnblad:2001iq}. In fact, any recursive cluster-based jet algorithm is an example of such a map, but to be useful, \eq{eq:PSprojection} has to respect the symmetry and singularity structure of QCD, which means that the clustering procedure should respect flavor, and the scale $\bar{\mu}$ should roughly determine the ``soft-collinear distance''. Such maps give a sequence of resolution scales, and in the past, it has been mainly the ability to obtain such a sequence of scales which has been used in the literature. An example is the CKKW procedure \cite{Catani:2001cc}, which uses a flavor-aware variant of the $k_T$ algorithm to identify the scales needed to implement leading-logarithmic improvements to tree-level calculations. In \GenEvA, the actual four-momenta determined by \eq{eq:PSprojection} are also important, and we crucially assume that the projection \eq{eq:PSprojection} returns a set of \emph{on-shell} four-momenta, such that partonic calculations can be defined in terms of
%%%
\begin{equation}
\lvert \cM_n(\Phi_n, \mu_n; \Phi_m, \bar{\mu}_m; \ldots) \rvert^2
\,.\end{equation}
%%%

In the \geneva\ algorithm \cite{genevatechnique}, we will show that we can use the evolution variable of the parton shower itself as the resolution variable. Thus, the projection in \eq{eq:PSprojection} will effectively be the inverse of the parton shower map in \eq{eq:MC_map}.\footnote{Strictly speaking, \eq{eq:PSprojection} will be a pseudo-left-inverse of \eq{eq:MC_map}, meaning that running the shower and applying the projection will yield the identity map up to discrete ambiguities.} This reversal property of the projection means that we have a consistent way to not just increase but also \emph{decrease} the matching scale at will. This could potentially be very important because it allows one to \emph{raise} the value of $n_\max$ in \eq{eq:master_generic} and thus use additional partonic calculations after the events have been generated or even passed through a detector simulation. In fact, a reversable projection can be used to \emph{define} $\df\MC(\mu)$, as any given phase space integration region $\df \Phi_n$ can be mapped to the proper $\df\MC_m(\mu_m)$ region as needed in \fig{fig:phasecartoonb}. The presence of a reversable phase space projection is one of the reasons why the \geneva\ algorithm is efficient and versatile, but this reversibility property is not strictly needed for the \geneva\ framework discussed here.

%===============================================================================
\subsection{Strategy for Partonic Calculations}
%===============================================================================

In the language of \eq{eq:mastergeneva}, the strategy to systematically improve Monte Carlo is simply to eliminate phenomenological models by pushing the scale $\mu$ as low as possible (ideally to $\Lambda_{\rm QCD}$) with improved QCD partonic calculations $\lvert\cM(\mu)\rvert^2$. Of course, one would also like to refine and tune phenomenological models based on experimental and theoretical input, but that is not the focus of the present work, where we are concerned primarily with improving the perturbative description of QCD.

How does one determine a partonic calculation $\lvert\cM(\mu)\rvert^2$? As discussed, the choice can be made by physics considerations alone without worrying about algorithmic implications,\footnote{In practice, the availability of fast numerical methods to evaluate $\lvert\cM(\mu)\rvert^2$ might affect the precise form of the partonic calculation, especially in the presence of loop expressions.} and the physics goal is to determine an expression for $\lvert\cM(\mu)\rvert^2$ that is formally correct both in a perturbative expansion in $\alpha_s$ and in a kinematic expansion about $r$, where $r$ denotes the ratio of two kinematic scales, such as the invariant mass between two partons compared to the center-of-mass energy. The details of how to define partonic calculations are discussed in subsequent sections, and we give just a schematic overview here.

As is well known, an expansion to fixed order in $\alpha_s$ does not yield a good description of perturbative QCD over all of phase space. The presence of double-logarithmic terms $(\alpha_s \log^2 r)^n$ at each order in the perturbative expansion invalidates the fixed-order expansion for small values of $r$. These double-logarithmic terms can be resumed to all orders in perturbation theory by doing a soft-collinear expansion of QCD. In this paper we will only work to leading-logarithmic order, leaving a treatment of subleading logarithms for future work.

The most naive partonic calculation is one that completely avoids the issue of logarithmic resummation and only uses tree-level (LO) matrix elements
%%%
\begin{equation}
\lvert\cM^\LO_n(\mu)\rvert^2 \simeq \lvert\cM_n^{\rm tree}\rvert^2
\,.\end{equation}
%%%
Note that the infrared divergences in the tree-level diagrams are regulated because phase space has a matching scale $\mu$ that imposes a phase space restriction.

Because of the double-logarithmic sensitivity to the scale $\mu$ from running the phenomenological model, we want to supplement $\lvert\cM^{\rm LO}(\mu)\rvert^2$ with the correct leading-logarithmic (LL) $\mu$ dependence. As observed by CKKW~\cite{Catani:2001cc}, this can be accomplished by multiplying the tree-level diagram with a $\mu$-dependent Sudakov factor $\Delta$,
%%%
\begin{equation}
\lvert\cM^{\rm LO/LL}_n(\mu)\rvert^2 \simeq \lvert\cM_n^{\rm tree}\rvert^2 \, \Delta(E_{\rm CM}, \mu)
\,.\end{equation}
%%%
Because the Sudakov factor can be formally expanded as $\Delta = 1 + \ord(\alpha_s)$, this LO/LL partonic cross section is simultaneously correct to leading order in an $\alpha_s$ expansion and to leading-logarithmic order in the soft-collinear limit.

The story becomes more complicated at next-to-leading order (NLO) because of the need to cancel infrared divergences between $n$-body virtual and $(n+1)$-body real-emission diagrams. Though in practice, one typically uses Catani-Seymour subtractions~\cite{Catani:1996jh, Catani:1996vz} to calculate NLO observables, for the moment, we will use a slicing method~\cite{Fabricius:1981sx, Kramer:1986mc, Bergmann:1989zy, Giele:1991vf, Giele:1993dj, Harris:2001sx}, which is conceptually simpler. Schematically, one writes
%%%
\begin{equation}
\lvert\cM^{\rm NLO}_n(\mu)\rvert^2 \simeq \lvert\cM_n^{\rm real}\rvert^2 + \lvert\cM_n^{\rm virtual}\rvert^2 + \int_{\mu} \, \lvert\cM_{n+1}^{\rm real} \rvert^2
\,,\end{equation}
%%%
where the integral $\int_{\mu}$ represents slicing part of the $(n+1)$-body real-emission phase space to cancel the infrared divergences in the $n$-body virtual diagrams. Any divergences in the $n$-body real emission are regulated by the matching scale $\mu$ as in the tree-level case, but the effect of those divergences can be used again in a slicing scheme to cancel infrared divergences in the $(n-1)$-body virtual diagrams.

Adding leading-logarithmic information to the NLO calculation is much more involved. A naive approach is simply incorrect,
%%%
\begin{equation}
\lvert\cM^{\rm NLO/LL}_n(\mu)\rvert^2 \neq \lvert\cM^{\rm NLO}_n(\mu)\rvert^2 \, \Delta(E_{\rm CM}, \mu)
\,,\end{equation}
%%%
because the $\ord(\alpha_s)$ pieces in the Sudakov factor change the $\ord(\alpha_s)$ behavior of the fixed-order calculation. An achievement of MC@NLO~\cite{Frixione:2002ik} was to figure out an expression for $\lvert\cM^{\rm NLO/LL}(\mu)\rvert^2$ that simultaneously has NLO and LL accuracy. In Ref.~\cite{Frixione:2002ik}, a specific algorithmic method to implement $\lvert\cM^{\rm NLO/LL}(\mu)\rvert^2$ was used that generated events with sometimes negative weights, but in the context of \GenEvA\ the same $\lvert\cM^{\rm NLO/LL}(\mu)\rvert^2$ expression can be used to generate NLO/LL accurate events with manifestly positive weights.

At this point, \GenEvA\ is simply reproducing known results from PS/ME merging (LO/LL) and PS/NLO merging (NLO/LL) in the language of partonic calculations $\lvert\cM(\mu)\rvert^2$. The advantage of this language is that it makes it straightforward to merge these two results into a partonic calculation that has combined NLO/LO/LL accuracy. As already mentioned in \sec{subsec:PSwithmatching}, and as we will explain in more detail in \sec{subsec:importancetruncation}, one can define a scale $\bar{\mu}_m > \mu$ at which only $m < n$ partons are resolved. This can be achieved using the phase space projection map of \eq{eq:PSprojection}. If one has LO/LL information for $n$ partons, and additional NLO/LL information for $m$ partons,\footnote{Here, NLO/LL information for $m$ partons means that one has performed an $(m-1)$-body one-loop calculation.} one can supplement the LO/LL partonic cross section with this higher-order information, using
%%%
\begin{equation}
\lvert\cM^{\rm NLO/LO/LL}_n(\mu)\rvert^2
= \lvert\cM^{\rm LO/LL}_n(\mu)\rvert^2 \, \frac{\displaystyle \lvert\cM^{\rm NLO/LL}_m(\bar{\mu}_m)\rvert^2}{\displaystyle \lvert\cM^{\rm LO/LL}_m(\bar{\mu}_m)\rvert^2}
\,.\end{equation}
%%%
This result describes NLO observables correct to NLO, LO observables correct to LO, all with the correct LL behavior.

Though we do not demonstrate an NNLO/NLO/LO/NLL/LL merged sample in the present work, whether or not such a sample can be built in principle only depends on whether it is possible to derive an expression for $\lvert\cM(\mu)\rvert^2$ that simultaneously allows N$^i$LO observables to be correct to N$^i$LO and N$^j$LL observables to be correct to N$^j$LL. Of course, since we will interface with a phenomenological model that might not have the correct NLL behavior, we now have to push $\mu$ as low as possible to make sure that our N$^i$LO/N$^j$LL calculation describes as much of phase space as possible. Using the language of partonic calculations, we see that how far Monte Carlo can be improved on the perturbative side is in principle a pencil-and-paper question that can be separated from algorithmic issues.

%===============================================================================
\subsection{Summary of GenEvA}
\label{subsec:summary}
%===============================================================================

\begin{figure}
\includegraphics[scale=0.7]{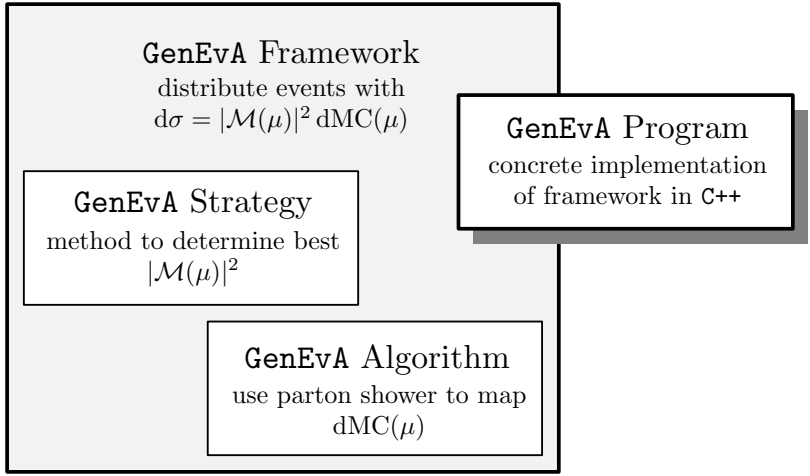}
\caption{The four different meanings of ``\GenEvA''. The \GenEvA\ \textbf{framework} is an umbrella for the idea of distributing partonic calculations with a matching scale, $\lvert\cM(\mu)\rvert^2$, across phase space with a matching scale, $\df\MC(\mu)$. The \GenEvA\ \textbf{strategy} asserts that different QCD expansions can be merged into a single partonic quantity $\lvert\cM(\mu)\rvert^2$. The \GenEvA\ \textbf{algorithm} is a method for using a parton shower as a phase space generator. The \GenEvA\ \textbf{program} is a concrete implementation of the \GenEvA\ framework employing the \GenEvA\ algorithm.}
\label{fig:definition}
\end{figure}

Just as ``fixed-order calculation'' and ``parton shower'' referred to various concepts, \GenEvA\ itself refers to four different concepts, summarized in \fig{fig:definition}. \GenEvA\ is:
%%%
\begin{enumerate}
\item A Monte Carlo \textbf{framework} based on the idea of generating events according to
%%%
\begin{equation}
\df \sigma = \lvert\cM(\mu)\rvert^2 \, \df\MC(\mu)
\end{equation}
%%%
by distributing a generic partonic calculation with a matching scale, $\lvert\cM(\mu)\rvert^2$, using a generic phase space generator with a matching scale, $\df\MC(\mu)$. This framework could be implemented using traditional Monte Carlo tools without ever referencing the \GenEvA\ phase space algorithm.
\item A \textbf{strategy} for improving Monte Carlo based on merging different QCD expansions to determine the best partonic calculations $\lvert\cM(\mu)\rvert^2$. This strategy will be the subject of the remaining sections and yields a formula for an NLO/LO/LL merged calculation and should be generalizable to NNLO or NLL.
\item An \textbf{algorithm} for generating phase space with a variable matching scale, $\df\MC(\mu)$, by analytically reweighting a parton shower to fixed-multiplicity phase space. This algorithm is detailed in Ref.~\cite{genevatechnique}.
\item A computer \textbf{program} that gives a concrete implementation of the \GenEvA\ framework by using the \GenEvA\ algorithm. An alpha version of this software is available from the authors upon request, and is used here to show the results of the NLO/LO/LL calculation.
\end{enumerate}
%%%
We emphasize that while the \geneva\ algorithm for generating phase space is quite novel, the important physics behind the \GenEvA\ framework is captured not by the phase space integration $\df\MC(\mu)$ but by the \geneva\ strategy to define the partonic calculation $\lvert\cM(\mu)\rvert^2$. Though we need a phase space generator with a variable matching scale in order to create an NLO/LO/LL merged sample, the physics behind this merging comes from finding a suitable form of $\vert\cM(\mu)\rvert^2$.

\begin{figure}
\includegraphics[scale=0.58]{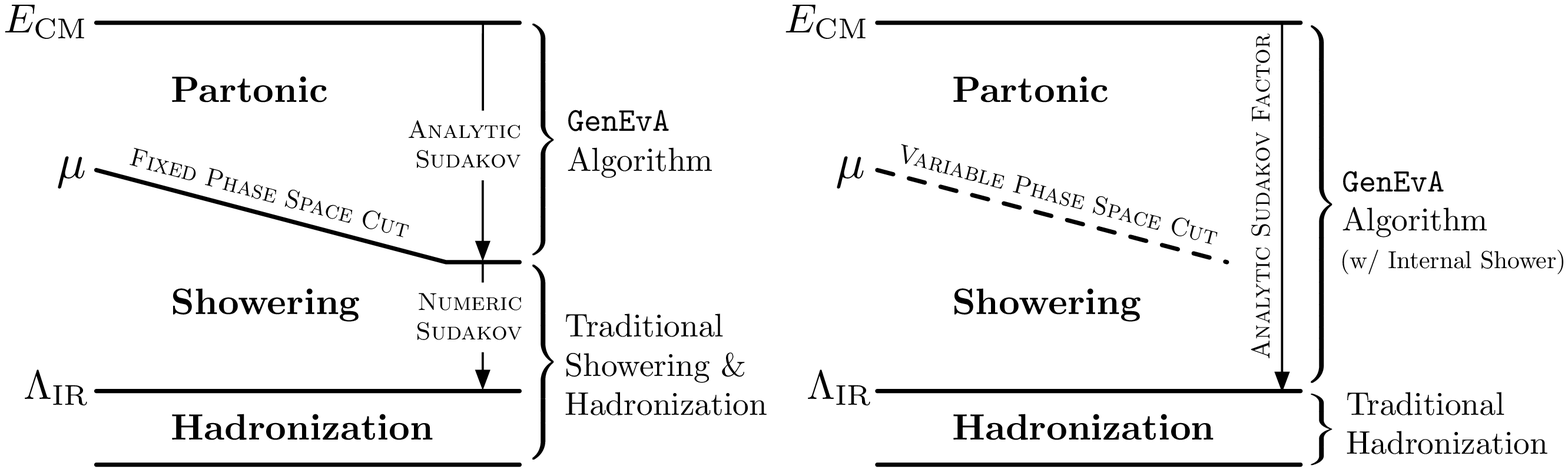}
\caption{The three regimes of event generation. Left panel: The \GenEvA\ algorithm \cite{genevatechnique} could be used only in the partonic regime, with the showering regime covered by a traditional, already tuned, showering program. Right panel: Alternatively, both partonic and showering regimes can be covered by the \GenEvA\ algorithm using its internal parton shower. In this case the matching scale between the partonic and showering regimes remains freely adjustable through the reversable phase space projection. While these two approaches differ in how Sudakov factors are calculated, they both give results that are accurate to leading-logarithmic order.}
\label{fig:regimes}
\end{figure}

To summarize the complete \GenEvA\ approach to Monte Carlo, it is instructive to separate the phenomenological model into a parton shower component and a hadronization component. If we now divide event generation into three regimes as in \fig{fig:regimes}---partonic, showering, and hadronization---then the \GenEvA\ framework improves our ability to describe the partonic regime through improved partonic calculations $\lvert\cM(\mu)\rvert^2$. The showering regime can either be described by a traditional showering program such as \texttt{Pythia}~\cite{Sjostrand:2000wi, Sjostrand:2006za} or \texttt{Herwig}~\cite{Corcella:2000bw, Gieseke:2003hm, Gieseke:2006ga} or by the internal parton shower used in the \GenEvA\ algorithm.\footnote{As we discuss more in the companion paper \cite{genevatechnique}, the reason to use a traditional algorithm is that they have already been tuned to data, though there are some technical issues regarding evolution variables that would have to be overcome to use this option. If the \GenEvA\ algorithm is used in the showering regime, then the user gains the freedom to adjust the scale $\mu$ used in the partonic calculations even after detector simulation.} Regardless of which method is used, there will be no leading-logarithmic $\mu$ dependence in the event generation as long as $\rvert\cM(\mu)\vert^2$ contains the correct Sudakov factors and the same definition of $\mu$ is used in the partonic and showering regimes. Finally, the hadronization regime uses some model of the strong interactions to hadronize the obtained partons into the hadrons observed in the collider, and \GenEvA\ inherits the same smooth showering/hadronization interface as traditional frameworks.

%%%%%%%%%%%%%%%%%%%%%%%%%%%%%%%%%%%%%%%%%%%%%%%%%%%%%%%%%%%%%%%%%%%%%%%%%%%%%%%%
\section{GenEvA in a Toy Example}
\label{sec:toy}
%%%%%%%%%%%%%%%%%%%%%%%%%%%%%%%%%%%%%%%%%%%%%%%%%%%%%%%%%%%%%%%%%%%%%%%%%%%%%%%%

To see how the \GenEvA\ framework works in practice, it is instructive to first consider a toy example that includes all of the issues involved in constructing partonic calculations $\lvert\cM(\mu)\rvert^2$, without the technical complications introduced by full QCD. In particular, the toy theory will only have a single unambiguous scale. In actual QCD, the scale at which an emission occurs will be ambiguous, both because of the ambiguity in the choice of evolution variable and the ambiguity of how to pair together daughter particles to form mother particles. We will see how to deal with these additional complications in \sec{sec:QCDfirst}.

To build the analogy with QCD, we need an example theory with both analytic cross-section information as well as a separate phenomenological description. We will use a slightly modified version of the toy example introduced in Ref.~\cite{Frixione:2002ik}. After reviewing the toy theory, we will describe ``Monte Carlo space'' $\df\MC(\mu)$ and the kinds of partonic calculations $\vert\cM(\mu)\vert^2$ that can be distributed across Monte Carlo space.

%===============================================================================
\subsection{Review of Toy Theory}
%===============================================================================

\begin{figure}
\includegraphics[scale=0.7]{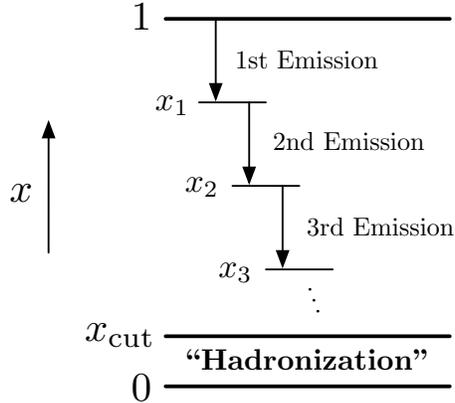}
\caption{Structure of the toy theory. Starting at the scale $x=1$, the system can generate ``radiation'' at scales $x_1$, $x_2$, $\ldots$, with the $x_i$ values always decreasing. At the scale $x_\cut$, the perturbative radiation matches onto a ``hadronization'' scheme.} \label{fig:toytheory}
\end{figure}

Consider a system that can radiate off ``photons'' as in \fig{fig:toytheory}. After emission of a photon, the system has energy $x$ left for further emissions with
%%%
\begin{equation}
0 \leq x < x_\start \leq 1
\,,\end{equation}
%%%
where $x_\start$ was the energy of the system before the emission (with $x_\start = 1$ initially), so the energy of the radiated photon is $x_\start - x$.\footnote{%
Unlike Ref.~\cite{Frixione:2002ik}, we take $x$ to be the energy of the system available for radiation, rather than the energy of the emitted photon. Thus, $x$ is continuously decreasing and plays the role of a parton shower evolution variable. Note that this means that this model has the somewhat unusual property of having singularities if the photons radiate with maximum energy, rather than with minimum energy.}
The different phase space integration ranges for $n$ emissions are given by
%%%
\begin{equation}
\begin{aligned}
\label{eq:toyphasespace}
\df\Phi_0 &= 1
\,,\qquad
\df\Phi_1 = \df x_1
\,,\qquad
\df\Phi_2 = \df x_1 \, \df x_2 \, \theta(x_1 > x_2)
\,,\qquad \ldots \,,
\\
\df\Phi_n &= \df x_1\, \prod_{i=2}^n \df x_i \, \theta(x_{i-1} > x_i)
\,,\end{aligned}
\end{equation}
%%%
and the restriction $0 \leq x_i \leq 1$ is implied.

The Born ``cross section'' corresponds to no emissions at zeroth order in perturbation theory. At first order in perturbation theory, there is a contribution to the cross section from real and virtual photon emissions. We define
%%%
\begin{equation}
\label{eq:toycrosssection}
\frac{\df \sigma_B}{\df x} = B \, \delta(x)
\,,\qquad
\frac{\df \sigma_V}{\df x} = \a \left( \frac{B}{2\epsilon} + V \right) \delta(x)
\,, \qquad
\frac{\df \sigma_R}{\df x} = \a B \,\frac{R(x)}{x^{1+2\epsilon}}
\,,\end{equation}
%%%
where we are doing a perturbative expansion in $\a$, and $\epsilon$ is the dimensional regularization parameter in $d=1-2\epsilon$ dimensions.
We need
%%%
\begin{equation}
\label{eq:Rlimit}
\lim_{x \to 0} R(x) = 1
\,,\end{equation}
%%%
such that the integral over the real contribution cancels the infrared divergence in the virtual contribution,
%%%
\begin{equation}
\int_0^1 \! \df x \, \frac{R(x)}{x^{1+2\epsilon}}
= \int_0^1 \! \df x \, \frac{1}{x^{1+2\epsilon}} + \int_0^1 \! \df x \, \frac{R(x)-1}{x}
= -\frac{1}{2\epsilon} + \mathrm{finite}
\,,\end{equation}
%%%
yielding the total cross section to $\ord(\a)$
%%5
\begin{equation}
\label{eq:toysigmaNLO}
\sigma_\mathrm{NLO} = B + \a V + \a B \int_0^1 \! \df x\, \frac{R(x) - 1}{x}
\,.\end{equation}
%%%

Next, we define a function $Q(x)$ which plays the role of the splitting function in QCD and reproduces the full radiation in the singular limit $x \to 0$. That is, from \eq{eq:Rlimit} we require
%%%
\begin{equation}
\label{eq:Qlimit}
\lim_{x \to 0} Q(x) = 1
\,,\end{equation}
%%%
such that
%%%
\begin{equation}
\label{eq:toyRQfinite}
\lim_{x \to 0} \frac{R(x) - Q(x)}{x} = \mathrm{finite}
\,.\end{equation}
%%%
The simplest choice would be $Q(x) =1$. In general, a nontrivial dependence on $x$ is possible, in which case we require $Q(x)$ to be positive between $x=0$ and $x=1$. The function $Q(x)$ can be used to define
%%%
\begin{equation}
\label{eq:toysudakov}
\Delta_Q(x_1,x_2) = \exp \biggl[ - \a \int_{x_2}^{x_1} \! \df x\, \frac{Q(x)}{x} \biggr]
\,,\end{equation}
%%%
which plays the role of the Sudakov factor, and which we will often just call $\Delta$ for simplicity.
Finally, the splitting function together with the Sudakov factor can be used to define a parton shower with the differential probability to branch (emit a photon) at $x$ given by
%%%
\begin{equation}
\label{eq:toyshower}
\df \mathcal{P}(x) = \a\, \frac{Q(x)}{x}\, \Delta(x_\start,x)\, \df x
\,.\end{equation}
%%%
As before, $x_\start$ is the energy of the system prior to the branching. We also define a lower cutoff $x_\cut$ on the parton shower, which corresponds to the scale at which a QCD parton shower would be matched onto a hadronization model. In the following, the parton shower defined by \eq{eq:toyshower} will serve as the phenomenological model in our toy theory.

%===============================================================================
\subsection{Phase Space with a Matching Scale}
\label{subsec:toyphasespace}
%===============================================================================

\begin{figure}
\includegraphics[scale=0.7]{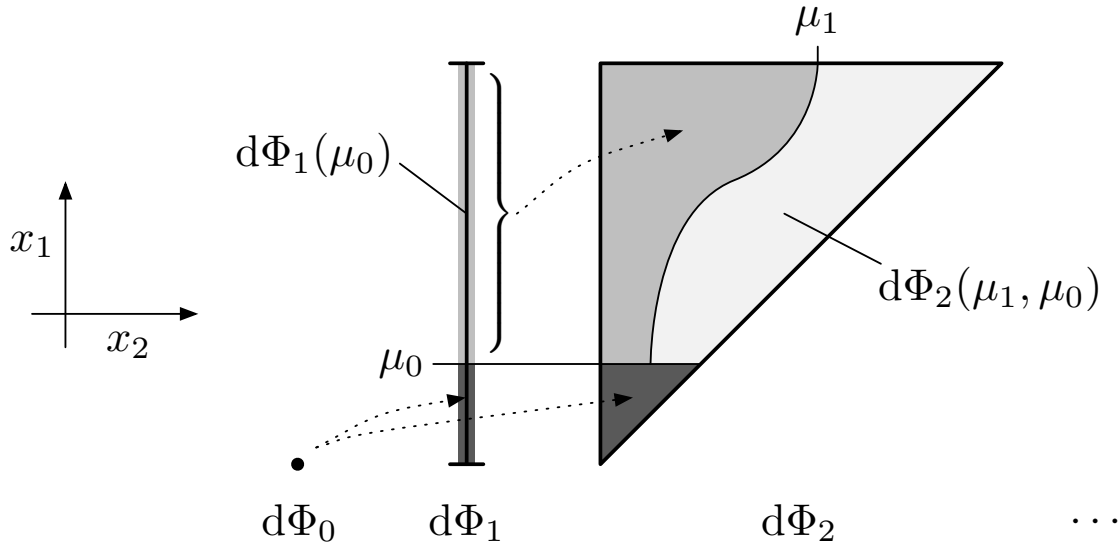}
\caption{The scale-dependent phase space $\df\Phi_n(\{\mu_i\}_{i<n})$ in the toy theory. The scale $\mu_0$ is a fixed number, whereas $\mu_1 \equiv \mu_1(x_1)$ can be a function of $x_1$. $\df\Phi_1(\mu_0)$ is given by $x_1 > \mu_0$, and $\df\Phi_2(\mu_0,\mu_1)$ is given by $x_1 > \mu_0$ and $\mu_1(x_1) < x_2 < x_1$.}
\label{fig:toyphasespacematch}
\end{figure}

As discussed in \sec{subsec:newphasespace}, to avoid any double-counting between the parton shower started from a system with $n$ emissions and one with $m>n$ emissions, we want to define phase space in the presence of matching scales. Because we are working with a single-scale system, it is straightforward to amend the ordinary phase space $\df\Phi_n$ in \eq{eq:toyphasespace} with a set of matching scales $\{\mu_i\}_{i<n}$. We have
%%%
\begin{equation}
\label{eq:toyphasespacewithmatch}
\df\Phi_n(\{\mu_i\}_{i<n})
= \df x_1\,\theta(x_1 > \mu_0) \prod_{i=2}^n \df x_i \, \theta(x_{i-1} > x_i) \, \theta(x_i > \mu_{i-1})
\,,\end{equation}
%%%
where the restriction $0 \leq x_i \leq 1$ is again implied. An illustration of \eq{eq:toyphasespacewithmatch} for $n \leq 2$ is shown in \fig{fig:toyphasespacematch}. Before going on to explain \eq{eq:toyphasespacewithmatch} in more detail, we want to highlight a few main points:
%%%
\begin{enumerate}
\item For each $n$, the scale $\mu_n$ is the scale at which we start the ``parton shower'' when acting on $n$-emission phase space $\df\Phi_n$. In general, $\mu_n \equiv \mu_n(\Phi_n)$ can be a function of $\Phi_n \equiv \{ x_i \}_{i \leq n}$. This can lead to some counter-intuitive but still well-defined situations. The example shown in \fig{fig:toyphasespacematch} has a seemingly pathological situation where $\mu_1(x_1) < \mu_0$ for certain values of $x_1$, but this is perfectly consistent and even potentially useful. The only restriction on the functional form of $\mu_n(\Phi_n)$ for \eq{eq:toyphasespacewithmatch} to make sense is that $\mu_n(\Phi_n) \leq x_n$.
\item In accordance with our general discussion in \sec{subsec:newphasespace}, the restriction $x_i>\mu_{i-1}$ in \eq{eq:toyphasespacewithmatch} cuts out the region of $\df\Phi_n$ for any $n\geq i$ that gets populated by the parton shower when acting on $\df\Phi_{i-1}$, thus guaranteeing that there is no phase space double-counting.

\item If we have only calculations for up to $n_\max$ emissions, then to avoid dead zones in phase space, we have to start the parton shower at $\mu_{n_\max}(\Phi_{n_\max}) = x_{n_\max}$ when acting on the $n_\max$-emission contribution, such that $\df\Phi_n$ for $n > n_\max$ is completely covered by the parton shower. The easiest way to get full phase space coverage with no dead zones is to simply use a single fixed matching scale $\mu$, with $\mu_n(\Phi_n) = \mu$ for $n < n_\max$ and $\mu_{n_\max}(\Phi_{n_\max}) = x_{n_\max}$. This also avoids some of the seeming pathologies mentioned above.
\end{enumerate}
%%%

To explain \eq{eq:toyphasespacewithmatch} in more detail, we take a closer look at how the toy parton shower works. Let $\MC_{x_1,x_2, \ldots}(x)$ denote an event for which emissions have occurred at $x_1$, $x_2$, $\ldots$, and the system has energy $x$ left to radiate. An event in its initial state is denoted by $\MC(1)$, which means that no photons have been emitted, and the total energy available for radiation is $x = 1$. The event then evolves according to the parton shower \eq{eq:toyshower}, giving
\begin{align}
\label{eq:toyMCdef}
\MC(1)
&\to \Delta(1, x_\cut) \,\MC(x_\cut)
+ \int_{x_\cut}^1\!\df x_1\, \biggl[\a\, \frac{Q(x_1)}{x_1}\, \Delta(1, x_1) \biggr] \MC_{x_1}(x_1)
\nn\\
&\to \Delta(1, x_\cut) \,\MC(x_\cut)
+ \int_{x_\cut}^1\!\df x_1\, \biggl[\a\, \frac{Q(x_1)}{x_1}\, \Delta(1, x_1) \biggr] \Delta(x_1, x_\cut) \,\MC_{x_1}(x_\cut)
\nn\\ &\quad
+ \int_{x_\cut}^1 \! \df x_1 \int_{x_\cut}^{x_1} \! \df x_2
\biggl[\a\, \frac{Q(x_1)}{x_1}\, \Delta(1, x_1) \biggr] \biggl[\a\, \frac{Q(x_2)}{x_2}\, \Delta(x_1, x_2) \biggr] \MC_{x_1, x_2}(x_2)
\nn\\
&\to \cdots
\,.\end{align}
The first term denotes the possibility that no emissions happen down to the ``hadronization scale'' $x_\cut$, which occurs with the no-branching probability $\Delta(1,x_\cut)$ and leaves an event with energy $x_\cut$ left for radiation. The second term denotes the possibility that exactly one emission happens above the scale $x_\cut$ and so on. The Sudakov factors are defined in such a way that the total probability is conserved,
%%%
\begin{equation}
\int_{x_\cut}^1\!\df x\,\a\,\frac{Q(x)}{x}\,\Delta(1,x) + \Delta(1,x_\cut) = 1
\,,\end{equation}
%%%
so if we start with some number of events $\MC(1)$, the number of events after showering will still be the same.

Now imagine running the parton shower starting at the scale $\mu_n(\Phi_n)$ on an event for which $n$ emissions have occurred at $\Phi_n = \{x_1,\ldots, x_n\}$. In this case
%%%
\begin{align}
\label{eq:toymcpoint}
\MC_{x_1,\ldots, x_n}(\mu_n)
&\to \Delta(\mu_n, x_\cut)\, \MC_{x_1,\ldots, x_n}(x_\cut)
\nn \\ &\quad
+ \int_{x_\cut}^{\mu_n} \! \df x_{n+1}
\biggl[\a \frac{Q(x_{n+1})}{x_{n+1}}\, \Delta(\mu_n, x_{n+1}) \biggr]
\MC_{x_1,\ldots, x_n, x_{n+1}}(x_{n+1})
\nn\\
\to \cdots
\,.\end{align}
%%%
By assumption, the emissions are ordered with $x_{n+1} < x_n$, so for \eq{eq:toymcpoint} to be consistent, we need to choose $\mu_n(\Phi_n) \leq x_n$. If in addition we consider an $(n+1)$-emission event $\MC_{x_1,\ldots, x_n, x_{n+1}}(\mu_{n+1})$, then to avoid double-counting with the $n$-emission event, we have to restrict the range of $x_{n+1}$ to $x_{n+1} > \mu_n$. Similarly, considering the effect of all of the $0$- through $n$-emission samples, to avoid double-counting regions of phase space when running the parton shower, we always need to take $x_{i} > \mu_{i-1}$, as anticipated in \eq{eq:toyphasespacewithmatch}.

\begin{figure}
\includegraphics[scale=0.7]{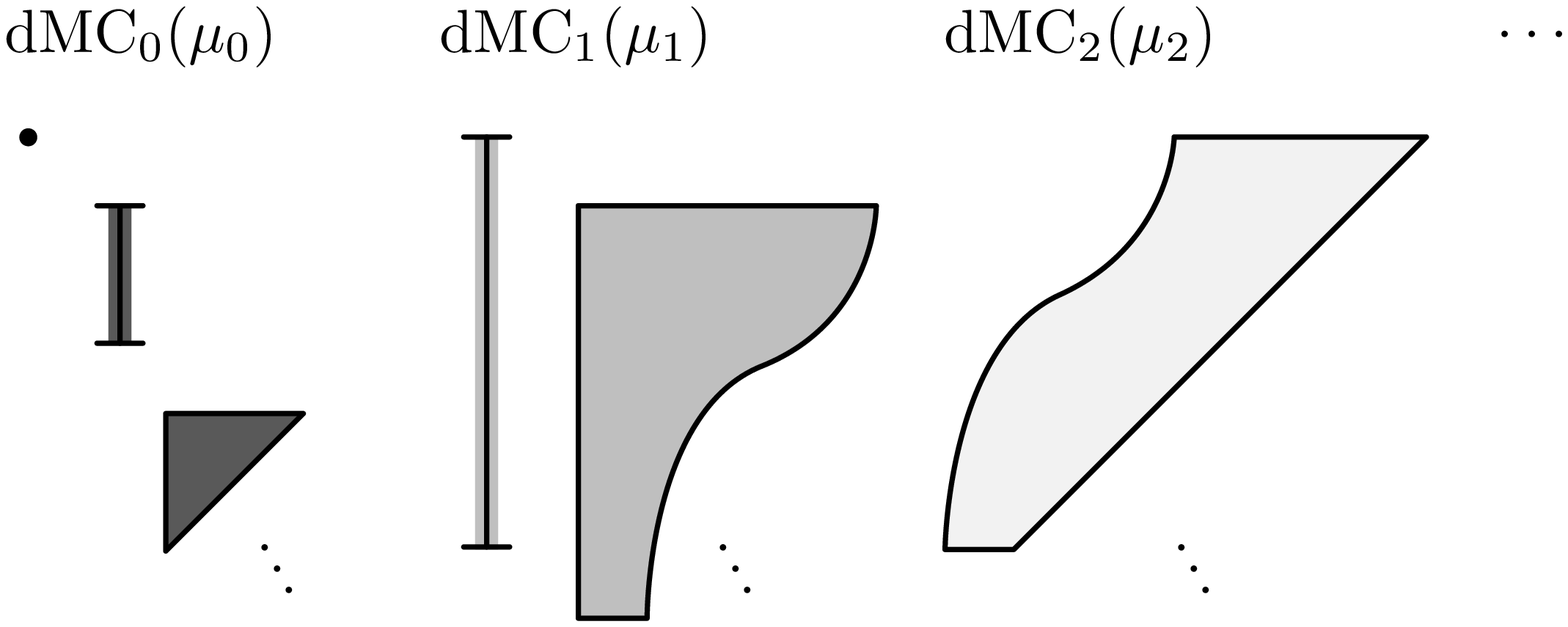}
\caption{Definition of ``Monte Carlo space'' $\df\MC_n(\mu_n)$ in the toy theory, which is obtained by collecting the various parts of $\df\Phi_n$ in \fig{fig:toyphasespacematch}.}
\label{fig:toymontecarlospace}
\end{figure}

Putting together the phase space restriction from \eq{eq:toyphasespacewithmatch} with the parton shower evolution of \eq{eq:toyMCdef}, we can now define differential ``Monte Carlo space'' $\df\MC_n(\mu_n)$ as the result of running the parton shower on $n$-emission phase space $\df\Phi(\{\mu_i\}_{i<n})$:
%%%
\begin{align}
\label{eq:toyMCspace}
\df\MC_n(\{\mu_i\}_{i\leq n})
&= \df\Phi_n(\{\mu_i\}_{i<n})\, \biggl\{ \Delta(\mu_n,x_\cut)
\nn\\ &\quad
+ \df x_{n+1} \, \theta(\mu_n > x_{n+1} > x_\cut)
\biggr[\a\, \frac{Q(x_{n+1})}{x_{n+1}}\, \Delta(\mu_n,x_{n+1}) \biggr] \Delta(x_{n+1}, x_\cut)
\nn\\ &\quad
+ \df x_{n+1} \, \df x_{n+2} \, \theta(\mu_n > x_{n+1} > x_{n+2} > x_\cut )
\biggr[\a\, \frac{Q(x_{n+1})}{x_{n+1}}\, \Delta(\mu_n,x_{n+1}) \biggr]
\nn \\ &\quad\quad
\times \biggl[\a\, \frac{Q(x_{n+2})}{x_{n+2}}\, \Delta(x_{n+1}, x_{n+2}) \biggr] \Delta(x_{n+2}, x_\cut)
\nn\\ &\quad
+ \cdots \biggr\}
\,,\end{align}
%%%
where the restriction $0 \leq x_i \leq 1$ is again implied. Note that the Monte-Carlo space $\df\MC_n(\{\mu_i\}_{i\leq n})$ has variable multiplicity. An illustration of \eq{eq:toyMCspace} for $n\leq 2$ is shown in \fig{fig:toymontecarlospace}.

With the appropriate choice of $\mu_n$, as discussed above,
%%%
\begin{equation}
\sum_{n=0}^{n_\max}\df\MC_n(\mu_n) \to \sum_{n=0}^{\infty} \df\Phi_n(x_\cut)
\end{equation}
%%%
gives a covering of phase space with no dead zones and no double-counting. Here, $\df\Phi_n(x_\cut)$ represents running the hadronization model on the partons at the scale $x_\cut$. The arrow indicates that $\df\MC_n(\mu_n)$ includes additional weight information from the splitting functions and Sudakov factors.

%===============================================================================
\subsection{The Master Formula}
%===============================================================================

A compact master formula for the fully differential cross section coming from our as-of-yet undefined partonic calculation interfaced with the parton shower is given in analogy with \eq{eq:master_generic}:
%%%
\begin{equation}
\label{eq:toymaster}
\df\sigma = \sum_{n=0}^{n_\max} \lvert \cM_n(\mu_n)\rvert^2 \, \df\MC_n(\mu_n)
\,.\end{equation}
%%%
As already mentioned, $\vert\cM_n(\mu_n)\rvert^2$ is a function of $\Phi_n$. More generally, $\lvert\cM_n(\mu_n)\rvert^2$ can also depend on the scales $\{\mu_i\}_{i<n}$ that are implicit in $\df\MC_n(\mu_n)$.

It is at this point that the different concepts behind the parton shower come into play. As an instructive exercise, imagine we only know the Born cross section $B$ and the splitting function $Q(x)$. In that case, the best approximation to the differential cross section we can write down is
%%%
\begin{equation}
\label{eq:toyborn}
\df\sigma = \lvert\cM_0(\mu_0)\rvert^2 \, \df\MC_0(\mu_0)
\qquad\text{with}\qquad
\mu_0 = 1
\,, \quad
\lvert\cM_0(\mu_0 = 1)\rvert^2 = B
\,.\end{equation}
%%%
However, there is an entirely equivalent description of the same physics in terms of the differential cross section
%%%
\begin{equation}
\label{eq:toyfreedom}
\df\sigma
= \lvert\cM_0(\mu_0)\rvert^2 \, \df\MC_0(\mu_0) + \lvert\cM_1(\mu_1)\rvert^2 \, \df\MC_1(\mu_1)
\,,\end{equation}
%%%
where we now have to choose $\mu_1 = x_1$ to avoid dead zones, and the partonic calculations are defined as
\begin{center}
\bf Parton Shower (LL):
\end{center}
\begin{align}
\label{eq:toyshowerinamplitude}
\lvert\cM_0(\mu_0)\rvert^2 &= B\, \Delta(1, \mu_0)
\,, \nn\\
\lvert\cM_1(x_1)\rvert^2 &= \a B\, \frac{Q(x_1)}{x_1}\, \Delta(1, x_1)
\,.\end{align}
%%%
In \eq{eq:toyborn}, we would say that the boundary between the partonic calculation and the parton shower is $\mu = 1$, whereas in \eq{eq:toyshowerinamplitude}, we would say that the boundary is some scale $\mu = \mu_0$. This is an example of the ambiguity of the parton shower having a meaning above or below the scale $\mu$, as discussed in \fig{fig:boxes}. This flexibility in changing the value of $\mu$ is a feature and not a bug, and it comes from the fact that the parton shower is both a phenomenological model based on QCD and a well-defined expansion of QCD. The only difference between the two descriptions are the words and perhaps algorithms we use above and below $\mu$. But the freedom to incorporate a Sudakov factor $\Delta$ into a partonic calculation $\lvert\cM(\mu)\rvert^2$ as needed is what makes it possible to define partonic calculations that include leading-logarithmic resummation.

%===============================================================================
\subsection{Partonic Matrix Elements for the First Emission}
\label{subsec:toyxsec}
%===============================================================================

\begin{table}
\begin{tabular}{r||l|l}
& $\lvert\cM_0(\mu_0)\rvert^2$ & $ x \lvert\cM_1(x)\rvert^2$ \\
\hline\hline
Parton Shower (LL) & $B\, \Delta_Q(1,\mu_0)$ &$\alpha B\, Q(x)\, \Delta_Q(1,x)$ \\
\hline
Tree-Level (LO) & $B$ & $\alpha B\, R(x)$ \\
Sudakov-Improved (LO/LL) & $B\, \Delta_Q(1,\mu_0)$ & $\alpha B\,R(x)\, \Delta_Q(1,x)$ \\
\hline
NLO Slicing (NLO) & $\tilde{B}(\mu_0)$ &$\alpha B\, R(x)$ \\
NLO Subtraction (NLO/LL) & $\bar{B}(\mu_0)\, \Delta_Q(1,\mu_0)$ & $\alpha B\, (R(x) - Q(x))+ \alpha \bar{B}(\mu_0)\, Q(x)\, \Delta_Q(1,x)$ \\
NLO Elegant (NLO/LL) & $\sigma_\NLO\, \Delta_T(1,\mu_0)$ & $\alpha B\, R(x)\, \Delta_T(1,x)$ \\
\hline
\end{tabular}
\caption{Summary of the various possible one-emission partonic calculations in the toy theory. At the one-emission level, all of the theory information can be summarized by a zero-emission calculation $\lvert\cM_0(\mu_0)\rvert^2$ with a matching scale $\mu_0$ and a one-emission calculation $\lvert\cM_1(x)\rvert^2$ with a matching scale $\mu_1 = x$. The definitions of the samples and symbols are given in the text.}
\label{tab:firstemission}
\end{table}

With the toy master formula \eq{eq:toymaster} in hand, we simply need to define the best partonic calculations $\lvert\cM(\mu)\rvert^2$ we can with the available information from the toy theory. Apart from issues of Monte Carlo efficiency, this definition can be dictated by physics considerations alone, as the \GenEvA\ framework can distribute any $\lvert\cM(\mu)\rvert^2$ across $\df\MC(\mu)$. We will start in this subsection by only considering the first emission in our toy theory, writing for simplicity $x\equiv x_1$. The generalization to multiple emission will be given in the next subsection. Keeping in mind the freedom offered by \eq{eq:toyfreedom}, we can define everything in terms of
%%%
\begin{equation}
\lvert\cM_0(\mu_0)\rvert^2 \qquad \text{and} \qquad \vert\cM_1(\mu_1 = x)\rvert^2
\,.\end{equation}
%%%
A summary of the partonic calculations in this subsection is given in Table~\ref{tab:firstemission}.

The most naive use of the analytic cross section information from \eq{eq:toycrosssection} is to choose
%%%
\begin{center}
\bf Tree Level (LO):
\end{center}%
\begin{align}
\label{eq:toytree}
\lvert\cM_0(\mu_0)\rvert^2 &= B
\,,\nn\\
\lvert\cM_1(x)\rvert^2 &= \a B\, \frac{R(x)}{x}
\,.\end{align}
%%%
The total and one-emission cross sections resulting from \eq{eq:toytree} are\footnote{In what follows, $\df\sigma/\df x$ always refers to the inclusive differential cross section obtained by integrating over all values of $x_i$ with $i \ge 2$.}
%%%
\begin{align}
\label{eq:toytree_xsecs}
\sigma &= B + \a B \int_{\mu_0}^1 \!\df x \, \frac{R(x)}{x}
\,,\nn\\
\frac{\df\sigma}{\df x} &= \a B \,\frac{1}{x} \left\{
\begin{aligned}
&R(x) && \qquad \text{for} \quad x > \mu_0 \\
&Q(x)\,\Delta(\mu_0, x) && \qquad \text{for} \quad x < \mu_0
\,,\end{aligned}\right.
\end{align}
%%%
which is what one would obtain if events from a tree-level event generator were passed to a parton shower program.\footnote{%
Actually, \eq{eq:toytree} is slightly better than what one would get by blindly running a tree-level event generator through a parton shower program, because implicit in the definition of $\lvert\cM_0(\mu_0)\rvert^2$ is that the parton shower will run from $\mu_0$ to avoid double-counting. Out of the box, an ordinary parton shower program would start the parton shower at $x=1$ for the $0$-emission piece, which would lead to double-counting.}

The main problem with the choice in \eq{eq:toytree} is that it does not sum the leading-logarithmic contributions, which although technically of higher order in perturbation theory, can be numerically very important. This leads to the logarithmic dependence on the unphysical matching scale $\mu_0$ in both $\sigma$ and $\df\sigma/\df x$ in \eq{eq:toytree_xsecs}. The reason is of course that since the parton shower does include this resummation, it has logarithmic dependence on the starting scale $\mu_0$. Thus, if the tree-level results are combined with the parton shower using \eq{eq:toymaster}, we are left with this logarithmic dependence on $\mu_0$ in \eq{eq:toytree_xsecs}.

To improve the situation, we can resum the leading logarithms in the partonic calculation by supplementing the naive tree-level result with Sudakov factors:
%%%
\begin{center}
\bf Sudakov-Improved (LO/LL):
\end{center}
\begin{align}
\label{eq:toyLOLL}
\lvert\cM_0(\mu_0)\rvert^2 &= B\, \Delta(1,\mu_0)
\,,\nn\\
\lvert\cM_1(x)\rvert^2 &= \a B\, \frac{R(x)}{x}\, \Delta(1,x)
\,.\end{align}
%%%
At leading order in $\a$, this result is identical to the tree-level result \eq{eq:toytree}. The resulting total and one-emission cross sections are now given by
%%%
\begin{align}
\label{eq:toyLOLL_xsecs}
\sigma &= B + \a B \int_{\mu_0}^1 \!\df x \, \frac{R(x)-Q(x)}{x}\, \Delta(1,x)
\,,\nn\\
\frac{\df\sigma}{\df x} &= \a B \,\frac{1}{x} \left\{
\begin{aligned}
&R(x)\,\Delta(1,x) && \qquad \text{for} \quad x > \mu_0 \\
&Q(x)\,\Delta(1,\mu_0)\, \Delta(\mu_0, x) && \qquad \text{for} \quad x < \mu_0
\,.\end{aligned}\right.
\end{align}
%%%
The factor of $\Delta(1, \mu_0)$ in $\df\sigma/\df x$ for $x < \mu_0$ comes from $\lvert\cM_0(\mu_0)\rvert^2$ and cancels the $\mu_0$ dependence in $\Delta(\mu_0, x)$, which is generated by the parton shower. Furthermore, since we know from \eq{eq:toyRQfinite} that $(R(x)-Q(x))/x$ is finite as $x \to 0$, the logarithmic dependence on $\mu_0$ in the total cross section cancels, too.

Since \eq{eq:toyLOLL_xsecs} does not exhibit leading-logarithmic dependence on the matching scale $\mu_0$, \eq{eq:toyLOLL} provides a well-behaved merging between matrix elements and the parton shower. This result is consistent with the observation made in Ref.~\cite{Catani:2001cc} that matrix element results need to be supplemented with Sudakov factors to allow for a merging with parton shower algorithms. What is important, however, is that we did not need a special algorithm to implement \eq{eq:toyLOLL}, because this is just a choice of what partonic matrix elements to use in the master formula \eq{eq:toymaster}. Including the correct Sudakov factors is required by physics considerations alone and, for example, has nothing to do with the fact that the \geneva\ algorithm happens to have an efficient mechanism to implement this LO/LL merged sample.

The total and one-emission cross sections in \eqs{eq:toytree_xsecs}{eq:toyLOLL_xsecs} are correct to $\ord(1)$ and $\ord(\a)$, respectively, which is of course because we only used tree-level information so far. To also get the total cross section correct at $\ord(\a)$ (NLO) we need to improve the partonic calculation by including the virtual corrections to the zero-emission cross section, given by $\df\sigma_V/\df x$ in \eq{eq:toycrosssection}. There are numerous strategies one could consider. If we do not wish to reproduce the leading-logarithmic resummation, and only want to get observables accurate to NLO, then we can implement a naive slicing scheme \cite{Giele:1991vf,Fabricius:1981sx,Kramer:1986mc,Bergmann:1989zy,Giele:1993dj,Harris:2001sx}:
%%%
\begin{center}
\bf NLO Slicing (NLO):
\end{center}
\begin{align}
\label{eq:toyslicing}
\lvert\cM_0(\mu_0)\rvert^2 &= \tilde{B}(\mu_0)
\,,\nn\\
\lvert\cM_1(x)\rvert^2 &= \a B\,\frac{R(x)}{x}
\,,
\end{align}
%%%
where
%%%
\begin{align}
\tilde{B}(\mu_0) &=
B + \a V + \a B \lim_{\epsilon\to0} \biggr[\frac{1}{2\epsilon}
+ \int_0^{\mu_0} \! \df x\, \frac{R(x)}{x^{1+2\epsilon}} \biggl]
\nn\\
&= B + \a V + \a B \int_0^{\mu_0} \! \df x\, \frac{R(x)-1}{x} - \a B\int_{\mu_0}^1 \!\df x\, \frac{1}{x}
\,.\end{align}
%%%
The cross sections following from \eq{eq:toyslicing} are
%%%
\begin{align}
\label{eq:toyslicing_xsecs}
\sigma &= \tilde{B}(\mu_0) + \a B \int_{\mu_0}^1 \!\df x\, \frac{R(x)}{x} = B + \a V + \a B \int_0^1 \! \df x \, \frac{R(x)-1}{x}
= \sigma_\mathrm{NLO}
\,,\nn\\
\frac{\df\sigma}{\df x} &= \a\,\frac{1}{x} \left\{
\begin{aligned}
&B\,R(x) && \qquad \text{for} \quad x > \mu_0 \\
&\tilde{B}(\mu_0)\, Q(x)\, \Delta(\mu_0, x) && \qquad \text{for} \quad x < \mu_0
\,.\end{aligned}\right.
\end{align}
%%%
The total cross section now equals the correct NLO cross section $\sigma_\NLO$ from \eq{eq:toysigmaNLO}
and has no $\mu_0$ dependence. Furthermore, all observables are accurate to NLO in the $\mu_0 \to 0$ limit. However, as in the tree-level result \eq{eq:toytree}, the leading logarithmic dependence is not resummed, meaning that the one-emission cross section $\df\sigma/\df x$ does not have the correct Sudakov suppression for small $x$ and has large $\mu_0$ dependence from both $\tilde{B}(\mu_0)$ and $\Delta(\mu_0, x)$.

To include the resummation, Sudakov factors must be included in the partonic calculation in such a way that all observables are still accurate to NLO when they are expanded in $\a$. This can be achieved using a subtraction method similar to the one proposed by Frixione and Webber in MC@NLO~\cite{Frixione:2002ik}. For the purpose of this discussion, we do not need to know the actual algorithm used in that paper to generate their results, only the corresponding partonic calculation itself:
%%%
\begin{center}
\bf NLO Subtraction (NLO/LL):
\end{center}
\begin{align}
\label{eq:toysubtraction}
\lvert\cM_0(\mu_0)\rvert^2
&= \bar{B}(\mu_0)\, \Delta_Q(1,\mu_0)
\,,\nn \\
\lvert\cM_1(x, \mu_0)\rvert^2
&= \a B\, \frac{R(x) - Q(x)}{x} + \a \bar{B}(\mu_0)\, \frac{Q(x)}{x}\, \Delta_Q(1,x)
\,,\end{align}
%%%
where we made the dependence of $\Delta$ on $Q(x)$ explicit, and we defined
%%%
\begin{align}
\label{eq:toysubtraction2}
\bar B(\mu_0)
&=\tilde{B}(\mu_0) + \a B \int_{\mu_0}^1 \! \df x\, \frac{Q(x)}{x}
\nn\\
&= B + \a V + \a B \int_0^{\mu_0} \! \df x\, \frac{R(x)-Q(x)}{x}-\a B \int_0^1\!\df x\, \frac{1-Q(x)}{x}
\,.\end{align}
%%%
Note that we have used the fact that the one-emission piece can in general depend on $\mu_0$. The cross sections predicted by the substraction method in \eq{eq:toysubtraction} are
%%%
\begin{align}
\sigma &= \bar{B}(\mu_0) + \a B \int_{\mu_0}^1 \!\df x\, \frac{R(x) - Q(x)}{x}
= \sigma_\mathrm{NLO}
\,,\nn\\
\frac{\df\sigma}{\df x} &= \a\,\frac{1}{x} \left\{
\begin{aligned}
&B\,R(x) - Q(x)[B - \bar{B}(\mu_0)\Delta_Q(1,x)] && \qquad \text{for} \quad x > \mu_0
\\
&\bar{B}(\mu_0)\, Q(x)\,\Delta_Q(1,\mu_0)\Delta_Q(\mu_0, x) && \qquad \text{for} \quad x < \mu_0
\,.\end{aligned}\right.
\end{align}
%%%
By expanding the above expressions to $\ord(\a)$, one can easily show that the predictions of the subtraction method are equal at NLO to the predictions of the slicing method in \eq{eq:toyslicing_xsecs}. Furthermore, in addition to producing the correct NLO total cross section, the one-emission cross section in the subtraction method has no leading-logarithmic $\mu_0$ dependence and exhibits the correct Sudakov suppression for small $x$.

In Ref.~\cite{Frixione:2002ik}, the form of \eq{eq:toysubtraction} was selected mainly for its algorithmic simplicity and the lack of explicit $\epsilon$ dependence. The important difference is that in Ref.~\cite{Frixione:2002ik} each of the two terms in $\lvert\cM_1(x, \mu_0)\rvert^2$ is generated by a separate event sample. This can be inefficient and lead to events with negative weights, since in general, the relative size of the two terms can be very different, and the first term can also become negative. In contrast, in our approach, the form of $\lvert\cM_1(x, \mu_0)\rvert^2$ in \eq{eq:toysubtraction} is just another choice of a partonic calculation, so it can be used in the master formula \eq{eq:toymaster} to generate a single, positive-weight event sample.

Although we could in principle use \eq{eq:toysubtraction} in our approach, it still has some drawbacks. First, due to the rather complicated structure of \eqs{eq:toysubtraction}{eq:toysubtraction2}, they are tedious to generalize to more emissions or higher orders in $\a$. Second, since $\lvert\cM_1(x, \mu_0)\rvert^2$ depends on the subtraction function $Q(x)$, it is in general not positive definite, and for pathological choices of $Q(x)$ can lead to negative cross sections.\footnote{Of course, for any reasonable choice of $Q(x)$, \eq{eq:toysubtraction} will lead to a positive weight NLO/LL event sample. Pathologies occur when, say, $R(x)-Q(x) \sim \mathcal{O}(\alpha^{-1})$.}

However, there is no deep conceptual reason why subtractions must be used to generate Monte Carlo. Indeed, there are many different ways to get results that are accurate at NLO after expanding in $\a$, that have no leading-logarithmic $\mu_0$ dependence, and that have the correct Sudakov suppression in the small $x$ limit. One particularly elegant method proposed by Nason in Ref.~\cite{Nason:2004rx} in a slightly different context is
%%%
\begin{center}
\bf NLO Elegant (NLO/LL):
\end{center}
\begin{align}
\label{eq:toyelegant}
\lvert\cM_0(\mu_0)\rvert^2 &=
\sigma_\mathrm{NLO} \, \Delta_T(1,\mu_0)
\,,\nn\\
\lvert\cM_1(x)\rvert^2 &=
\a\, \sigma_\mathrm{NLO}\, \frac{T(x)}{x}\, \Delta_T(1,x)
\,,\end{align}
%%%
where $\sigma_\mathrm{NLO}$ is the total NLO cross section from \eq{eq:toysigmaNLO}, $T(x)$ is an effective ``splitting function''
%%5
\begin{equation}
T(x) = \frac{B}{\sigma_\mathrm{NLO}}\, R(x)
\,,\end{equation}
%%%
and $\Delta_T$ is the Sudakov factor obtained from $T(x)$ in analogy to \eq{eq:toysudakov}. The cross sections are now
%%%
\begin{align}
\sigma &= \sigma_\mathrm{NLO} \biggl[
 \Delta_T(1,\mu_0) + \int_{\mu_0}^1 \!\df x\, \a\, \frac{T(x)}{x}\, \Delta_T(1,x) \biggr]
= \sigma_\mathrm{NLO}
\,,\nn\\
\frac{\df\sigma}{\df x} &= \a\, \sigma_\mathrm{NLO}\,\frac{1}{x} \left\{
\begin{aligned}
&T(x) \, \Delta_T(1,x) && \qquad \text{for} \quad x > \mu_0
\\
&Q(x)\, \Delta_T(1,\mu_0)\,\Delta_Q(\mu_0, x) && \qquad \text{for} \quad x < \mu_0
\,.\end{aligned}\right.
\end{align}
%%%
As in the subtraction method, the total cross section is identical to the NLO result, and the one-emission cross section has the correct Sudakov suppression with no leading logarithmic $\mu_0$ dependence. Note that the latter again relies on the fact that $R(x)$ and $Q(x)$ have the same singularities for $x\to 0$, so $\Delta_T$ and $\Delta_Q$ resum the same leading logarithms.

The form of \eq{eq:toyelegant} is identical to the merged LO/LL example of \eq{eq:toyLOLL} by replacing $Q(x) \to T(x)$ and $B \to \sigma_\mathrm{NLO}$. The nice feature of \eq{eq:toyelegant} is that it is not only conceptually simple, but it also uses only functions that are completely well defined in an NLO calculation without ever needing to introduce an ad-hoc subtraction function. It is also clear from the above expressions that the cross sections are always guaranteed to be positive. In \sec{sec:QCDfirst}, we show how to generalize this method to the realistic case of QCD. On the other hand, there is also no reason why subtractions could not be used in \GenEvA\ if a subtraction method is more convenient for theoretical or numerical reasons.

%===============================================================================
\subsection{Multiple Emissions}
%===============================================================================

The results from the previous subsection can be extended straightforwardly to multiple emissions in the partonic regime. At tree level, the partonic cross sections are given by
%%%
\begin{center}
\bf Tree Level (LO):
\end{center}
\begin{equation}
\lvert\cM^{\rm LO}_n(\mu_n)\rvert^2 = \frac{\a^n f_n^{\rm tree}(x_1,x_2, \ldots, x_n)}{x_1 x_2 \cdots x_n}
\,,\end{equation}
%%%
where the symbol $f_n^{\rm tree}$ represents an ordinary tree-level calculation in the toy theory with the $\a$ dependence and $x_i$ singularities made manifest. For example, $f_0^{\rm tree} = B$ and $f_1^{\rm tree}(x_1) = B R(x_1)$. We assume that in the singular regions of phase space
%%%
\begin{equation}
\label{eq:multiemissionfiniteness}
\lim_{x_n \to 0} \frac{f_n^{\rm tree}(x_1,\ldots, x_{n-1},x_n) - f_{n-1}^{\rm tree}(x_1,\ldots, x_{n-1}) Q(x_n)}{x_n}= \mathrm{finite}
\,,\end{equation}
%%%
the analog of which is indeed true in the collinear limit of QCD as long as $Q(x)$ is generalized to include flavor information.

As discussed before, the tree-level results do not resum the leading-logarithmic behavior, which can be fixed by including Sudakov factors appropriately:
%%%
\begin{center}
\bf Sudakov-Improved (LO/LL):
\end{center}
\begin{align}
\lvert\cM^{\rm LO/LL}_n(\mu_n)\rvert^2
&= \frac{\a^n f_n^{\rm tree}(x_1,x_2, \ldots, x_n)}{x_1 x_2 \cdots x_n} \, \Delta_Q(1,x_1)\, \Delta_Q(x_1,x_2) \cdots \Delta_Q(x_{n},\mu_n) \label{eq:toymultisudimproved}
\,.\end{align}
%%%
This is precisely the result that was advocated in Ref.~\cite{Catani:2001cc} in order to merge matrix elements and parton showers.\footnote{%
Ref.~\cite{Catani:2001cc} also advocated to evaluate the couplings $\a$ at successive intermediate $x_i$ scales in the problem, which in QCD corresponds to including extra subleading-logarithmic information.} To see that \eq{eq:toymultisudimproved} is well behaved, note that the difference across the boundary $x_n = \mu_{n-1}$ is\footnote{\eq{eq:toydiffacrossboundary} is again the inclusive differential cross section, where we have integrated over all values of $x_i$ with $i > n$.}
%%%
\begin{multline}
\lim_{\varepsilon\to 0} \biggl(
\frac{\df^n \sigma}{\df x_1\cdots \df x_n} \biggr\rvert_{x_n = \mu_{n-1} + \varepsilon}
- \frac{\df^n \sigma}{\df x_1 \cdots \df x_n} \biggr\rvert_{x_n = \mu_{n-1} - \varepsilon} \biggr)
\\
= \alpha^n \frac{f_n^{\rm tree}(x_1,\ldots, x_{n-1}, x_n) - f_{n-1}^{\rm tree}(x_1,\ldots, x_{n-1}) Q(x_n)}{x_1\cdots x_{n-1} x_n} \,\Delta_Q(1,x_n) \biggr\rvert_{x_n = \mu_{n-1}}
\!\!\! + \ord(\alpha^{n+1}) \label{eq:toydiffacrossboundary}
\,,\end{multline}
%%%
which has no large logarithmic dependence on $\mu_{n-1}$, assuming \eq{eq:multiemissionfiniteness} holds.

As the example most relevant for creating our NLO/LO/LL merged sample, we finally consider the combination of the NLO results with higher-multiplicity tree-level matrix elements. This will give a partonic calculation that reproduces the exact NLO results, but also uses the full tree-level quantum interference for multiple emissions. Using the previous definitions of $\sigma_\mathrm{NLO}$ and $T(x)$, we find:
\pagebreak
%%%
\begin{center}
\bf Best Combination (NLO/LO/LL):
\end{center}
\begin{align}
\label{eq:toybest}
\lvert\cM^{\rm NLO/LO/LL}_0(\mu_0)\rvert^2
&= \sigma_\mathrm{NLO} \,\Delta_T(1,\mu_0)
\,,\nn\\
\lvert\cM^{\rm NLO/LO/LL}_1(\mu_1)\rvert^2
&= \frac{\a f_1^{\rm tree}(x_1)}{x_1} \,\Delta_T(1,x_1)\, \Delta_Q (x_1, \mu_1)
\,,\nn\\&\,\,\,\vdots\nn\\
\lvert\cM^{\rm NLO/LO/LL}_n(\mu_n)\rvert^2
&= \frac{\a^n f_n^{\rm tree}(x_1,x_2, \ldots, x_n)}{x_1 x_2 \cdots x_n} \, \Delta_T(1,x_1) \,\Delta_Q(x_1,x_2) \cdots \Delta_Q(x_{n},\mu_n)
\,.\end{align}
%%%
For $n = 1$ and $\mu_1 = x_1$ this reduces to \eq{eq:toyelegant}.

The factor of $\Delta_T(1,x_1)$ appears in $\lvert\cM_n(\mu_n)\rvert^2$ to ensure that the total cross section has no large logarithmic dependence on $\mu_1$.\footnote{We cannot replace every $\Delta_Q$ with $\Delta_T$ because in QCD there is additional flavor structure, which would make this choice ill-defined.} In fact, this is the only difference between the LO/LL and NLO/LO/LL expressions for $\lvert\cM_n(\mu_n)\rvert^2$, but it is crucial to ensure that all NLO observables (including the total cross section) are accurate to NLO with no large logarithms. In particular, it is straightforward to integrate \eq{eq:toybest} to show that
%%%
%%%
\begin{equation}
\sigma = \sigma_\NLO + \alpha^2 \int_{\mu_0}^1 \!\df x_1\, \int_{\mu_1}^{x_1} \!\df x_2\,
\frac{f_2^{\rm tree}(x_1,x_2) - f_1^{\rm tree}(x_1) Q(x_2)}{x_1 x_2} \,\Delta_T(1,x_1) \,\Delta_Q(x_1,x_2) + \mathcal{O}(\alpha^3)
\end{equation}
%%%
%%%
which has no large logarithmic dependence in $\mu_0$ or $\mu_1$ as long as \eq{eq:multiemissionfiniteness} holds.

%===============================================================================
\subsection{The Importance of Phase Space Projection}
\label{subsec:importancetruncation}
%===============================================================================

In the above discussion, we never needed to talk about the phase space projection feature from \eq{eq:PSprojection} to write down our partonic calculations $\lvert\cM(\mu)\rvert^2$. The reason is that the toy theory is a single-scale theory, so phase space projection acts trivially. Given a phase space point $\{x_1,x_2,\ldots, x_n\}$, we can simply define the intermediate resolution scales as $\bar{\mu}_{n-1} = x_{n-1}$, $\bar{\mu}_{n-2} = x_{n-2}$, and so on, such that the projected phase space points are
%%%
\begin{equation}
\{x_1,x_2,\ldots, x_n\} \, \mathop{\longrightarrow}^{~\bar{\mu}_{n-1}~}\, \{x_1,x_2,\ldots, x_{n-1}\} \, \mathop{\longrightarrow}^{~\bar{\mu}_{n-2}~}\, \{x_1,x_2,\ldots, x_{n-2}\} \, \longrightarrow \, \cdots
\,.\end{equation}
%%%
In full QCD, there are many more scale choices one could make for $\bar{\mu}$, and the difference between two different choices will yield logarithms (hopefully not large) of $\bar{\mu}_A / \bar{\mu}_B$. More importantly, in QCD there is an important requirement of final states being on shell, \emph{e.g.} if we project two gluons into one mother gluon, we have to somehow shuffle momenta to put that mother gluon on shell.

So why use phase space projection in QCD if it is so hard? In the language of \eq{eq:toybest}, the reason is that we need some way of getting the analog of the factors of $\Delta_T$ to appear in $\lvert\cM_n(\mu_n)\rvert^2$. In some sense, $\Delta_T$ is a property of the NLO calculation that we want the LO tree-level matrix elements to inherit, but the tree-level matrix elements do not have enough information to properly calculate $\Delta_T$ themselves. But if we take an $n$-emission matrix element and project it to a one-emission matrix element, then we can use our existing NLO machinery to NLO-improve the LO calculations. In formulas, we can write the $n$-emission cross section in \eq{eq:toybest} as
%%%
\begin{equation}
\label{eq:toymatch}
\lvert\cM^{\rm NLO/LO/LL}_n(\mu_n)\rvert^2
= \lvert\cM^{\rm LO/LL}_n(\mu_n)\rvert^2 \times
\frac{\lvert\cM^{\rm NLO/LL}_1(\bar{\mu}_1 = x_1)\rvert^2}
{\lvert\cM^{\rm LO/LL}_1(\bar{\mu}_1 = x_1)\rvert^2}
\,.\end{equation}
%%%
The ratio on the right-hand side equals precisely $\Delta_T(1,x_1) / \Delta_Q(1,x_1)$ and can be thought of as the ``matching coefficient'' between the NLO/LL calculation and the other Sudakov improved tree-level calculations.

\begin{figure}
\centering
\includegraphics[scale=0.7]{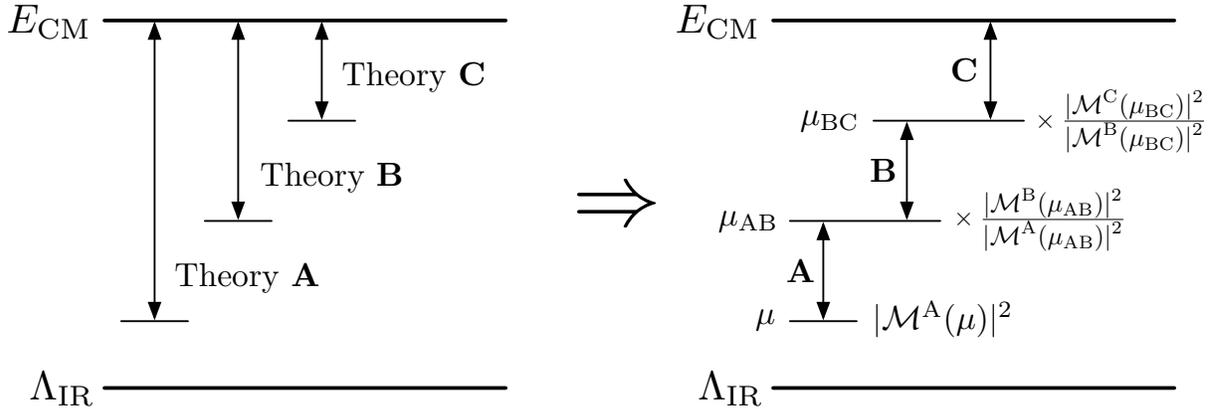}
\caption{Using nested descriptions to define the best partonic calculations. Given a theory $\mathbf{A}$ valid down to the scale $\mu$, the partonic calculation $|\cM^{\rm A}(\mu)|^2$ can be improved using information from a presumably better theory $\mathbf{B}$ which is however only valid down to a scale $\mu_{\rm AB} > \mu$. The ``matching coefficient'' $\lvert\cM^{\rm B}(\mu_{\rm AB})\rvert^2 / \lvert\cM^{\rm A}(\mu_{\rm AB})\rvert^2$ allows the full event sample to capture all the physics of theory $\mathbf{B}$, while still retaining information about theory $\mathbf{A}$ below the scale $\mu_{\rm AB}$. The same logic can now be repeated with a third theory $\mathbf{C}$ which is only valid down to a scale $\mu_{\rm BC} > \mu_{\rm AB}$.}
\label{fig:nestedtheories}
\end{figure}

In fact, \eq{eq:toymatch} is just a specific case of a more general version of matching different partonic descriptions and running between them. Imagine, we have a set of nested partonic descriptions $\mathbf{A}$, $\mathbf{B}$, $\mathbf{C}$, and so on as in \fig{fig:nestedtheories}, then the best partonic calculation we can construct with the available information is
%%%
\begin{center}
\boldmath
\bf Best Combination ($\mathbf{A}$/$\mathbf{B}$/$\mathbf{C}$/$\cdots$):
\end{center}
\begin{equation}
\label{eq:toybestgeneral}
\lvert\cM^\mathrm{A/B/C/\cdots}(\mu)\rvert^2 = \lvert\cM^{\rm A}(\mu)\rvert^2
\times \frac{\lvert\cM^{\rm B}(\mu_{\rm AB})\rvert^2}{\lvert\cM^{\rm A}(\mu_{\rm AB})\rvert^2}
\times \frac{\lvert\cM^{\rm C}(\mu_{\rm BC})\rvert^2}{\lvert\cM^{\rm B}(\mu_{\rm BC})\rvert^2}
\times \cdots
\,,\end{equation}
%%%
where the various $\mu$ scales encode the multiplicities and energy scales at which the matching occurs.

Though obscured in the $\df\MC(\mu)$ notation, but clear from \eq{eq:toyshowerinamplitude}, the NLO/LO/LL merged sample is equivalent to taking
%%5
\begin{equation}
\mathbf{A} = \mathrm{LL}
\,, \qquad
\mathbf{B} = \mathrm{LO/LL}
\,, \qquad
\mathbf{C} = \mathrm{NLO/LL}
\,,\end{equation}
%%%
where the LL description is just the parton shower written in the spirit of \eq{eq:toyshowerinamplitude}. The factors of $\lvert\cM^{\rm LL}(\mu)\rvert^2$ do not appear in \eq{eq:toymatch} because we use the parton shower not as partonic calculation but as phenomenological model. Though we will not pursue this direction in the present work, we expect that creating an N$^i$LO/N$^j$LL merged sample will just require finding appropriate descriptions $\bf A$, $\bf B$, $\bf C$, and so on.

As a side note, the reversable phase space projection in the \GenEvA\ algorithm \cite{genevatechnique} is equivalent to the statement that we have an analytic way of calculating $\lvert\cM^{\rm LL}(\mu)\rvert^2$ for the parton shower at all scales $\mu$. One of the reasons that the \GenEvA\ algorithm is efficient comes from the realization that the Sudakov factors in
%%%
\begin{equation}
\frac{\lvert\cM^{\rm LO/LL}(\mu)\rvert^2}{\lvert\cM^{\rm LL}(\mu)\rvert^2}
\end{equation}
%%%
cancel, reducing the number of calculations necessary to merge an LO/LL calculation with the parton shower.

The next two sections deal with the technical complications introduced by full QCD. However, the main ideas are captured by the toy example just presented, so the reader not interested in the technical details can safely skip directly to \sec{sec:results} for the results of the NLO/LO/LL merged calculation in QCD.

%%%%%%%%%%%%%%%%%%%%%%%%%%%%%%%%%%%%%%%%%%%%%%%%%%%%%%%%%%%%%%%%%%%%%%%%%%%%%%%%
\section{The First Emission in QCD}
\label{sec:QCDfirst}
%%%%%%%%%%%%%%%%%%%%%%%%%%%%%%%%%%%%%%%%%%%%%%%%%%%%%%%%%%%%%%%%%%%%%%%%%%%%%%%%

Having investigated the toy model in detail, we now move on to the realistic case of QCD. In this section, we will study how to implement the QCD cross section for $e^+ e^- \to n \text{ jets}$ up to $\ord(\alpha_s)$, thus including the processes $e^+ e^- \to q \bar q$ and $e^+ e^- \to q \bar q g$. The extension to final states with more than three partons will be discussed in the next section, though most of the important physics considerations necessary to construct the NLO/LO/LL merged sample appear here.

The essential complication for $e^+ e^- \to n \text{ jets}$ compared to the toy example is the presence of multiple well-defined energy scales. This same complication appears in the process $pp \text{ or } p \bar{p} \to n \text{ jets}$, so we anticipate that the same solutions present in the leptonic case should have a generalization to the hadronic case. At the end of the day, these complications are resolved by a thorough understanding of phase space, so we will begin with 3-body phase space before moving on to the relevant QCD calculations and the definition of the partonic matrix elements.

%===============================================================================
\subsection{Three-Body Phase Space}
%===============================================================================

At the level of phase space, there are two complications in QCD compared with the toy model studied in the previous section. First, a single emission in QCD is specified by three independent variables, as opposed to the single variable in the toy model. Second, on top of these three variables specifying the hadronic kinematics, there are two additional variables describing the orientation of the hadronic system relative to the $e^+ e^-$ beamline.

There are many equivalent ways of choosing these five variables. For our discussion, it will be convenient to use the thrust axis to define the orientation of the hadronic system relative to the beamline, and we use $\Omega_2^T$ to describe this orientation. The remaining three variables describe the kinematics of the hadronic system relative to the thrust axis, which we choose to be the invariant mass $t_a$ between the quark and the gluon, the invariant mass $t_b$ between the antiquark and the gluon, as well as the azimuthal angle $\phi$ of the hadronic system relative to the thrust axis.
In other words, we decompose Lorentz-invariant $3$-body phase space as
%%%
\begin{equation}
\label{eq:QCD3body}
\df\Phi_3 = \frac{1}{(4\pi)^5}\, E_{\rm CM}^2\, \df \Omega_2^T\, \df t_a\, \df t_b\, \df \phi
\,,\end{equation}
%%%
where $E_{\rm CM}^2 = (p_{e^+} + p_{e^-})^2$ and
%%%
\begin{equation}
\label{tdef}
t_a = \frac{(p_q + p_g)^2}{ E_{\rm CM}^2}\,, \qquad t_b = \frac{(p_{\bar q} + p_g)^2}{ E_{\rm CM}^2}
\,.\end{equation}
%%%
For the remainder of the discussion, we will mainly ignore the $\Omega_2^T$ and $\phi$ dependence.

The key to constructing the analog of the master formula in \eq{eq:toymaster} is to have an unambiguous definition of the scale $\mu$ that will separate our QCD calculations from the phenomenological parton shower. For simplicity, we will assume that the parton shower uses virtuality
%%%
\begin{equation}
t \equiv \frac{Q^2}{E_{\rm CM}^2}
\end{equation}
%%%
as the evolution variable. We discuss the issues with extending our results to more modern $p_\perp$-ordered showers in the companion paper \cite{genevatechnique}. To keep the notion of $\mu$ as a mass scale, we also define
%%%
\begin{equation}
\hat{\mu}^2 = \frac{\mu^2}{E_{\rm CM}^2}
\,,\end{equation}
%%%
and use dimensionless variables in the remainder of this discussion. We anticipate running the \GenEvA\ framework from $1 = E_{\rm CM}^2/E_{\rm CM}^2$ until the scale $\hat{\mu}^2$, attaching a parton shower starting from $\hat{\mu}^2$ going down to $t_\cut \simeq \Lambda_{\rm IR}^2/ E_{\rm CM}^2$, and finally interfacing with a hadronization model at $t_\cut$.

\begin{figure}
\vspace{-1ex}
\includegraphics[scale=0.7]{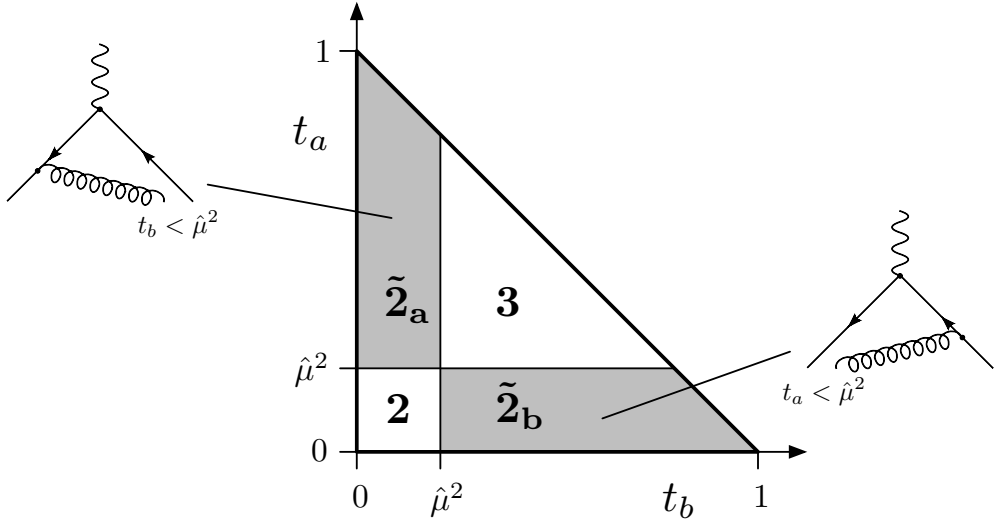}
\vspace{-2ex}
\caption{The available phase space $0 \leq t_a \leq 1-t_b \leq 1$ for $e^+ e^- \to q \bar q g$, where $t_{a,b}$ are defined in \eq{tdef}. The separation between the 2-jet region and the 3-jet region is defined by a matching scale $\hat{\mu}^2$. Only region $\mathbf{3}$ has both $t_a$ and $t_b$ above the matching scale $\hat{\mu}^2$ and therefore corresponds to the 3-jet region of phase space. The ambiguous $\mathbf{\tilde{2}}$ regions could be generated by a parton shower cut off at $\hat{\mu}^2$ by ``accidentally'' populating a region closer to the QCD collinear singularity than $\hat{\mu}^2$. However, that part is also generated by emissions below $\hat{\mu}^2$ and is therefore covered by $\df\MC_2(\hat{\mu}^2)$ and must be vetoed in $\df\MC_3(\hat{\mu}_3^2)$.}
\label{fig:phasepace}
\end{figure}

The  total 3-parton phase space $0 \leq t_a \leq 1-t_b \leq 1$ is shown in \fig{fig:phasepace}. The 3-jet region $\mathbf{3}$ is defined as the region where both $t_a$ and $t_b$ are larger than $\hat{\mu}^2$. The 2-jet region $\mathbf{2}$ corresponds to both $t_a$ and $t_b$ being less than $\hat{\mu}^2$. Finally, there is an ambiguous $\mathbf{\tilde{2}_a}$ region where $t_a$ is greater than $\hat{\mu}^2$ and $t_b$ is less than $\hat{\mu}^2$, and a similar $\mathbf{\tilde{2}_b}$ region where the roles of $t_a$ and $t_b$ are reversed. Formally, these phase space integration regions are defined as
%%%
\begin{align}
\int_{\mathbf{3}} \!\df t_a \, \df t_b
&\equiv
\int_{\hat{\mu}^2}^{1-\hat{\mu}^2} \! \df t_a \int_{\hat{\mu}^2}^{1-t_a} \!\df t_b
\,,
&\int_{\mathbf{2}} \! \df t_a \,\df t_b &\equiv \int_{0}^{\hat{\mu}^2} \! \df t_a \int_{0}^{\hat{\mu}^2} \! \df t_b
\,,\nn\\
\int_{\mathbf{\tilde{2}_a}} \!\df t_a \,\df t_b
&\equiv \int_{0}^{\hat{\mu}^2} \! \df t_b \int_{\hat{\mu}^2}^{1-t_b} \! \df t_a
\,,
&\int_{\mathbf{\tilde{2}_b}}\! \df t_a \,\df t_b
&\equiv \int_{0}^{\hat{\mu}^2} \! \df t_a \int_{\hat{\mu}^2}^{1-t_a} \! \df t_b
\,.\end{align}
%%%

By  definition, the regions $\mathbf{2}$, $\mathbf{\tilde{2}_a}$, $\mathbf{\tilde{2}_b}$
are covered by running the phenomenological parton shower on
2-body phase space starting at $\hat{\mu}^2$, with the $\mathbf{\tilde{2}_a}$ and $\mathbf{\tilde{2}_b}$ regions arising from small virtuality, small angle gluon radiation. Hence, they belong to $\df\MC_2(\hat{\mu}^2)$. The remaining part of $\df\Phi_3$ is $\mathbf{3} \equiv \df\Phi_3(\hat{\mu}^2)$, and running the shower on $\mathbf{3}$ defines $\df\MC_3(\hat{\mu}_3^2)$. Since $n_\max = 3$, we need to pick an appropriate scale $\hat{\mu}_3^2$ to avoid dead zones, similar to our discussion in the toy model in \sec{subsec:toyphasespace}. What we need is the QCD-version of the condition $\mu_{n_\max} = x_{n_\max}$ we found in the toy model. Hence, we have to decide whether to take $\hat{\mu}_3^2$ equal to $t_a$ or $t_b$, and in fact, either choice would leave no dead zones.

For  the moment, imagine using the parton shower to cover all of phase space. Given the hard scattering process $e^+ e^- \to q \bar{q}$, we can run the parton shower from $t_{\rm start} = 1$. The first emission can either come from a gluon being radiated from the quark at $t_{\rm emit} = t_a$ or from the antiquark at $t_{\rm emit} = t_b$. Regardless at what scale $t_{\rm emit}$ the emission occurs, the parton shower continues to run from that scale down, filling out the rest of phase space. Thus, knowing $t_a$ and $t_b$ alone does not permit a definition of the proper scale $\hat \mu^2 = t_{\rm emit}$, and a wrong choice could lead to potentially large logarithms $\alpha_s \log^2(t_a/t_b)$. The challenge is that a point $\Phi_3$ gives us only access to $t_a$ and $t_b$, and we have to decide how to properly account for the logarithms of the ratio of these two scales.

We  will discuss several ways to resolve this ambiguity in the definition of $\hat \mu^2$, and for now we only provide a framework that allows the $\hat{\mu}^2 = t_a \text{ vs. } t_b$ problem to be solved in principle. In particular, for now we will simply have two samples of 3-parton events, 3-parton events where we start the shower at $t_a$ and 3-parton events where we start the shower at $t_b$, and let the user decide what partonic calculations to use for each sample. Thus, we will have three \GenEvA\ event samples
%%%
\begin{equation}
\label{eq:QCDoneeventsamples}
\df \MC_2(\hat{\mu}^2)
\,,\qquad
\df \MC_3(\hat{\mu}_3^2 = t_a)
\,,\qquad
\df \MC_3(\hat{\mu}_3^2 = t_b)
\,,\end{equation}
%%%
which give a complete covering of phase space. To not introduce double-counting in the cross section, we now have to be careful to remember that there are two $\df \MC_3$ samples. We stress that this potential double-counting is conceptually very different from the double-counting between different phase space algorithms discussed earlier. Here, it arises from a scale ambiguity in QCD which has to be resolved by calculational means.

The $\mathbf{\tilde{2}}$ regions are yet another example of a scale ambiguity and how the notion of ``above $\mu$'' and ``below $\mu$'' can be so confusing. Since the $\df \mathrm{MC}_2$ sample covers the $\mathbf{\tilde{2}_a}$ region, it has to be vetoed in $\df \mathrm{MC}_3$, despite the fact that it could have been generated by the phenomenological model with a large angle emission above $\hat{\mu}^2$. As seen in \fig{fig:phasepace}, a gluon emission could have occurred at a large angle at a large scale, while for the purposes of phase space this is part of $\df\MC_2(\hat{\mu}^2)$, and so the actual scale of emission should be considered below $\hat{\mu}^2$.

%===============================================================================
\subsection{The Master Formula}
%===============================================================================

With the event samples of \eq{eq:QCDoneeventsamples}, we can write the differential cross section in terms of the master formula
%%%
\begin{equation}
\label{eq:QCDonemaster}
\df\sigma
= \frac{\df\sigma_{2}(\hat{\mu}^2)}{\df \Phi_2} \, \df \MC_2(\hat{\mu}^2)
+ \frac{\df\sigma_{3a}(t_a)}{\df \Phi_3} \, \df \MC_3(t_a)
+ \frac{\df\sigma_{3b}(t_b)}{\df \Phi_3} \, \df \MC_3(t_b)
\,,\end{equation}
%%%
where we are using the notation $\df\sigma_{i}/ \df \Phi$ instead of $\lvert\cM_i\rvert^2$ because we anticipate integrating over some of the phase space variables to simplify expressions. The argument in the $\df\sigma_{i}$ functions reminds us of the scale at which we start running the parton shower for those samples. To avoid double-counting regions of phase space, the cross sections $\df\sigma_{3a}$ and $\df\sigma_{3b}$ must satisfy
%%%
\begin{equation}
\label{eq:sumalphacomponents}
\frac{\df\sigma_{3}}{\df \Phi_3}
= \frac{\df\sigma_{3a}(t_a)}{\df \Phi_3} + \frac{\df\sigma_{3b}(t_b)}{\df \Phi_3}
\,,\end{equation}
%%%
where $\df\sigma_3$ is the full 3-parton cross section. If we are not worried about large $\log^2 (t_a/t_b)$ logarithms, then a simple definition of the $\df\sigma_{3i}$ partonic cross sections is
%%%
\begin{equation}
\label{eq:alphaseparation}
\frac{\df\sigma_{3a}(t_a)}{\df \Phi_3} = \frac{\df\sigma_{3}}{\df \Phi_3}\, \alpha_a (\Phi_3)
\,, \qquad
\frac{\df\sigma_{3b}(t_b)}{\df \Phi_3} = \frac{\df\sigma_{3}}{\df \Phi_3}\,  \alpha_b (\Phi_3)
\,,\end{equation}
%%%
where $\alpha_a (\Phi_3) + \alpha_b (\Phi_3) = 1$. In the absence of leading-logarithmic information, any choice of $\alpha_a$ and $\alpha_b$ is equally valid, however in the presence of leading-logarithmic information the $\alpha_i$'s have to be chosen to correctly treat logarithms of $t_a/t_b$.

With \eq{eq:QCDonemaster} in hand, we can now try to find the best definitions for the partonic cross sections $\df\sigma_{i}$, but first we have to review the relevant QCD calculations.

%===============================================================================
\subsection{Relevant QCD Calculations}
%===============================================================================

The process $e^+e^- \to q\bar{q} g$ is singular as $t_i \to 0$. To understand this singularity structure, consider the differential cross section in $d = 4-2\epsilon$ dimensions, and integrated over the three angular variables $\Omega^T_2$ and $\phi$. One finds the well-known result \cite{Ellis:1991qj}
%%%
\begin{align}
\frac{\df\sigma_B}{\df t_a \df t_b} &= B \, \delta(t_a)\, \delta(t_b)
\,, \nn\\
\frac{\df\sigma_V}{\df t_a \df t_b} &= \frac{\alpha_sC_F}{2\pi}\,
\biggl(\frac{4 \pi \mu_\epsilon^2}{E_{\rm CM}^2} \biggr)^\epsilon B \,\biggl( -\frac{2}{\epsilon^2} - \frac{3}{\epsilon} + V \biggr) \, \delta(t_a)\, \delta(t_b)
\,, \nn\\
\frac{\df\sigma_R}{\df t_a \df t_b} &= \frac{\alpha_sC_F}{2\pi}\,
\biggl(\frac{4 \pi \mu_\epsilon^2}{E_{\rm CM}^2} \biggr)^\epsilon
\frac{1-2\epsilon}{\Gamma(2-2\epsilon)}\, B \, \frac{R(t_a,t_b)-2\epsilon(t_a+t_b)^2}{(1-t_a-t_b)^{2\epsilon} t_a^{1+2\epsilon} t_b^{1+2\epsilon}}
\,,\end{align}
%%%
where $\mu_\epsilon$ is the standard renormalization scale of dimensional regularization, and has nothing to do with the scale $\mu$ that separates QCD calculations from the parton shower. Here, $B$ denotes the Born cross section
%%%
\begin{equation}
B = N_c\, Q_q^2\, \frac{4 \pi \alpha_{\rm em}^2}{3 E_{\rm CM}^2}
\,,\end{equation}
%%%
with $N_c = 3$, $Q_q$ the charge of the quark, and $\alpha_{\rm em}$ the fine-structure constant.\footnote{For simplicity, we are only including $e^+ e^- \to \gamma^* \to {\rm partons}$. An intermediate $Z$-boson can easily be included by the appropriate change in the Born cross section $B$.} We have also defined
%%%
\begin{equation}
R(t_a,t_b) = (1-t_a)^2+(1-t_b)^2
\,,\qquad
V = -8 + \frac{7 \pi^2}{6}
\,.\end{equation}
%%%

The virtual diagram only contributes at $t_a = t_b = 0$, but is infrared divergent in $4$ dimensions. However, the real emission also diverges as $t_i \to 0$. Integrating these expressions over the allowed phase space $0 \leq t_a \leq 1-t_b \leq 1$ we find for the various contributions to the total cross section
%%%
\begin{align}
\sigma_B &= B
\,,\nn\\
\sigma_V &= \frac{\alpha_sC_F}{2\pi} \biggl( \frac{4 \pi \mu_\epsilon^2}{E_{\rm CM}^2} \biggr)^\epsilon B \biggr( -\frac{2}{\epsilon^2} - \frac{3}{\epsilon} + V \biggr)
\,,\nn\\
\sigma_R &= \frac{\alpha_sC_F}{2\pi} \biggl(\frac{4 \pi \mu_\epsilon^2}{E_{\rm CM}^2} \biggr)^\epsilon B \biggr( \frac{2}{\epsilon^2}+ \frac{3}{\epsilon} + \frac{19}{2} - \frac{7 \pi^2}{6} \biggr)
\,,\end{align}
%%%
such that the infrared $1/\epsilon$ divergences cancel and the total cross section gives the well-known result
%%%
\begin{equation}
\label{eq:QCDsigmatotal}
\sigma_{\rm NLO} = \sigma_B + \sigma_V + \sigma_R
= B \biggl[ 1 + \frac{\alpha_s C_F}{2 \pi} \biggl( V + \frac{19}{2} - \frac{7 \pi^2}{6} \biggr) \biggr]
= B \biggl( 1 + \frac{3}{2}\, \frac{\alpha_s C_F}{2 \pi} \biggr)
\end{equation}
%%%
in the $\epsilon \to 0$ limit.

The singularity structure of the real emission in QCD is reproduced by the well-known Altarelli-Parisi~\cite{Altarelli:1977zs}  splitting function
%%%
\begin{equation}
f(t_i,z_i) = \frac{1}{t_i}\, \frac{1+z_i^2}{1-z_i}
\,,\end{equation}
%%%
where $z_i$ are defined as the energy fractions
%%%
\begin{equation}
z_a = \frac{E_q}{E_q+E_g} = \frac{1-t_b}{1+t_a}
\,, \qquad
z_b = \frac{E_{\bar q}}{E_{\bar q}+E_g} = \frac{1-t_a}{1+t_b}
\,.\end{equation}
%%%
Rewriting the splitting functions in terms of the variables $t_a$ and $t_b$, taking into account the relevant Jacobian factors, we can define the analog of the $Q$ functions in the toy example
%%%
\begin{equation}
\frac{Q_a(t_a,t_b)}{t_at_b}
= \frac{1}{t_a(t_a+t_b)}\, \biggl[ 1 + \biggl( \frac{1-t_b}{1+t_a}\biggr)^2 \biggr]
\,, \quad
\frac{Q_b(t_a,t_b)}{t_at_b}
= \frac{1}{t_b(t_a+t_b)}\, \biggl[ 1 + \biggl(\frac{1-t_a}{1+t_b} \biggr)^2 \biggr]
\,,\end{equation}
%%%
where $Q_a$ is the splitting function for $q \to qg$ and $Q_b$ for $\bar{q} \to \bar{q}g$. One can easily verify that the sum of the two splitting functions reproduces all the singular behavior of the function $R(t_a,t_b)$
%%%
\begin{equation}
\label{eq:QCDRtoQ}
\frac{R(t_a,t_b) - Q_a(t_a,t_b) - Q_b(t_a,t_b)}{t_a t_b} = \text{finite as } t_i \to 0
\,.\end{equation}
%%%
For later convenience, we also define a Sudakov factor
%%%
\begin{equation}
\Delta_Q(t_1,t_2)
= \exp\biggl[ - \frac{\alpha_sC_F}{2\pi} \int_{t_2}^{t_1} \! \df t_a \int_0^{1-t_a} \! \df t_b \, \frac{Q_a(t_a,t_b)}{t_at_b} \biggr]
\,,\end{equation}
%%%
which is symmetric in $a \leftrightarrow b$. The $t_b$ integration range goes down to zero because this corresponds to the proper integration range in the original $z_a$-dependent splitting function.

For technical reasons relating to the conservation of four-momentum, the parton shower in the companion paper \cite{genevatechnique} actually uses the splitting function
%%%
\begin{equation}
\frac{\tilde{Q}_a(t_a,t_b)}{t_at_b}
= \frac{t_b}{\sqrt{1-t_a}}\, \frac{1 + z_*^2}{1-z_*}
\,, \qquad
z_* = \frac{1}{2} \biggl(1 - \frac{2 t_b + t_a - 1}{\sqrt{1-t_a}}\biggr)
\,,\end{equation}
%%%
and $\tilde{Q}_b(t_a,t_b) = \tilde{Q}_a(t_b,t_a)$. These $\tilde{Q}$ functions are perfectly consistent with the condition of \eq{eq:QCDRtoQ} as longer as the simultaneous limit $t_a, t_b \to 0$ is avoided. This double-soft divergence leads to extra subleading logarithms in the Sudakov factor. Because we only ever work to leading-logarithmic order, this is not a problem, though the specific distributions obtained using $Q$ vs.\ $\tilde{Q}$ will of course differ. Eventually, one would want to tune the $\tilde{Q}$ functions to minimize this subleading effect.  In addition, one can use the difference as an estimate of subleading-logarithmic effects.

%===============================================================================
\subsection{Partonic Cross Sections at LO}
%===============================================================================

In analogy with \sec{subsec:toyxsec}, we can define partonic cross sections for the master formula \eq{eq:QCDonemaster} using the above QCD calculations. For simplicity, we will largely ignore the dependence on $\Omega^T_2$ and $\phi$ and focus only on $t_a$ and $t_b$.

If we only had access to a parton shower, then it might distribute events according to:
%%%
\begin{center}
\bf Parton Shower (LL):
\end{center}
\begin{align}
\label{eq:ll}
\sigma_{2}(\hat{\mu}^2)
&=B\, \biggl\{ \Delta_Q^2(1,\hat{\mu}^2)
\nn\\ &\quad
+ \frac{\alpha_s C_F}{2 \pi}\, \biggl[
\int_{\mathbf{\tilde{2}_a}} \! \df t_a\, \df t_b\, \frac{Q_a(t_a,t_b)}{t_a t_b}\, \Delta_Q^2(1,t_a)
+ \int_{\mathbf{\tilde{2}_b}} \! \df t_a \, \df t_b\, \frac{Q_b(t_a,t_b)}{t_a t_b}\, \Delta_Q^2(1,t_b) \biggr]\biggr\}
\,,\nn\\
\frac{\df\sigma_{3a}(t_a)}{\df t_a \df t_b}
&= \frac{\alpha_s C_F}{2 \pi}\, B\, \frac{Q_a(t_a,t_b)}{t_a t_b}\, \Delta_Q^2(1,t_a)
\,,\nn\\
\frac{\df\sigma_{3b}(t_b)}{\df t_a \df t_b}
&= \frac{\alpha_s C_F}{2 \pi}\, B\, \frac{Q_b(t_a,t_b)}{t_a t_b}\, \Delta_Q^2(1,t_b)
\,.\end{align}
%%%
The details of the parton shower algorithm that yield this kind of distribution are given in the companion paper \cite{genevatechnique}, though we note that this shower treats the gluon emission from the quark and antiquark symmetrically. The fact that the Sudakov factors are always squared comes from the fact that in a symmetric shower, the no-branching probability for the quark and the antiquark are tied together.

For the present purposes, the most relevant information about \eq{eq:ll} is that these functions include the correct leading-logarithmic resummation through the Sudakov factors while reproducing the Born cross section:
%%%
\begin{equation}
\sigma_{2}(\hat{\mu}^2) + \int_{\mathbf{3}} \! \df t_a \, \df t_b\, \frac{\df\sigma_{3a}(t_a)}{\df t_a \df t_b} + \int_{\mathbf{3}} \!\df t_a\, \df t_b\, \frac{\df\sigma_{3b}(t_b)}{\df t_a \df t_b} = B
\,.\end{equation}
%%%
The integrations over the $\mathbf{\tilde{2}}$ regions in $\sigma_{2}$ are necessary to get the total cross section correct, and correspond to the fact that in the shower, a gluon can split off at such a large angle from the quark that it becomes singular with the antiquark and vice verse, effectively yielding 2-jet events, as shown in \fig{fig:phasepace}.

If we now have the tree-level real-emission calculation and ignore leading logarithms, then the most naive partonic cross section is
%%%
\begin{center}
\bf Tree Level (LO):
\end{center}
\begin{align}
\label{eq:tree}
\sigma_{2}(\hat{\mu}^2) &= B
\,,\nn\\
\frac{\df\sigma_{3}}{\df t_a \df t_b}
&= \frac{\alpha_s C_F}{2 \pi}\, B\, \frac{R(t_a,t_b)}{t_a t_b}
\,.\end{align}
%%%
At this point, we do not have enough information to define separate $\df\sigma_{3a}$ and $\df\sigma_{3b}$ cross sections and will rely on some unspecified $\alpha_i$ function to determine the relative weight of the two different $\df \MC_3$ samples, as in \eq{eq:sumalphacomponents}.

While \eq{eq:tree} gives the correct leading order behavior for 3-jet differential distributions, the total cross section has a strong dependence on the unphysical matching scale, which can be seen by expanding the resulting expression for the total cross section around small values of $\hat{\mu}^2$.
This gives
%%%
\begin{equation}
\label{eq:sigma_doublelog}
\sigma = B \biggl[ 1 + \frac{\alpha_s C_F}{2 \pi}\, \biggl( 4 \log^2 \hat{\mu} + 6 \log \hat{\mu} + \frac{5}{2} - \frac{\pi^2}{3} \biggr) + {\cal O}(\hat{\mu}^2) \biggr]\,.
\end{equation}
%%%
This double-logarithmic dependence on the matching scale is of course due to having used a fixed-order calculation, which by construction does not sum any of the leading-logarithmic behavior in the cross section.

The most precise way to determine the properly resummed expressions is to use renormalization group evolution in an effective field theory setup to determine the partonic cross sections. However, one can use the fact that parton showers do sum the leading-logarithmic behavior correctly to derive expressions at this order. Just as in the toy model, we can Sudakov-improve the matrix element by finding a suitable merging of \eqs{eq:ll}{eq:tree}.

Since there are two event samples describing the single gluon emission, care has to be taken to minimize the effect of large $\log^2 (t_a/t_b)$ logarithms. The simplest way to do so is to find an expression that reduces to \eq{eq:ll} if we were to take $R(t_a,t_b) = Q_a(t_a,t_b) + Q_b(t_a,t_b)$, which guarantees the Sudakov-improved result will have the same singularity structure as the leading-logarithmic result. We choose
%%%
\begin{center}
\bf Sudakov-Improved (LO/LL):
\end{center}
\begin{align}
\label{eq:sudakovimproved}
\sigma_{2}(\hat{\mu}^2)
&= B\, \biggl\{ \Delta_Q^2(1,\hat{\mu}^2)
+ \frac{\alpha_s C_F}{2 \pi}\, \biggl[\int_{\mathbf{\tilde{2}_a}} \!\df t_a\, \df t_b\, \frac{Q_a}{t_a t_b}\, \Delta_Q^2(1,t_a)
+ \int_{\mathbf{\tilde{2}_b}} \!\df t_a\,\df t_b\, \frac{Q_b}{t_a t_b}\, \Delta_Q^2(1,t_b) \biggr] \biggr\}
\,,\nn\\
\frac{\df\sigma_{3a}(t_a)}{\df t_a \df t_b}
&= \frac{\alpha_s C_F}{2 \pi}\, B\, \frac{R(t_a,t_b)}{t_a t_b}\, \biggl[ \frac{Q_a}{Q_a+Q_b }\, \Delta_Q^2(1,t_a) \biggr]
\,, \nn \\
\frac{\df\sigma_{3b}(t_b)}{\df t_a \df t_b}
&= \frac{\alpha_s C_F}{2 \pi}\, B\, \frac{R(t_a,t_b)}{t_a t_b}\, \biggl[ \frac{Q_b}{Q_a+Q_b}\, \Delta_Q^2(1,t_b) \biggr]
\,,\end{align}
%%%
where we have omitted the dependence of the splitting functions on the phase space variables. The Sudakov factors in $\df\sigma_3$ are weighted by the relative splitting probabilities of their respective shower history, which is equivalent to the approach of Ref.~\cite{Lonnblad:2001iq}.
By integrating these expressions and expanding around small values of $\hat{\mu}^2$, one can easily verify that the leading-logarithmic dependence on $\hat{\mu}^2$ has vanished,\footnote{The fact that the subleading single-logarithmic terms $\alpha_s \log \hat{\mu}$ vanished as well is an accident, and will not in general be true for the $\tilde{Q}$ splitting functions or for gluon splitting functions.} and to leading order in $\alpha_s$ one finds
%%%
\begin{equation}
\sigma = B \biggl[ 1 + \frac{\alpha_s C_F}{4 \pi} (5-12\log 2) + \ord(\alpha_s^2) \biggr]
\,.\end{equation}
%%%
Note that the Sudakov-improved and tree-level results for $\df\sigma_{3}$ agree to leading order in $\alpha_s$.

\eq{eq:sudakovimproved} is by no means unique. For example, the dominant shower history used in the CKKW prescription \cite{Catani:2001cc} is morally equivalent to
%%%
\begin{center}
\bf Dominant History Sudakov-Improved (LO/LL):
\end{center}
\begin{align}
\label{eq:ckkwsudakovimproved}
\sigma_{2}(\hat{\mu}^2)
&= B\, \Delta_Q^2(1,\hat{\mu}^2)
\,,\nn\\
\frac{\df\sigma_{3a}(t_a)}{\df t_a \df t_b}
&= \frac{\alpha_s C_F}{2 \pi}\, B\, \frac{R(t_a,t_b)}{t_a t_b}\, \theta(t_b > t_a)\, \Delta_Q^2(1,t_a) \,,\nn \\
\frac{\df\sigma_{3b}(t_b)}{\df t_a \df t_b}
&= \frac{\alpha_s C_F}{2 \pi} \,B\, \frac{R(t_a,t_b)}{t_a t_b}\, \theta(t_a > t_b)\, \Delta_Q^2(1,t_b)
\,.\end{align}
%%%
The reason that this expressions works is that
%%%
\begin{equation}
\frac{R(t_a,t_b)\, \theta(t_b > t_a) - Q_a(t_a,t_b)}{t_a t_b} = \text{finite as }t_a \to 0
\,.\end{equation}
%%%
The lack of the $\mathbf{\tilde{2}}$ integrations in $\sigma_{2}$ can contribute at most at the subleading-logarithmic level. As expected, there is no leading-logarithmic dependence in the total cross section for the dominant history method\footnote{Again, the lack of subleading single-logarithmic terms $\alpha_s \log \hat{\mu}$ is an accident.}
%%%
\begin{equation}
\sigma = B \biggl[ 1 + \frac{\alpha_s C_F}{4 \pi} \biggl(5 - \frac{2 \pi^2}{3} - 12 \log 2 \biggr) + {\cal O}(\alpha_s^2) \biggr]\,.
\end{equation}
%%%

%===============================================================================
\subsection{Partonic Cross Sections at NLO}
%===============================================================================

We now want to include the virtual corrections to the 2-jet rate to obtain partonic expressions that give both differential distributions as well as the total cross section correct to $\ord(\alpha_s)$. As in the toy model, we will slowly build up towards an answer that has all the desired properties. Ignoring the leading logarithmic dependence of the partonic cross sections, we can use an NLO slicing method~\cite{Fabricius:1981sx, Kramer:1986mc, Bergmann:1989zy, Giele:1991vf, Giele:1993dj, Harris:2001sx}
%%%
\begin{center}
\bf NLO Slicing (NLO):
\end{center}
\begin{align}
\label{eq:QCDslicing}
\sigma_{2}(\hat{\mu}^2) &= \tilde{B}(\hat{\mu}^2)
\,,\nn\\
\frac{\df\sigma_{3}}{\df t_a \df t_b}
&= \frac{\alpha_s C_F}{2 \pi}\, B\, \frac{R(t_a,t_b)}{t_at_b}
\,,\end{align}
%%%
where
%%%
\begin{equation}
\tilde{B}(\hat{\mu}^2)
= B\, \biggl[1 + \frac{\alpha_s C_F}{2 \pi}\, \lim_{\epsilon\to0} \biggl( -\frac{2}{\epsilon^2}
- \frac{3}{\epsilon} + V + \int_{\mathbf{2}+\mathbf{\tilde{2}_a}+\mathbf{\tilde{2}_b}} \!\! \df t_a\, \df t_b \, \frac{R(t_a,t_b)}{t_a^{1+2\epsilon} \, t_b^{1+2\epsilon}}\biggr) \biggr]
\,.\end{equation}
%%%
Note the cancellation between the terms that are divergent as $\epsilon \to 0$, such that $\tilde B$ is a finite expression, which only depends on $\hat{\mu}^2$. It is a simple exercise to show that the total cross section reproduces the NLO cross section in \eq{eq:QCDsigmatotal}. As for the LO calculation, we do not have enough information to separately determine $\df\sigma_{3a}$ and $\df\sigma_{3b}$ at this point.

While \eq{eq:QCDslicing} does reproduce both the differential and total cross sections correctly to $\ord(\alpha_s)$, it does not sum logarithms and has large $\hat{\mu}^2$ dependence. The leading logarithmic resummation can be included employing a subtraction method similar to MC@NLO~\cite{Frixione:2002ik}, and we find
%%%
\begin{center}
\bf NLO Subtraction (NLO/LL):
\end{center}
\begin{align}
\label{eq:QCDsubtraction}
\sigma_{2}(\hat{\mu}^2)
&= \bar{B}(\hat{\mu}^2) \biggl\{ \Delta_Q^2(1,\hat{\mu}^2) + \phantom{\frac{1}{1}}
\nn\\ &\quad
+ \frac{\alpha_s C_F}{2 \pi} \biggl[\int_{\mathbf{\tilde{2}_a}}\!\df t_a\, \df t_b\, \frac{Q_a(t_a,t_b)}{t_a t_b}\, \Delta_Q^2(1,t_a)
+ \int_{\mathbf{\tilde{2}_b}}\!\df t_a\, \df t_b\, \frac{Q_b(t_a,t_b)}{t_a t_b}\, \Delta_Q^2(1,t_b) \biggr] \biggr\}
\,,\nn\\
\frac{\df\sigma_{3}}{\df t_a \df t_b}
&= \frac{\alpha_s C_F}{2 \pi} \, \biggl[ B \, \frac{R(t_a,t_b) - Q_a(t_a,t_b)-Q_b(t_a,t_b)}{t_a,t_b}
\nn\\ &\quad
+ \bar{B}(\hat{\mu}^2)\biggl( \frac{Q_a(t_a,t_b)}{t_a t_b }\, \Delta_Q^2(1,t_a)+ \frac{Q_b(t_a,t_b)}{t_at_b}\, \Delta_Q^2(1,t_b) \biggr) \biggr]
\,,\end{align}
%%%
where
%%%
\begin{equation}
\bar{B}(\hat{\mu}^2) = B \biggl[ 1 + \frac{\alpha_s C_F}{2 \pi} \biggl(\frac{3}{2} - \int_{\mathbf{3}}\!\df t_a\, \df t_b, \frac{R(t_a,t_b) - Q_a(t_a,t_b)-Q_b(t_a,t_b)}{t_at_b} \biggr)\biggr]
\,.\end{equation}
%%%
Again, this reproduces the correct differential 3-jet distributions at $\ord(\alpha_s)$ as well as the total NLO cross section of \eq{eq:QCDsigmatotal}. \eq{eq:QCDsubtraction} only defines $\df\sigma_3/(\df t_a \df t_b)$, but not the individual $\df\sigma_{3i}/(\df t_a \df t_b)$. To avoid large $\log^2(t_a/t_b)$ logarithms, we have many options. The simplest is to use \eq{eq:alphaseparation} to define $\df\sigma_{3a}$ and $\df\sigma_{3b}$ via
%%%
\begin{equation}
\label{eq:QCDalphadef}
\alpha_a = \frac{Q_a}{Q_a + Q_b}, \qquad \alpha_b = \frac{Q_b}{Q_a+Q_b}
\,.\end{equation}
%%%
To see that this works, imagine that $R = Q_a + Q_b$, and note that
%%%
\begin{equation}
\frac{Q_a}{Q_a + Q_b} \to 1 \text{ as } t_a\to 0
\,,\end{equation}
%%%
which shows that the singularity structure will correspond to the right leading-logarithmic behavior. In fact, \eq{eq:QCDalphadef} is a general strategy for minimizing the effect of leading-logarithms between intermediate scales, and we will use a variant of this approach in the next section.

As was true in the case of the toy model, the particular form of the NLO/LL partonic cross section in \eq{eq:QCDsubtraction} is not unique. The criteria we wanted to satisfy is the proper resummation of the leading double-logarithms while simultaneously reproducing all observables to $\ord(\alpha_s)$ accuracy. As in the toy model, we can get a simple form for an NLO/LL partonic cross sections in analogy to Ref.~\cite{Nason:2004rx}:
%%%
\begin{center}
\bf NLO Elegant (NLO/LL):
\end{center}
\begin{align}
\label{eq:QCDelegant}
\sigma_{2}(\hat{\mu}^2)
&= \sigma_{\rm NLO}\, \Delta_T(\hat{\mu}^2)
\,,\nn \\
\frac{\df\sigma_{3}}{\df t_a \df t_b}
&= \frac{\alpha_s C_F}{2 \pi}\, \sigma_{\rm NLO}\, \frac{T(t_a,t_b)}{t_at_b}\, \Delta_T[\min(t_a,t_b)]
\,,\end{align}
%%%
where $\sigma_{\rm NLO}$ is given in \eq{eq:QCDsigmatotal}. The function $T(t_a,t_b)$ is defined by
%%%
\begin{equation}
T(t_a,t_b) = \frac{B}{\sigma_{\rm NLO}}\, R(t_a,t_b)
\,,\end{equation}
%%%
and has an accompanying ``Sudakov factor''
%%%
\begin{equation}
\Delta_T(t) = \exp \biggl[ -\frac{\alpha_s C_F}{2 \pi} \int_{\mathbf{3}(t)}\! \df t_a\, \df t_b\, \frac{T(t_a,t_b)}{t_at_b} \biggr]
\,,\end{equation}
%%%
where $\mathbf{3}(t)\equiv \df\Phi_3(t)$ denotes the region of phase space with $t_a,t_b > t$. Note that $\Delta_T(t)$ has the same double-logarithmic dependence as $\Delta_Q^2(1,t)$, ensuring that the leading-logarithmic dependence is reproduced correctly. To order $\alpha_s$, we are also clearly reproducing both the differential spectrum and the total rate. As before, $\df\sigma_{3a}$ and $\df\sigma_{3b}$ can be defined using \eqs{eq:alphaseparation}{eq:QCDalphadef} to minimize the effect of $\log^2(t_a/t_b)$ logarithms. \eq{eq:QCDelegant} will be the cornerstone for the NLO/LO/LL result we will present in the next section.

In our entire discussion above, we never talked about angular dependence. To keep the discussion simple, we have implicitly assumed that the angular dependence factorizes, but we do have to be mindful that in general, operator mixing can break this factorization. In any case, we can easily generalize \eq{eq:QCDelegant} to have angular dependence by
%%%
\begin{center}
\bf NLO Elegant with Angles (NLO/LL):
\end{center}
\begin{align}
\label{eq:QCDelegantangles}
\frac{\df\sigma_{2}(\hat{\mu}^2)}{d \Omega^T_2}
&= \sigma_{\rm NLO}\, f(\Omega^T_2)\, \Delta_T(\hat{\mu}^2)
\,,\nn \\
\frac{\df\sigma_{3}}{d \Omega^T_2 \df t_a \df t_b d\phi}
&= \frac{\alpha_s C_F}{2 \pi}\, B\, \frac{R(\Omega^T_2,t_a,t_b, \phi)}{t_at_b}\, \Delta_T[\min(t_a,t_b)]
\,,\end{align}
%%%
where the normalized $f(\Omega^T_2)$ function contains angular information about 2-jet distributions and we have introduced the generalized real emission function $R(t_a,t_b) \to R(\Omega^T_2, t_a, t_b,\phi)$. There are two different ways of defining the function $T$ that appears in the Sudakov factor,
%%%
\begin{equation}
T(\Omega^T_2,t_a,t_b, \phi)
= \frac{B}{\sigma_{\rm NLO}}\, R(\Omega^T_2, t_a, t_b,\phi)
\qquad \text {or} \qquad
\frac{B}{\sigma_{\rm NLO}}\, \frac{R(\Omega^T_2, t_a, t_b,\phi)}{f(\Omega^T_2)}
\,.\end{equation}
%%%
In either case, the Sudakov factor is given by
%%%
\begin{equation}
\Delta_T(t) = \exp \biggl[ -\frac{\alpha_s C_F}{2 \pi} \int_{\mathbf{3}(t)}\! \df t_a\, \df t_b \int \! \df \Omega^T_2\, \df \phi\, \frac{T(\Omega^T_2, t_a, t_b,\phi)}{t_at_b} \biggr]
\,.\end{equation}
%%%
The first choice is easier to implement and is the strategy adopted in this paper. The second option is likely to be more correct, but the difference can at most be at the subleading-logarithmic level.

%%%%%%%%%%%%%%%%%%%%%%%%%%%%%%%%%%%%%%%%%%%%%%%%%%%%%%%%%%%%%%%%%%%%%%%%%%%%%%%%
\section{NLO/LO/LL Merging}
\label{sec:NLO}
%%%%%%%%%%%%%%%%%%%%%%%%%%%%%%%%%%%%%%%%%%%%%%%%%%%%%%%%%%%%%%%%%%%%%%%%%%%%%%%%

For the first emission in QCD, we could understand everything analytically, but the calculations quickly become unwieldy with more than one emission. Both for simplicity of discussion and simplicity of the \GenEvA\ strategy, it is necessary to define more generally how to construct an LO/LL sample and how to merge together nested descriptions of QCD.

As we saw in the previous section, in order to properly treat the double-logarithms of $t_a/t_b$, we needed two event samples, $\df \MC_3(t_a)$ and $\df \MC_3(t_b)$. Since the parton shower has two possible histories to generate the 3-parton final state ($q\to qg$ or $\bar q \to \bar q g$ splitting), these two event samples can be interpreted as representing these two possible shower histories. This will generalize for more partonic final states, and we will have a master formula which contains different event samples for each possible shower history, in the process giving an unambiguous definition for the phase space projection in \eq{eq:PSprojection}. This will allow us to write the NLO/LO/LL sample by combining the NLO/LL, LO/LL, and LL descriptions, just as in the toy model.

%===============================================================================
\subsection{Generalized Master Formula}
%===============================================================================

To  build the generalized master formula, we first introduce the concept of a would-be shower history. Though phase space with a matching scale does not require us to ever talk about parton shower histories, they are convenient for resolving logarithmic ambiguities since shower histories track QCD singularities, so one should really think of these histories as ``singularity histories''. All of the possible leading-logarithmic ambiguities can be captured by introducing a separate event sample for every kind of allowed shower (or singularity) history. This gives the generalized master formula:
%%%
\begin{equation}
\label{eq:genmasterformula}
\df \sigma = \sum_{n = 2}^{n_\max} \sum_j \frac{\df\sigma_{nj}(\mu_{nj})}{\df \Phi_n}\, \df \MC_n(\mu_{nj})
\,,\end{equation}
%%%
where $j$ labels different shower histories for $n$ final states, and we start running the phenomenological shower at the scale $\mu_{nj}$ for each sample. Note that, similarly to \eq{eq:QCDonemaster}, \eq{eq:genmasterformula} does not introduce double-counting,  but merely reflects the scale ambiguity of QCD, and we anticipate defining $\df\sigma_{nj}$ in analogy with \eq{eq:sumalphacomponents}.

As anticipated in \sec{subsec:newphasespace}, $\df \MC_n(\mu_{nj})$ is defined by excising regions of $n$-body phase space that are covered by running the phenomenological model on $(m < n)$-body phase space starting from the various $\mu_{mj}$ scales. With so many matching scales, it is convenient to choose $\mu_{nj}$ equal to a common scale $\mu$, except for $\mu_{n_\max j}$, which has to be equal to a scale ``$t_j$'' in order to cover all of phase space without $\alpha_s \log^2 (t_j / t_k)$ ambiguities. Because scale-dependent $n$-body phase space does not depend on $\mu_n$, there is no need for special treatment of the $n_\max$-body phase space. Assuming that the phenomenological model is a virtuality-ordered shower, the excised phase space is
%%%
\begin{equation}
\df \Phi_n(\mu) = \df \Phi_n(p_{\rm CM}; p_1, p_2, \ldots, p_n)
\prod_{i \diamond j} \theta\bigl[(p_i + p_j)^2 > \mu^2 \bigr]
\,,\end{equation}
%%%
where $i \diamond j$ indicates pairs of partons that are associated with a QCD singularity.

Since we are choosing all of the $\mu_{n<n_\max}$ scales to be the same, the master formula in \eq{eq:genmasterformula} is highly redundant for $n<n_\max$, because many of the $\df \MC_n(\mu_{nj})$ event samples are multiply covered. Despite this redundancy, keeping the notion of a would-be shower history is convenient for the purposes of having a definition of phase space projection as in \eq{eq:PSprojection}. That is, a shower history defines a map between $n$-body phase space and a lower-dimensional $m$-body phase space
%%%
\begin{equation}
\label{eq:genweaktruncation}
\{ \Phi_n, (nj) \} \to \{\Phi_{m}, (mk) \} \qquad \text{ for } m < n\,,
\end{equation}
%%%
where the specifics of this map are determined by the details of a showering or clustering procedure. Note that this map does not necessarily give a unique map $\Phi_n \to \Phi_m$, but rather gives a series of different consistent maps depending which shower history $j$ is being selected. One could choose to implement a unique map $\Phi_n \to \Phi_m$ by only considering one type of shower history for a given phase space point.

Given a phase space point with a would-be shower history $\{ \Phi_n, (nj) \}$, we can always define the scale $t_j$ that we would need to start the phenomenological shower from to eliminate leading-logarithmic dependence. To do so, we simply ask at what scale the phenomenological shower would have generated the $n$-th branching. This is the scale we use to define $\mu_{n_\max j}$, and it guarantees that there are no phase space dead zones.

%===============================================================================
\subsection{Sudakov Improvements}
%===============================================================================

We now wish to extend the LO/LL result of \eq{eq:sudakovimproved} by adding additional tree-level emissions. As discussed above, an LO/LL result is obtained by supplementing a tree-level matrix element with appropriate Sudakov factors. That is, we wish to combine the $n$-body tree-level matrix element (LO)
%%%
\begin{equation}
\label{eq:multitreelevel}
\frac{\df\sigma^{\rm LO}_{n}}{\df \Phi_n}
\end{equation}
%%%
that has no notion of a matching scale $\mu$, with Sudakov factors taken from the equivalent partonic cross section for a set of $n$-body parton shower histories (LL)
%%%
\begin{equation}
\frac{\df\sigma^{\rm LL}_{nj}(\mu_{nj})}{\df \Phi_n},
\end{equation}
%%%
where $j$ labels the different shower histories for $n$ final states.

Note that schematically, the expression for the LL ``calculation'' can be written as a product of splitting functions and Sudakov factors,
%%%
\begin{equation}
\label{eq:trialshowern}
\frac{\df\sigma^{\rm LL}_{nj}(\mu_{nj})}{\df \Phi_n}
= B\, \prod_{r=1}^{n-2} Q_{j,r} \prod_{s=1}^{2n-2} \Delta_{j,s}
\equiv B\, Q_j\, \Delta_j(\mu_{nj})
\,,\end{equation}
%%%
where  $B$ is the initial hard scattering matrix element, $r$ labels the different $1 \to 2$ splitting function vertices, and $s$ labels the partons in the shower history. For convenience, we have introduced the notation
%%%
\begin{equation}
Q_j = \prod_{r=1}^{n-2} Q_{j,r}\,, \qquad \Delta_j(\mu_{nj}) = \prod_{s=1}^{2n-2} \Delta_{j,s}
\,,
\end{equation}
%%%
such that for the $j$-th shower history, $Q_j$ is the product of all splitting functions and $\Delta_j(\mu_{nj})$ is the product of all Sudakov factors. Since the parton shower correctly reproduces the singularity structure of QCD, the tree-level calculation should share the same singularity structure as the sum over all shower histories
%%%
\begin{equation}
\label{eq:singularregions}
\lim_{\Phi_n \to \mathrm{sing.}} \left[\frac{\df\sigma^{\rm LO}_{n}}{\df \Phi_n} - B \sum_i Q_i \right] = \text{finite}.
\end{equation}
%%%

Accounting for these singularities, one possible Sudakov-improved partonic cross section is
%%%
\begin{center}
\bf Sudakov-Improved (LO/LL):
\end{center}
\begin{equation}
\label{eq:QCDgensudakovimproved}
\frac{\df\sigma^{\rm LO/LL}_{nj}(\mu_{nj})}{\df \Phi_n}
= \frac{\df\sigma^{\rm LO}_{n}}{\df \Phi_n} \biggl( B \sum_i Q_i \biggr)^{-1} \frac{\df\sigma^{\rm LL}_{nj}(\mu_{nj})}{\df \Phi_n}
\,.\end{equation}
%%%
Expanding this result to leading order in $\alpha_s$, all of the $\Delta$ factors in $\df\sigma^{\rm LL}_{nj}$ are equal to unity, so summing over all of the $\df\sigma^{\rm LO/LL}_{nj}$ expressions yields the tree-level result as desired. By construction, this answer has the correct leading-logarithmic behavior because it has the same singularity structure as the parton shower.

\eq{eq:QCDgensudakovimproved} is a correct LO/LL answer, but there are various choices we can make for $\df\sigma^{\rm LL}_{nj}$. The first is to use the naive LL cross section from \eq{eq:trialshowern}, which includes the sum over all possible shower histories. One can simplify this result, by following the path taken by CKKW \cite{Catani:2001cc}, and use the dominant shower history, defined as the one with the largest value of $Q_j$. In QCD, there is never an ambiguity as to which shower history is dominant, and the dominant shower history always reproduces the QCD singularities. Thus,
%%%
\begin{equation}
\frac{Q_{\rm dom}}{\sum_j Q_j} \to 1
\,, \qquad \frac{Q_{\rm other}}{\sum_j Q_j} \to 0
\,.
\end{equation}
%%%
Taking these limits as equalities, the CKKW procedure essentially defines:
%%%
\begin{equation}
\text{\bf Dominant History Sudakov-Improved (LO/LL):} \nn
\end{equation}
%%%
%%%
\begin{equation}
\label{eq:QCDdomsudakovimproved}
\frac{\df\sigma^{\rm LO/LL}_{nj}(\mu_{nj})}{\df \Phi_n} = \left\{ \begin{array}{r l} \displaystyle \frac{\df\sigma^{\rm LO}_{n}}{\df \Phi_n} \Delta_j(\mu_{nj}) & \quad j = \mathrm{dom}, \\ 0 &\quad j = \mathrm{other}. \end{array} \right.
\end{equation}
%%%
\eqs{eq:QCDgensudakovimproved}{eq:QCDdomsudakovimproved} share the same singularity structure, so they are both valid LO/LL results.

A different route is to try to improve on \eq{eq:trialshowern}. As we saw in \eq{eq:ll}, there are additional $\mathbf{\tilde{n}}$ integrations that are necessary to get the total cross section correct in the parton shower. From the shower point of view, these integrations correspond to situations where the parton shower populates a region of phase space that has QCD singularities but uses a splitting function away from the singular region. That is, two partons are ``accidentally'' distributed closer than the cutoff of the shower. Schematically, we can write this as
%%%
\begin{equation}
\label{eq:tildenadded}
\frac{\df\sigma^{\rm LL}_{nj}(\mu_{nj})}{\df \Phi_n} = B \left( Q_j\, \Delta_j + \int_{\sum \mathbf{\tilde{n}_k}} \!\! Q_k\, \Delta_k \right)
\,.\end{equation}
%%%
These $\mathbf{\tilde{n}}$ regions do not exist as separate event samples, but their effect should be captured in the $n$-body event sample.

At this point, it is not at all clear how to figure out which $\mathbf{\tilde{n}_k}$ integrations correspond to which shower histories $j$.
At the leading-logarithmic level, these $\mathbf{\tilde{n}}$ integrations can simply be ignored because they are not associated with the dominant QCD singularities and give at most subleading-logarithmic corrections. However, for high enough multiplicity, the phase space volume of all the $\mathbf{\tilde{n}}$ regions is rather large, especially for moderate $\mu$, so numerically it is dangerous to ignore them. On the other hand, if the only reason we are keeping the $\mathbf{\tilde{n}}$ integrations is for overall normalization, then it would be simpler to just introduce \emph{ad hoc} $k$-factors to restore the normalization.

As discussed in the companion paper \cite{genevatechnique}, there is a numerically efficient way to accomplish the $\mathbf{\tilde{n}}$ integrations, so the user can address the $\mathbf{\tilde{n}}$ subtleties if desired. In particular, the \GenEvA\ algorithm has a way to figure out which ``wrong'' histories $k$ should be associated with which ``right'' history $j$, and it does this in a way that is reasonably efficient. Because the resulting differential distributions using \eqs{eq:trialshowern}{eq:tildenadded} are different, the two different choices can be used to test for systematics in the LO/LL sample. Note that for $n=3$, using \eq{eq:QCDgensudakovimproved} with \eq{eq:tildenadded} reproduces the result of \eq{eq:sudakovimproved}. To show our results in \sec{sec:results}, we will always use \eq{eq:tildenadded} because of the reduced cross section scale dependence.

%===============================================================================
\subsection{Merging Nested Descriptions}
%===============================================================================

In the case of LO/LL merging in \eq{eq:QCDgensudakovimproved} or NLO/LL merging in \eq{eq:QCDelegant}, we were trying to merge together two different descriptions of the physics that both give valid predictions at some scale $\mu$. More generally, we are interested in cases like in \fig{fig:nestedtheories} where there is one description that gives a better description of the physics at some high scale and one that gives a better description at a lower scale. To build an NLO/LO/LL merged sample, we want to use an NLO/LL result at high energies, but supplement it with additional LO/LL information for subsequent emissions.

From the toy model in \eq{eq:toybest} we learned that the NLO/LO/LL result for the partonic results with more than three partons are given by the LO/LL result, except that the first Sudakov factor changed to $\Delta_T$. How do we obtain this new Sudakov factor, which depends on the kinematics of the first emissions? Consider the differences between the $\df\sigma_{3}$ expressions for the LO/LL and NLO/LL merged samples in \eqs{eq:sudakovimproved}{eq:QCDelegant}. We will use the convention of \eq{eq:QCDalphadef} to define $\sigma^{\rm NLO/LL}_{3a}$ for the NLO/LL result. The ratio of the two answers is
%%%
\begin{equation}
\label{eq:NLO2LOcorrection}
{\frac{\df\sigma^{\rm NLO/LL}_{3a}(t_a)}{\df t_a \df t_b}} \Bigg/
{\frac{\df\sigma^{\rm LO/LL}_{3a}(t_a)}{\df t_a \df t_b}}
= \frac{\Delta_T[\min(t_a,t_b)]}{\Delta_Q^2(1,t_a)}
\,,\end{equation}
%%%
and therefore can be used to extract the Sudakov factor $\Delta_T$ from the NLO/LL calculation.\footnote{The fact that the numerator of \eq{eq:NLO2LOcorrection} is a function of $\min(t_a,t_b)$ instead of $t_a$ means that there are power suppressed logarithmic ambiguities of the form $(t_a/t_b) \log(t_a/t_b)$, which are beyond the order we are working.}

While \eq{eq:NLO2LOcorrection} is a function of 3-body phase space, we want to apply this correction factor to $n$-body matrix elements. This is straightforward, using the map from \eq{eq:genweaktruncation} to define the map
%%%
\begin{equation}
\label{eq:nto3weaktruncation}
\{ \Phi_n, (nj) \} \to \{\Phi_{3}, (3k) \} \qquad \text{for } n >3
\,.\end{equation}
%%%
Note that this implies that the specific value of the correction factor will depend on the particular would-be shower history and on the details of the $\Phi_n \to \Phi_3$ map. By assumption, the map in \eq{eq:genweaktruncation} respects QCD singularities, so any differences are formally beyond the order we are working to, but different choices will affect the specifics of the NLO/LO/LL merging.

Putting these pieces together, we find an expression for the best partonic calculation that \GenEvA\ can currently implement
%%%
\begin{center}
\bf GenEvA Best (NLO/LO/LL):
\end{center}
\begin{align}
\label{eq:QCDbest}
\frac{\df\sigma^{\rm NLO/LO/LL}_{2}(\mu_2)}{\df \Phi_2}
&= \frac{\df\sigma^{\rm NLO/LL}_{2}(\mu_2)}{\df \Phi_2}
\,,\nn \\
\frac{\df\sigma^{\rm NLO/LO/LL}_{3j}(\mu_{3j})}{\df \Phi_3}
&= \frac{\df\sigma^{\rm LO/LL}_{3j}(\mu_{3j})}{\df \Phi_3} \times \frac{\df\sigma^{\rm NLO/LL}_{3j}(t_j)}{\df \Phi_3} \Bigg/ \frac{\df\sigma^{\rm LO/LL}_{3j}(t_j)}{\df \Phi_3}
\,,\nn\\&\,\,\,\vdots\nn\\
\frac{\df\sigma^{\rm NLO/LO/LL}_{nj}(\mu_{nj})}{\df \Phi_n}
&= \frac{\df\sigma^{\rm LO/LL}_{nj}(\mu_{nj})}{\df \Phi_n} \times \frac{\df\sigma^{\rm NLO/LL}_{3k}(t_{k})}{\df \Phi_3} \Bigg/ \frac{\df\sigma^{\rm LO/LL}_{3k}(t_{k})}{\df \Phi_3}
\,,\end{align}
%%%
where the value of $n_\max$ is determined by the availability of high-multiplicity tree-level matrix elements, and $(3k)$ is determined uniquely from $(nj)$ by \eq{eq:nto3weaktruncation}. Note that \eq{eq:NLO2LOcorrection} is still needed to define $\df\sigma_{3j}$. This is analogous to \eq{eq:toybest}, where an additional Sudakov factor $\Delta_Q$ was needed once extra emissions were added. The reason is, that the NLO/LL sample had $n_\max = 3$ and therefore took $\mu_{3j=t_j}$, whereas in general the matching scale $\mu_{3j}$ will be lower than $t_j$. Taking $n_\max = 3$ and $\mu_{3j} = t_j$, \eq{eq:QCDbest} reduces to \eq{eq:QCDelegant}.

As in the toy model, \eq{eq:QCDbest} is a special case of a more general construction to merge together nested descriptions of QCD. In analogy with \eq{eq:toybest}, if we have partonic descriptions $\bf A$, $\bf B$, $\bf C$, $\ldots$, then these can be consistently merged via
%%%
\begin{center}
\bf Best Combination (A/B/C/$\mathbf{\cdots}$):
\end{center}
\begin{align}
\label{eq:aheadbest}
\frac{\df\sigma^\mathrm{A/B/C/\cdots}_{nj}(\mu_{nj})}{\df \Phi_n}
=\frac{\df\sigma^{\rm A}_{nj}(\mu_{nj})}{\df \Phi_n}
&\times
\biggl[\frac{\df\sigma^{\rm B}_{n'j'}(t_{j'})}{\df \Phi_{n'}} \Bigg/ \frac{\df\sigma^{\rm A}_{n'j'}(t_{j'})}{\df \Phi_{n'}} \biggr]
\nn \\ &\times
\biggl[\frac{\df\sigma^{\rm C}_{n''j''}(t_{j''})}{\df \Phi_{n''}} \Bigg/ \frac{\df\sigma^{\rm B}_{n''j''}(t_{j''})}{\df \Phi_{n''}} \biggr]
\times \dotsb
\,.\end{align}
%%%
This merging procedure does not spare the user from the need to first merge N$^i$LO and N$^j$LL descriptions that occupy the same $n$-body phase space. Rather, \eq{eq:aheadbest} gives the user one simple option how to supplement low-scale/high-multiplicity calculations with additional high-scale/low-multiplicity information.

%%%%%%%%%%%%%%%%%%%%%%%%%%%%%%%%%%%%%%%%%%%%%%%%%%%%%%%%%%%%%%%%%%%%%%%%%%%%%%%%
\section{Results}
\label{sec:results}
%%%%%%%%%%%%%%%%%%%%%%%%%%%%%%%%%%%%%%%%%%%%%%%%%%%%%%%%%%%%%%%%%%%%%%%%%%%%%%%%

In this section, we present the results obtained using the \geneva\ program, which implements the \geneva\ framework, as explained in \sec{subsec:summary}. As such, it uses the \geneva\ algorithm to generate Lorentz-invariant phase space, and then reweights the resulting events to the distributions discussed in this work. The current implementation of \geneva\ is capable of describing $e^+e^- \to n \text{ jets}$, but only describes massless quarks adequately. Therefore, we do not present results for the production of $b$-quark jets. A validation of the \geneva\ event generator is given in the companion paper~\cite{genevatechnique}.

For the fixed-order tree-level matrix elements we use the \texttt{HELAS}~\cite{Murayama:1992gi} Fortran routines generated by \madgraph~\cite{Stelzer:1994ta}, and \geneva\ currently utilizes matrix elements with up to $n_\max = 6$ final state partons. Final states with more than $n_\max$ partons are obtained by a subsequent parton shower, and in general any virtuality-ordered shower could be used for that purpose. For simplicity, here we only use the internal analytic parton shower of the \GenEvA\ algorithm, which is a virtuality-ordered shower that neither includes color coherence, $\alpha_s$ running, nor the hadronization of the final state. For this reason, the results presented here should be viewed as a proof of concept, and not as a prediction of the shown distributions. Unless otherwise noted, all results in this section use
%%%
\begin{equation}
E_{\rm CM} = 1000 \GeV
\,, \qquad
\mu = 50\GeV
\,, \qquad
\Lambda_{\rm IR} = 10 \GeV
\,.\end{equation}
%%%
As we will see in \fig{fig:sigma_mudep}, the matching scale $\mu = 50\GeV$ is rather low in the sense that the single-logarithmic dependence that we do not account for becomes important. However, we choose a low scale in order to accentuate the differences between different partonic calculations and leave enough phase space volume available for high-multiplicity partonic states.

\begin{table}[t]
 \begin{tabular}{r || c | l}
 Notation & Given by Eq. & Description \\
 \hline\hline
 LO$_{n_\max}$ & \eqref{eq:multitreelevel} & Tree-Level (LO)
 \\\hline
 LO$_{n_\max}$/LL & \eqref{eq:QCDgensudakovimproved} with \eqref{eq:tildenadded} & Sudakov-Improved (LO/LL)
 \\
 NLO$_2$/LO$_3$/LL & \eqref{eq:QCDelegant} & NLO Elegant (NLO/LL)
 \\\hline
 NLO$_2$/LO$_{n_\max}$/LL & \eqref{eq:QCDbest} & \geneva\ Best (NLO/LO/LL)
 \\\hline
 \end{tabular}
\caption{Notation and meaning for the various event samples. LO$_{n_\max}$ corresponds to tree-level matrix elements with $2$- through $n_\max$-emissions, and LL indicates the inclusion of leading-logarithmic information. In this work, we only consider virtual diagrams for 2-parton final states, denoted by NLO$_2$. Note that the 3-parton matrix elements in the LO$_{3}$/LL and NLO$_2$/LO$_3$/LL samples are \emph{not} the same, as \eqs{eq:sudakovimproved}{eq:QCDelegant} include different Sudakov factors. LO$_{n_\max}$/LL and NLO$_2$/LO$_3$/LL provide implementations of PS/ME merging and PS/NLO merging, respectively. The NLO$_2$/LO$_{n_\max}$/LL sample is the best partonic calculation implemented by \GenEvA, and simultaneously achieves PS/ME merging and PS/NLO merging.}
\label{tab:notation}
\end{table}

The notation for the various event samples used in this section is summarized in Table~\ref{tab:notation}. We use the notation LO$_{n_\max}$ for tree-level matrix elements with $2\leq n \leq n_\max$ partons in the final state, and LL indicates that leading-logarithmic Sudakov resummation is being used. Unlike in the companion paper \cite{genevatechnique}, we always use fully merged samples that give a complete covering of phase space. For example, LO$_4$ includes 2-, 3-, and 4-parton tree-level matrix elements and the remaining phase space is covered by the internal parton shower. The same is true for LO$_4$/LL, except that the matrix elements are now Sudakov-improved according to \eq{eq:QCDgensudakovimproved}. Note that even without the Sudakov improvement all divergences in LO$_n$ are regulated by the matching scale $\mu$. Note also, that since \GenEvA\ has no dead zones, the LO$_2$ sample is identical to LO$_2$/LL, because when only $2 \to 2$ matrix elements are used the parton shower always starts running at the center-of-mass energy. The matching scale $\mu$ can also affect the meaning of $n_\max$. For example, an LO$_4$/LL sample with $\mu = E_{\rm CM}$ would be identical to an LO$_2$/LL sample because in this case $\mu$ would be too high to allow for any additional emissions above $\mu$.

For one-loop corrected matrix elements with $n \leq n_\max$ final state partons we write NLO$_{n_\max}$, and we currently only use NLO$_2$. For example, an NLO$_2$/LL result corresponds to an LO$_2$/LL sample supplemented with an appropriate NLO $k$-factor, while NLO$_2$/LO$_3$/LL corresponds to fully consistent $\ord(\alpha_s)$ results, which contain all $\ord(\alpha_s)$ corrections from both virtual 2-parton and real-emission 3-parton diagrams, and including leading-log resummation according to \eq{eq:QCDelegant}. Note that the 3-parton matrix elements in the LO$_{3}$/LL and NLO$_2$/LO$_3$/LL samples include different Sudakov factors, as can be seen comparing \eqs{eq:sudakovimproved}{eq:QCDelegant}. An NLO$_2$/LO$_4$/LL sample additionally incorporates 4-parton tree-level matrix elements following \eq{eq:QCDbest}, and similarly for NLO$_2$/LO$_{n_\max}$/LL.

%===============================================================================
\subsection{Total Cross Section Scale Dependence}
%===============================================================================

\begin{figure}
\includegraphics[width=0.5\textwidth]{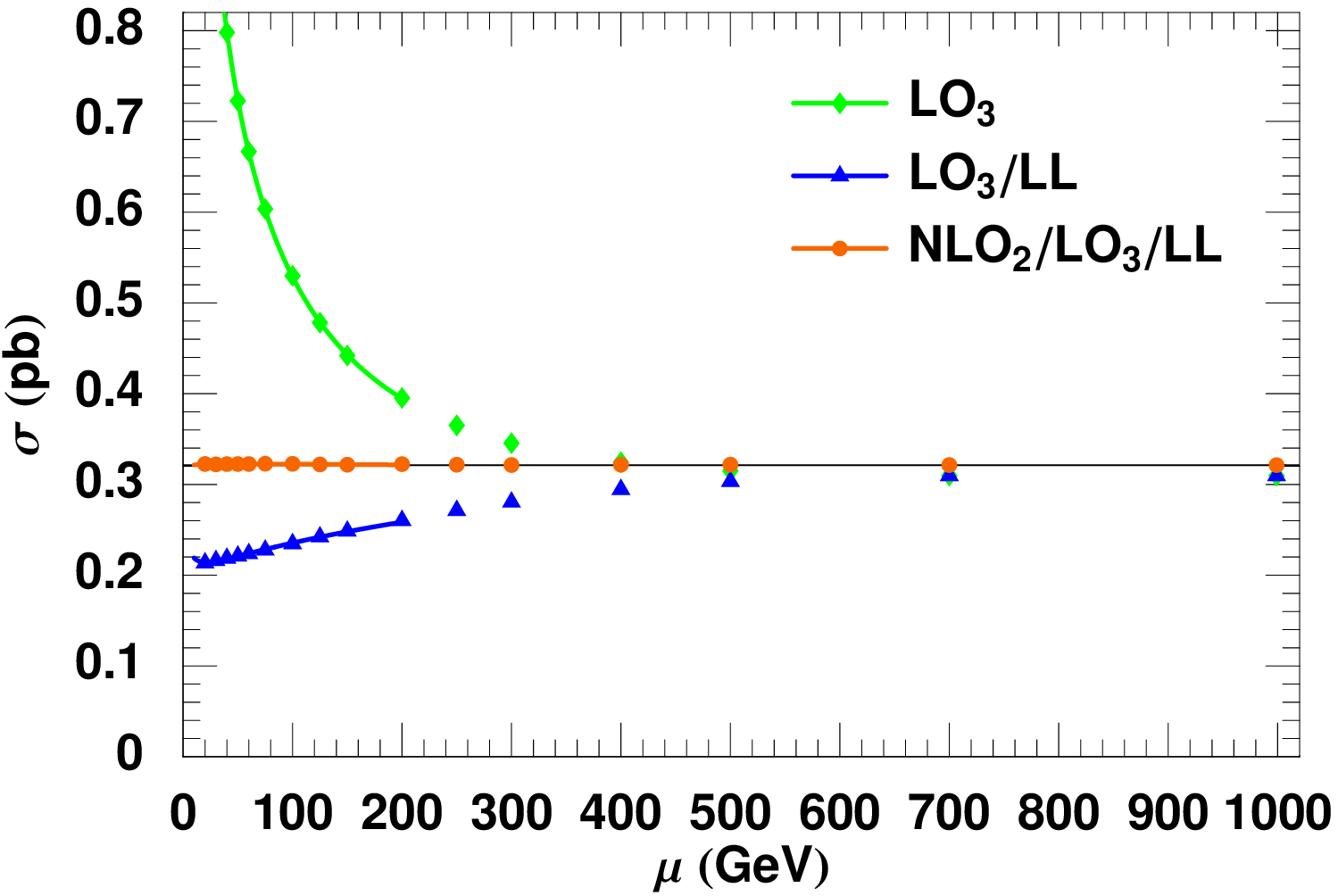}%
\hfill%
\includegraphics[width=0.5\textwidth]{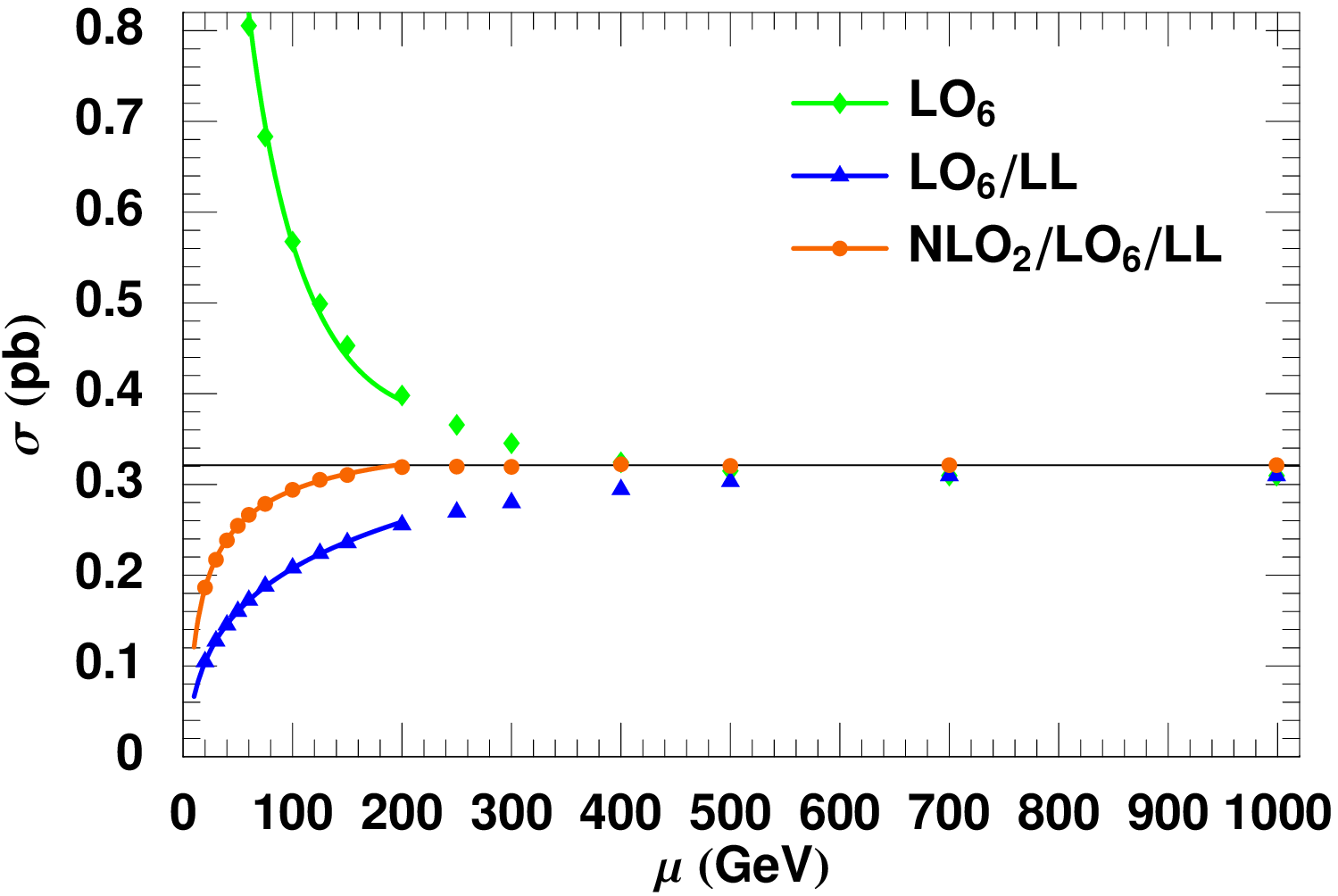}%
\caption{Dependence of the total cross section on the matching scale $\mu$ for $n_\max = 3$ (left panel) and $n_\max = 6$ (right panel) for $20\GeV\leq \mu \leq 1000 \GeV$. The horizontal line corresponds to $\sigma_\NLO$. The markers show the generated results and the solid lines the fit to \eq{eq:sigma_mudep}. For the LO and LO/LL samples $\sigma(E_{\rm CM}) = \sigma_\LO$, and for the NLO samples $\sigma(E_{\rm CM}) = \sigma_\NLO$. Lowering $\mu$, higher-multiplicity matrix elements are used in place of the parton shower to generate further emissions, causing $\sigma(\mu)$ for the LO samples to scale like $\alpha_s\log^2\mu$. With a proper LO/LL merging, the scale dependence reduces to $\alpha_s \log\mu$. Including NLO$_2$ information further reduces it to $(\alpha_s \log\mu)^2$ for NLO$_2$/LO$_6$/LL (right panel), while for NLO$_2$/LO$_3$/LL (left panel) the cross section becomes $\mu$ independent and identical to $\sigma_\mathrm{NLO}$ by construction.}
\label{fig:sigma_mudep}
\end{figure}

We begin by presenting the dependence of the total cross section on the matching scale $\mu$. On the left panel of \fig{fig:sigma_mudep} we show the results for $n_\max = 3$ and on the right panel for $n_\max = 6$. At $\mu = E_{\rm CM}$, the total cross section is given by $\sigma(E_{\rm CM}) = \sigma_\LO$ for the LO samples and $\sigma(E_{\rm CM}) = \sigma_\NLO$ for the NLO samples. As the matching scale $\mu$ is lowered, higher-multiplicity matrix elements are used in place of the parton shower to generate additional emissions. Using only tree-level matrix elements, we expect a double-logarithmic $\mu$ dependence $\alpha_s \log^2\mu$ for the LO samples, but only a single-logarithmic dependence $\alpha_s \log\mu$ for the Sudakov-improved LO/LL samples. For the NLO$_2$/LO$_{n_\max}$/LL samples, the full $\alpha_s$ dependence is included, and the $\mu$ dependence is thus expected to only start at $(\alpha_s \log\mu)^2$. For $n_\max = 3$, this term is absent because the total cross section is $\mu$-independent by construction. We can clearly see in \fig{fig:sigma_mudep} that the scale dependence is reduced as we go from LO to LO/LL to NLO/LO/LL.

\begin{table}[t]
\begin{tabular}{l||c|c|c||c|c|c}
& LO$_3$ & LO$_3$/LL & NLO$_2$/LO$_3$/LL & LO$_6$ & LO$_6$/LL & NLO$_2$/LO$_6$/LL
\\\hline \hline
$a_1$ & $0.713$ & $0.893$ & $-0.037$ & $7.40$ & $1.203$ &$-0.043$
\\\hline
$a_2$ & $0.840$ & $0.115$ & $-0.006$ & $2.74$ & $0.056$ & $-0.151$
\\\hline
\end{tabular}
\caption{Coefficients of the single-logarithmic (first row) and double-logarithmic (second row) $\mu$ dependence of the total cross section in \fig{fig:sigma_mudep}. Shown is a fit to $\sigma(\hat{\mu}) = a_0 + \frac{2 \alpha_s}{\pi} (a_1 \log \hat{\mu} + a_2 \log^2 \hat{\mu})$ in the range $0.02 < \hat{\mu} < 0.2$, where $\hat{\mu} = \mu/E_{\rm CM}$ and $E_{\rm CM} = 1000 \GeV$.}
\label{tab:sigma_mudep}
 \end{table}

To check the expected scaling with $\mu$ explicitly, we fit the result for the total cross section obtained by \geneva\ to the function
%%%
\begin{equation}
\label{eq:sigma_mudep}
\sigma(\hat{\mu}) = a_0 + \frac{2 \alpha_s}{\pi} (a_1 \log \hat{\mu} + a_2 \log^2 \hat{\mu} )
\,,\qquad\text{where}\qquad
\hat{\mu} = \frac{\mu}{E_{\rm CM}}
\,.\end{equation}
%%%
This functional form neglects any contributions from power corrections of the form $\hat\mu^n$, which dominate for $\hat\mu \sim 1$ so we only use the range $\hat \mu \leq 0.2$ in the fit. The results of the fit are shown by the solid lines in \fig{fig:sigma_mudep}, and the fitted parameters $a_1$ and $a_2$ are given in Table~\ref{tab:sigma_mudep}. As anticipated, the LO$_3$ result has both single- and double-logarithmic $\mu$ dependence. Taking into account the overall factors in \eq{eq:sigma_mudep} the values of $a_1$ and $a_2$ for the LO$_3$ sample are consistent with \eq{eq:sigma_doublelog}. The LO$_3$/LL result, has a much smaller $a_2$, indicating that the $\log^2\hat\mu$ term only starts at $\ord(\alpha_s^2)$, \ie\ is a single-logarithmic effect, while $a_1$ is of comparable size as for the LO$_3$ sample. Thus, as expected, LO$_3$/LL still has a single-logarithmic dependence at $\ord(\alpha_s)$. For NLO$_2$/LO$_3$/LL both $a_1$ and $a_2$ are small, indicating that there are no logarithmic terms present at $\ord(\alpha_s)$.

For the samples with $n_\max = 6$, the cancellation of the $\mu$ dependence becomes more dramatic. The LO$_6$ sample shows a much larger $\mu$ dependence than LO$_3$ due to the accumulated $\mu$ dependence of all matrix elements. Nevertheless, this large $\mu$ dependence cancels in the Sudakov-improved LO$_6$/LL, which has similar coefficients to LO$_3$/LL. Finally, for NLO$_2$/LO$_6$/LL the value of $a_2$ is consistent with a residual $\mu$ dependence of the size $(\alpha_s \log\hat\mu)^2$, as expected.

%===============================================================================
\subsection{Interpolating Kinematic Extremes}
%===============================================================================

Next, we study how differential distributions differ in the various implementations included in this work. We find the so-called $C$ parameter \cite{Ellis:1980wv} to be a particularly useful observable to highlight these effects. Given the linear sphericity tensor \cite{Parisi:1978eg, Donoghue:1979vi}
%%%
\begin{equation}
S^{\alpha \beta}
= \frac{\sum_i \frac{\mathbf{p}_i^\alpha \mathbf{p}_i^\beta}{|\mathbf{p}_i|}}{\sum_i |\mathbf{p}_i|}
\end{equation}
%%%
constructed out of the final state three-momenta $\mathbf{p}_i$, the $C$-parameter is defined as
%%%
\begin{equation}
C = 3 \left(\lambda_1\lambda_2 + \lambda_2\lambda_3 + \lambda_3\lambda_1 \right)
\,,\end{equation}
%%%
where $\lambda_i$ are the eigenvalues of $S^{\alpha \beta}$. Since $\sum_i \lambda_i = 1$ we have $0\leq C \leq 1$. Low values of the $C$ parameter are dominated by events which are mostly 2-jet-like, while larger values of $C$ indicate more and more final state jets. For planar events $\lambda_3 = 0$ forcing $C \leq 0.75$, so that 3-jet events are confined to $C \leq 0.75$, while events with four or more jets can contribute up to $C=1$.

\begin{figure}
\includegraphics[width=0.5\textwidth]{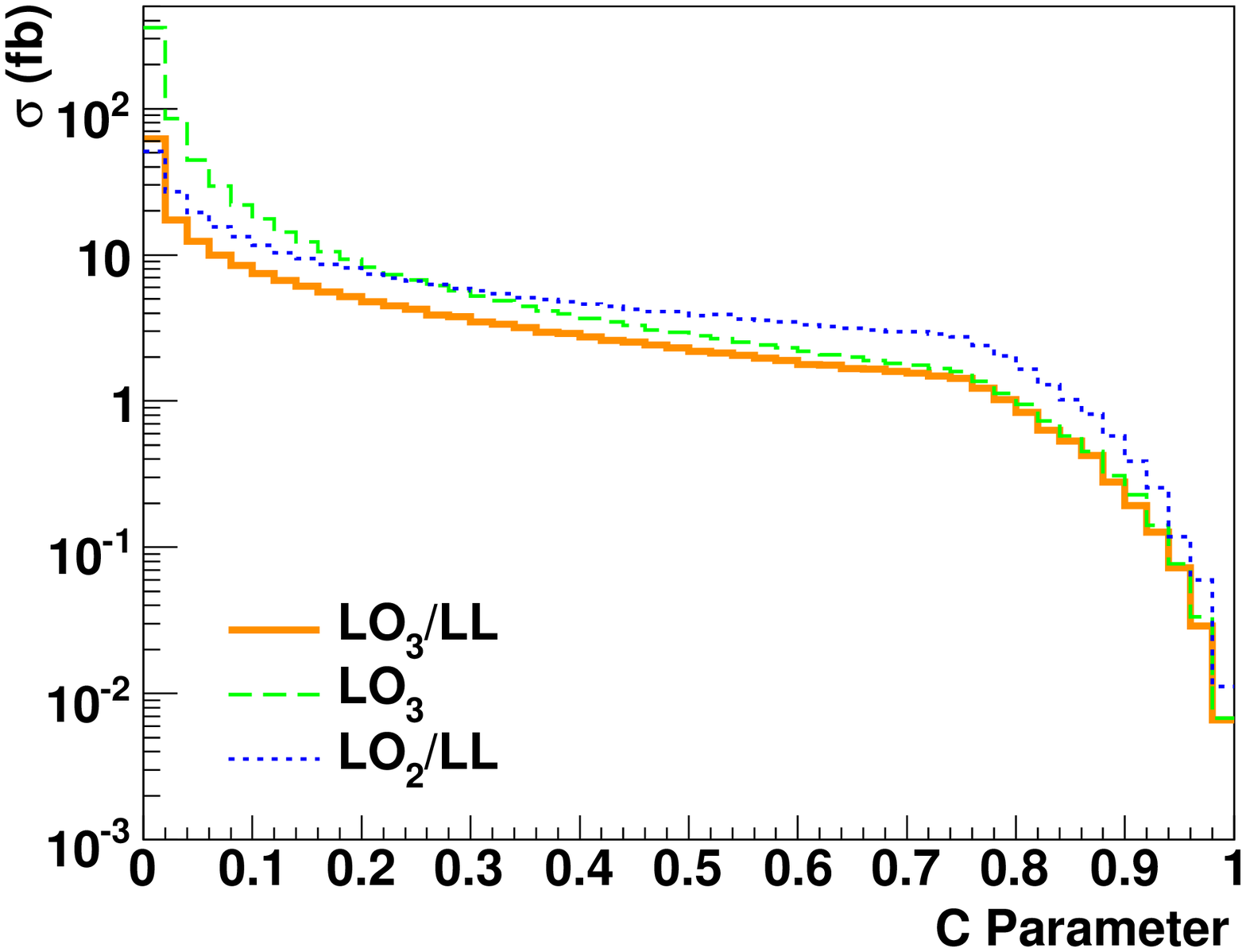}%
\hfill%
\includegraphics[width=0.5\textwidth]{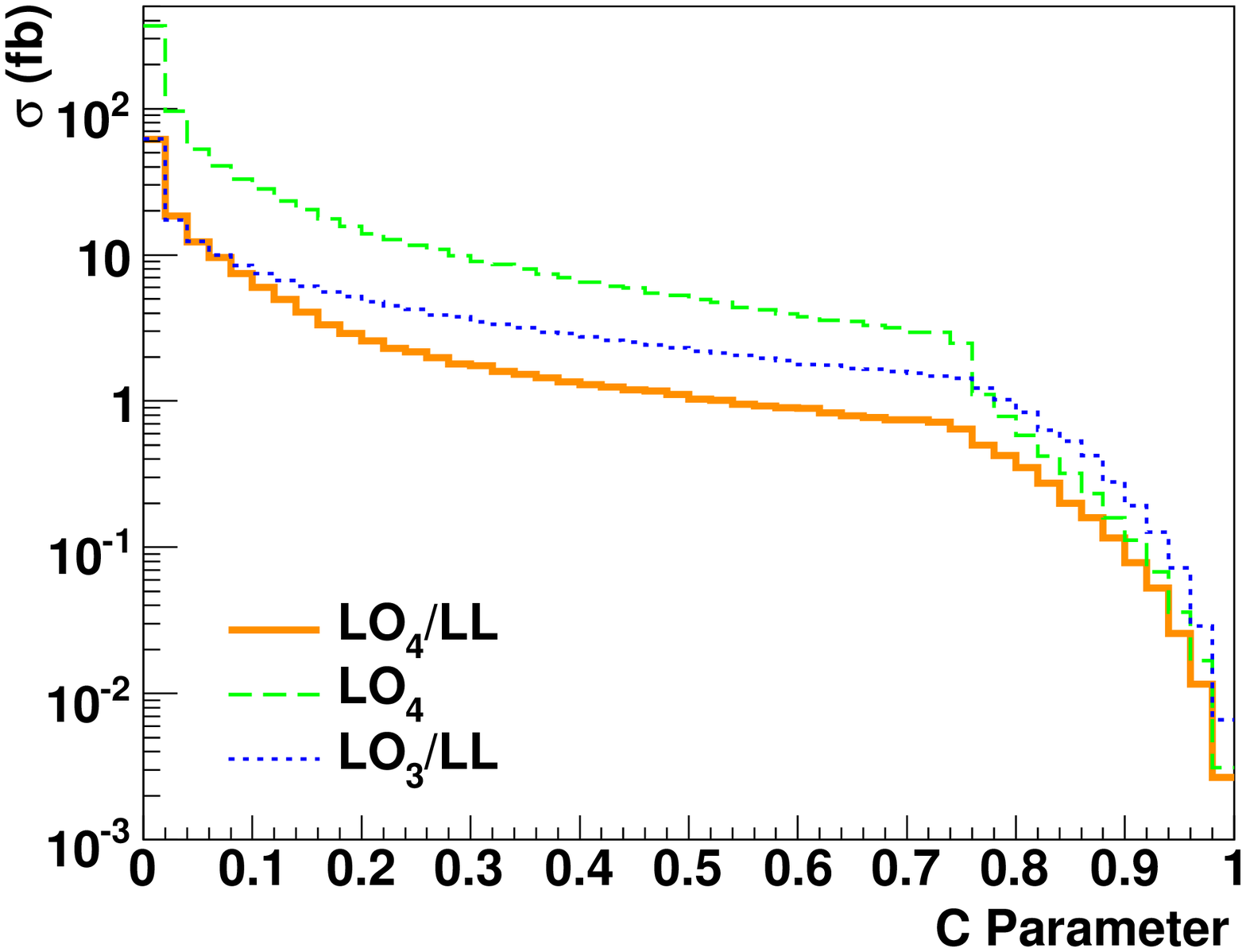}%
\caption{PS/ME merging as implemented in \GenEvA\ (LO/LL) with $E_{\rm CM} = 1000 \GeV$ and $\mu = 50 \GeV$. The $C$ parameter measures the ``jettiness'' of an event, with $C \sim 0$ giving the 2-jet region, $C < 0.75$ roughly giving the 3-jet region, and $C > 0.75$ roughly giving the 4-jet region. Left panel: The LO$_3$ sample has the soft-collinear divergence of the 3-parton tree-level matrix element and therefore becomes singular near $C=0$, while the LO$_3$/LL sample regulates that divergence with a Sudakov factor similar to LO$_2$/LL. At large $C$, the LO$_2$/LL sample lacks the quantum interference of the 3-parton matrix element, while the LO$_3$/LL sample contains the correct interference effects and therefore tracks LO$_3$. Right panel: The same interpolation for LO$_4$/LL, which captures the correct Sudakov-suppression of the LO$_3$/LL sample near $C\sim0$ and at the same time includes the extra interference effects of LO$_4$ near $C \sim 1$.}
\label{fig:cparaloll}
\end{figure}

We begin by studying the effect of Sudakov resummation on the tree-level matrix elements, which is the \GenEvA\ analog of PS/ME merging. It is well known that the leading-logarithmic resummation for the $C$ parameter reduces the cross section for small values of $C$, with the resummed expressions given in Ref.~\cite{Catani:1998sf}. It is also well known that the pure parton shower result overshoots the correct QCD result for large values of the $C$ parameter, since the interference between the two QCD diagrams contributing to the emission of a single gluon is destructive. On the left panel of \fig{fig:cparaloll}, we compare the results for the $C$-parameter obtained by running the pure parton shower (LO$_2$/LL), by using tree-level QCD matrix elements up to order $\alpha_s$ (${\rm LO}_3$), and the merged result with $n_\max = 3$ (LO$_3$/LL). We can see clearly how the LO$_2$/LL sample overshoots the LO$_3$ result for large values of $C$, while it suppresses the cross section for small values of $C$. The merged LO$_3$/LL result reproduces the result of the pure QCD matrix element for large values of $C$, where the leading logarithms are not important. The resummation of the double logarithmic terms becomes important for small values of $C$, and we can see the Sudakov suppression of the merged result compared to the pure matrix element for small values of $C$.

Note that the merged result has to approach the pure parton shower result for $C \to 0$, and we have checked that this is indeed the case. The reason this is not obvious from the figure is that the binning is too coarse for this effect to be visible, and the agreement does not happen until much smaller values of $C$.\footnote{Because of the single-logarithmic effect of the $\mathbf{\tilde{n}}$ integrations, the LO$_3$/LL sample first overshoots the LO$_2$/LL sample until finally asymptoting from above. With the $\mathbf{\tilde{n}}$ integrations turned off, the agreement happens at more moderate values of $C$, but still smaller than our bin size.} To see more easily that the desired interpolation does occur, we include one additional tree-level emission, and compare the results for LO$_4$, LO$_3$/LL and LO$_4$/LL on the right panel of Fig.~\ref{fig:cparaloll}. One can now clearly see that the LO$_4$/LL sample reproduces the LO$_3$/LL result for small values of $C$. The agreement between the LO$_4$/LL and the LO$_4$ sample for large values of $C$ is not expected to be perfect, since the Sudakov suppression has some effect all the way up to $C = 1$. However, we do see that the agreement is being obtained asymptotically.

\begin{figure}
\includegraphics[width=0.5\textwidth]{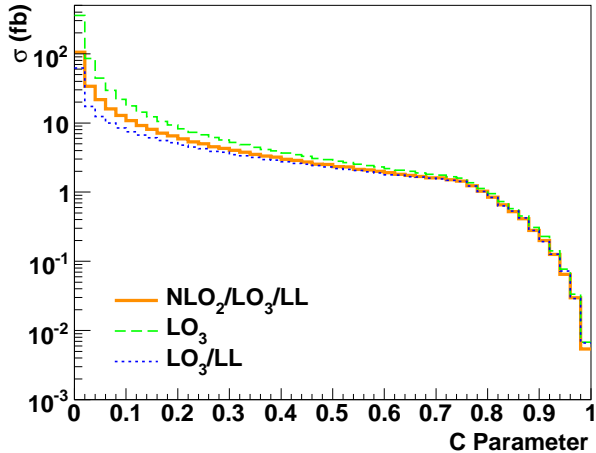}
\caption{PS/NLO merging as implemented in \GenEvA\ (NLO/LL). The NLO$_2$/LO$_3$/LL, LO$_3$, and LO$_3$/LL samples all share the same interference effects near $C=1$, given by the 3-parton tree-level matrix element. Near $C = 0$, LO$_3$ is too singular, while LO$_3$/LL is too Sudakov suppressed. The NLO$_2$/LO$_3$/LL sample has the correct NLO cross section information and therefore has the right normalization near $C=0$. Since the inclusion of NLO cross-section information affects the form of the Sudakov factor for the NLO/LL sample, the NLO$_2$/LO$_3$/LL sample is \emph{not} just a $k$-factor modification of the tree-level resummed LO$_3$/LL result.}
\label{fig:cparanloll}
\end{figure}

Next, we show the effect of including NLO information, which is \GenEvA's analog of PS/NLO merging. As we saw in \fig{fig:sigma_mudep}, the tree-level calculation contains double-logarithmic terms which results in a rising cross section as the scale $\hat \mu$ is lowered. For the $C$ parameter distribution, this implies that the tree-level result diverges as $-\log(C)/C$ for $C \to 0$. Resumming the leading-logarithmic terms removes this dominant singularity for small values of $C$, however we also see in \fig{fig:sigma_mudep} that the cross section of the resummed result undershoots the correct (NLO) result. In Fig.~\ref{fig:cparanloll}, we compare the tree-level matrix elements LO$_3$, the resummed matrix elements LO$_3$/LL, as well as the resummed NLO result NLO$_2$/LO$_3$/LL. For large values of $C$ all three results agree, as expected, because all three samples have the same interference terms from the tree-level 3-parton matrix element. However, for $C \to 0$, the NLO$_2$/LO$_3$/LL result has the correct $\ord(\alpha_s)$ normalization, while LO$_3$ is too singular, and LO$_3$/LL too Sudakov-suppressed. Note that the NLO$_2$/LO$_3$/LL result is \emph{not} simply a $k$-factor modification of the tree-level result, because as seen in \eqs{eq:sudakovimproved}{eq:QCDelegant}, the Sudakov factors in the 2- and 3-parton matrix elements must be modified to get the correct NLO cross section, while still incorporating leading-logarithmic results.

\begin{figure}
\includegraphics[width=0.5\textwidth]{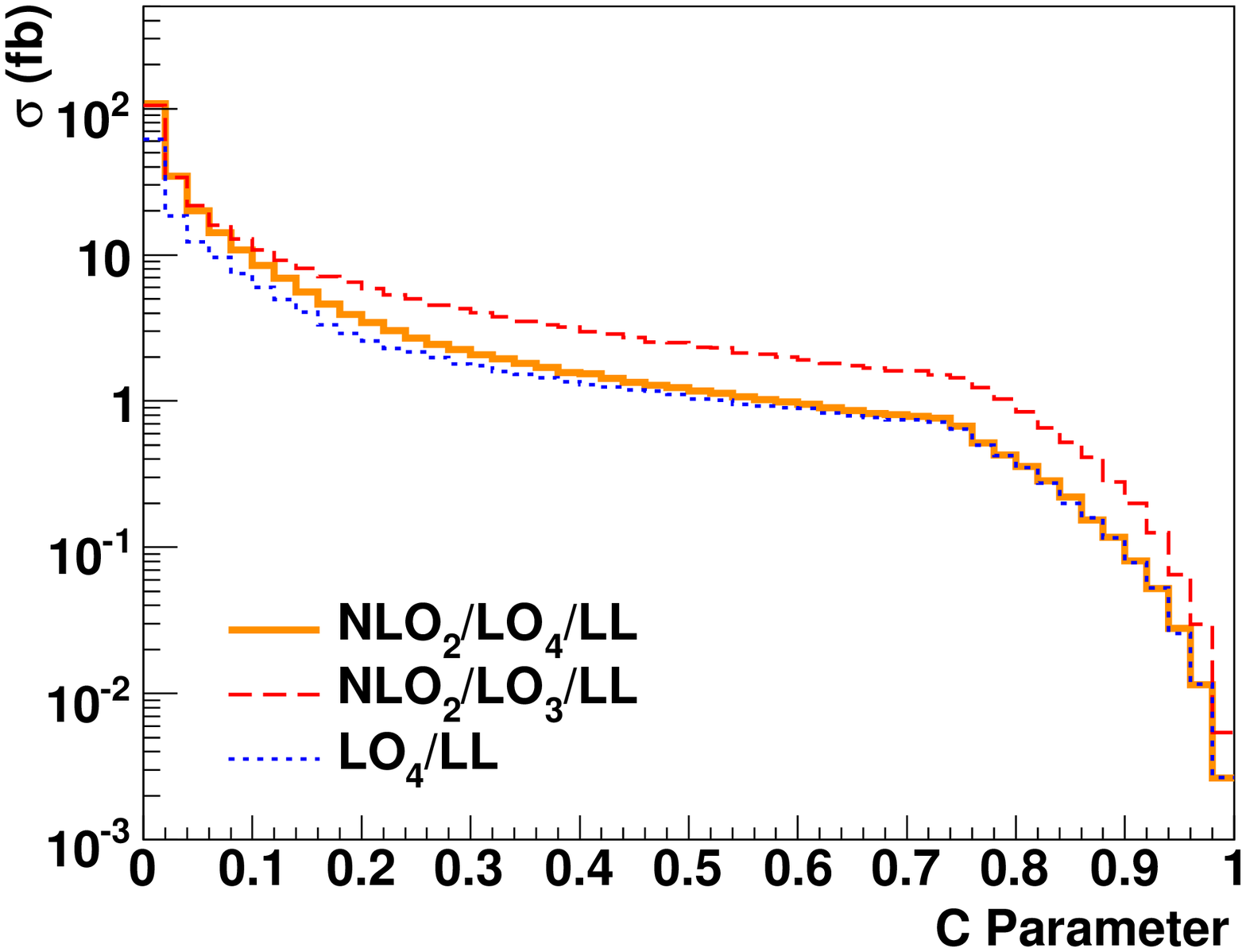}%
\hfill%
\includegraphics[width=0.5\textwidth]{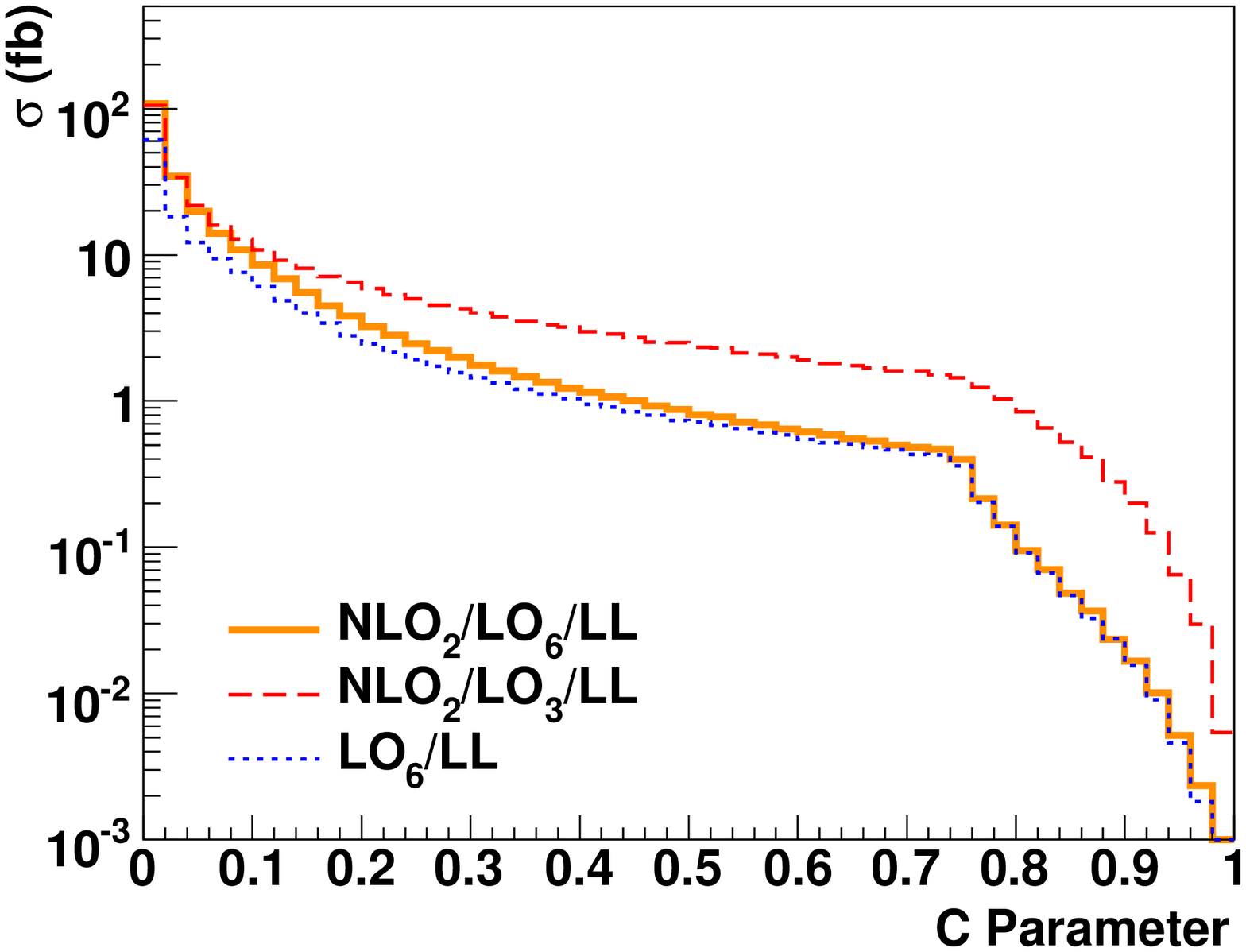}%
\caption{The combination of PS/ME merging and PS/NLO merging in \GenEvA\ (NLO/LO/LL).
Left panel: The LO$_4$/LL sample is the analog of PS/ME merging and contains the correct interference of the 4-parton matrix element near $C=1$, while the NLO$_2$/LO$_3$/LL sample is the analog of PS/NLO merging, and has the correct cross section information near $C=0$. The NLO$_2$/LO$_4$/LL sample smoothly interpolates between the two different regimes. Right panel: The same interpolation for $n_\max = 6$, where NLO$_2$/LO$_6$/LL smoothly interpolates between LO$_6$/LL at large $C$ and NLO$_2$/LO$_3$/LL at small $C$. The kinked behavior near $C= 0.75$ is a well-known physical effect \cite{Ellis:1980wv}, unrelated to the merging procedure. It occurs because $C \leq 0.75$ for planar events, meaning that 2- and 3-parton matrix elements cannot contribute much for $C > 0.75$. The NLO$_2$/LO$_6$/LL sample is the best partonic calculation currently implemented in \GenEvA.}
\label{fig:cparanlololl}
\end{figure}

We now consider the NLO/LO/LL partonic calculation which combines NLO cross section information with higher-order tree-level matrix elements, all Sudakov-improved. We expect that in the 4-jet region ($C > 0.75$), the NLO$_2$/LO$_4$/LL sample will agree well with the LO$_4$/LL result, since 3-parton states do not contribute much in this region. For small values of the $C$ parameter, however, we expect the NLO$_2$/LO$_4$/LL result to be close to the NLO$_2$/LO$_3$/LL result, since the NLO cross section information is important there. The results are shown on the left panel of \fig{fig:cparanlololl}, which confirms the expected interpolation. On the right panel of \fig{fig:cparanlololl}, we give the result of the best partonic calculation current available in \geneva: NLO$_2$/LO$_6$/LL. This sample includes additional interference terms from tree-level matrix elements with $n \leq 6$, while still maintaining NLO/LL accuracy. We clearly see that the NLO$_2$/LO$_6$/LL sample interpolates between the LO$_6$/LL (PS/ME merged) sample in the interference region ($C \sim 1$) and the NLO$_2$/LO$_3$/LL (PS/NLO merged) sample in the total cross section region ($C \sim 0$). In other words, the best implementation available in \geneva\ simultaneously achieves PS/ME merging and PS/NLO merging.

\begin{figure}
\includegraphics[width=0.5\textwidth]{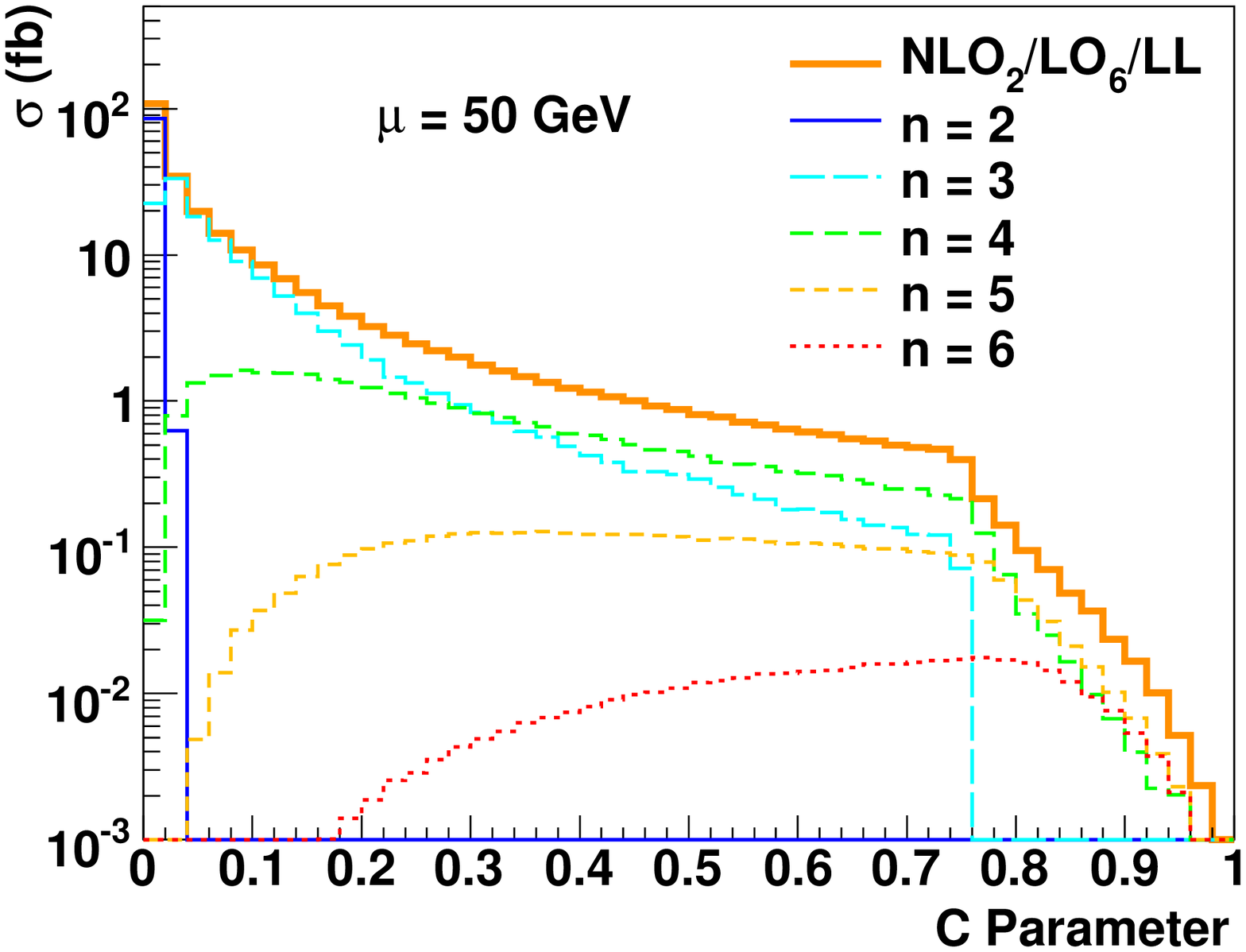}%
\hfill%
\includegraphics[width=0.5\textwidth]{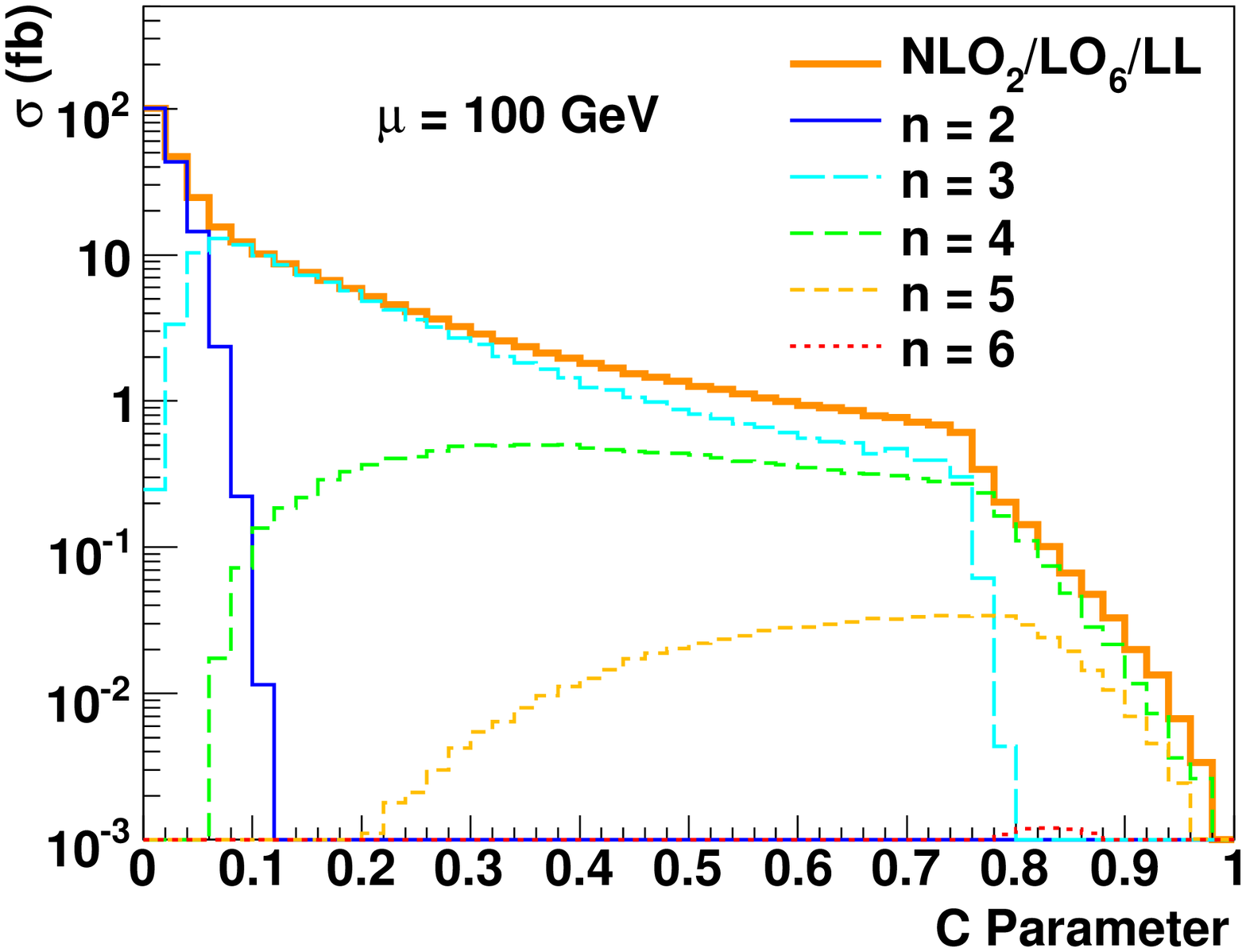}%
\caption{The five components of the \GenEvA\ Best NLO$_2$/LO$_6$/LL sample. The curves for different $n$ show the individual contributions of the $n$-parton matrix elements, where $n$ corresponds to the number of partons resolvable at the scale $\mu$.
Left panel: $\mu = 50\GeV$. As expected, the $n=2$ component dominates near $C=0$. Because $\mu$ is so low, the $n=3$ and $n=4$ matrix elements both contribute in the 3-jet ($C < 0.75$) region, where the crossover point $C= 0.3$ is directly related to the scale $\mu$. Above $C = 0.75$, the $n=3$ component can no longer contribute, and the rest of the $C$ distribution is filled out with additional contributions from the 4-, 5-, and 6-parton matrix elements. Right panel: $\mu = 100\GeV$. For this more reasonable $\mu$ scale, the various $n$-jet-like regions correspond more closely to the $n$-parton matrix element used. While the total $C$ parameter distribution is only single-logarithmically sensitive to changing the $\mu$ scale, the different matrix element components shift dramatically, with the $n=6$ matrix element now having very little available phase space.}
\label{fig:cparanlololliso}
\end{figure}

Finally, on the left panel of \fig{fig:cparanlololliso}, we illustrate the composition of the NLO$_2$/LO$_6$/LL sample, showing the individual contributions of the $n$-parton matrix elements, where $n$ corresponds to the number of partons resolvable at the scale $\mu$. Near the 2-jet region ($C \sim 0$), the $n=2$ matrix element dominates as expected. The $n=3$ matrix element fills out the region $0 < C < 0.3$, and since it gives mostly planar events, it turns off abruptly at $C=0.75$. Despite the fact that $C < 0.75$ is supposed to correspond to the 3-jet region, the $n=4$ matrix element dominates for $0.3 < C < 0.75$, because $\mu = 50 \GeV$ is so low, that the $n=4$ matrix element is really being used to determine some of the jet substructure in the 3-jet region. Above $C = 0.75$, the $n=4$ and $n=5$ matrix elements give roughly equal contributions, with the region near $C=1$ supplemented by the $n=6$ matrix element. This is not surprising as the $C$ parameter no longer resolves the difference between 4-jet and higher-jet event shapes. On the right panel of \fig{fig:cparanlololliso}, the same $C$ parameter is shown for $\mu = 100 \GeV$. With this matching scale, the separation of the $C$ parameter into $n$-jet-like regions roughly corresponds to the $n$-parton matrix elements. Note that the $n=6$ matrix element now barely has phase space available to contribute to the total distribution, so the remaining matrix elements shift to compensate.

%===============================================================================
\subsection{More Results From GenEvA Best}
%===============================================================================

To get more intuition about the \GenEvA\ Best NLO/LO/LL sample, it is helpful to use a jet algorithm to get more differential information than is available in an event shape measure like the $C$ parameter. We use the \texttt{FastJet} package \cite{Cacciari:2006sm} for that purpose, using the inclusive $k_T$ jet algorithm with $R = 1.0$ to identify jets, which are then ordered by their total energy.

\begin{figure}[t]
\includegraphics[width=0.5\textwidth]{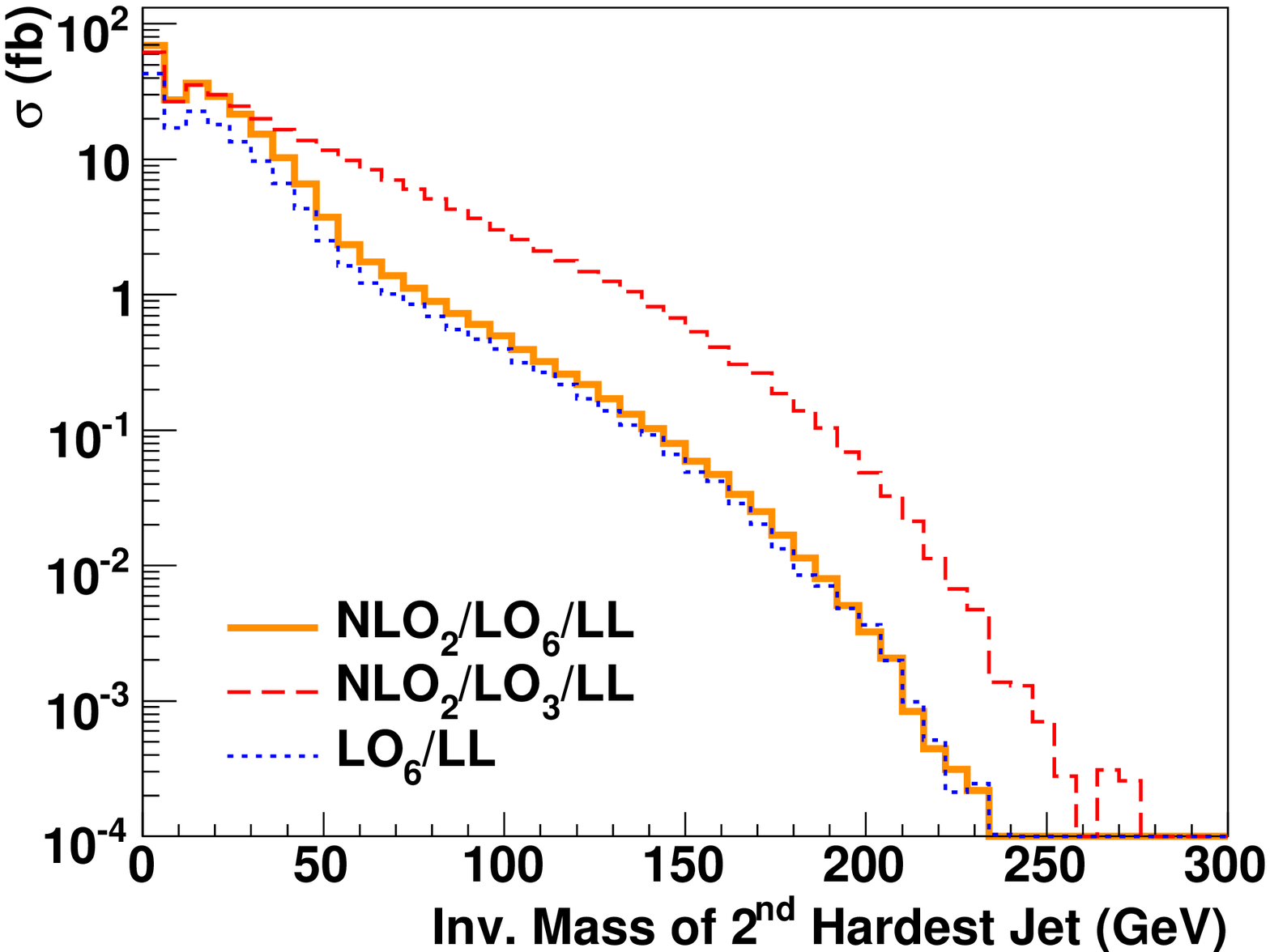}%
\hfill%
\includegraphics[width=0.5\textwidth]{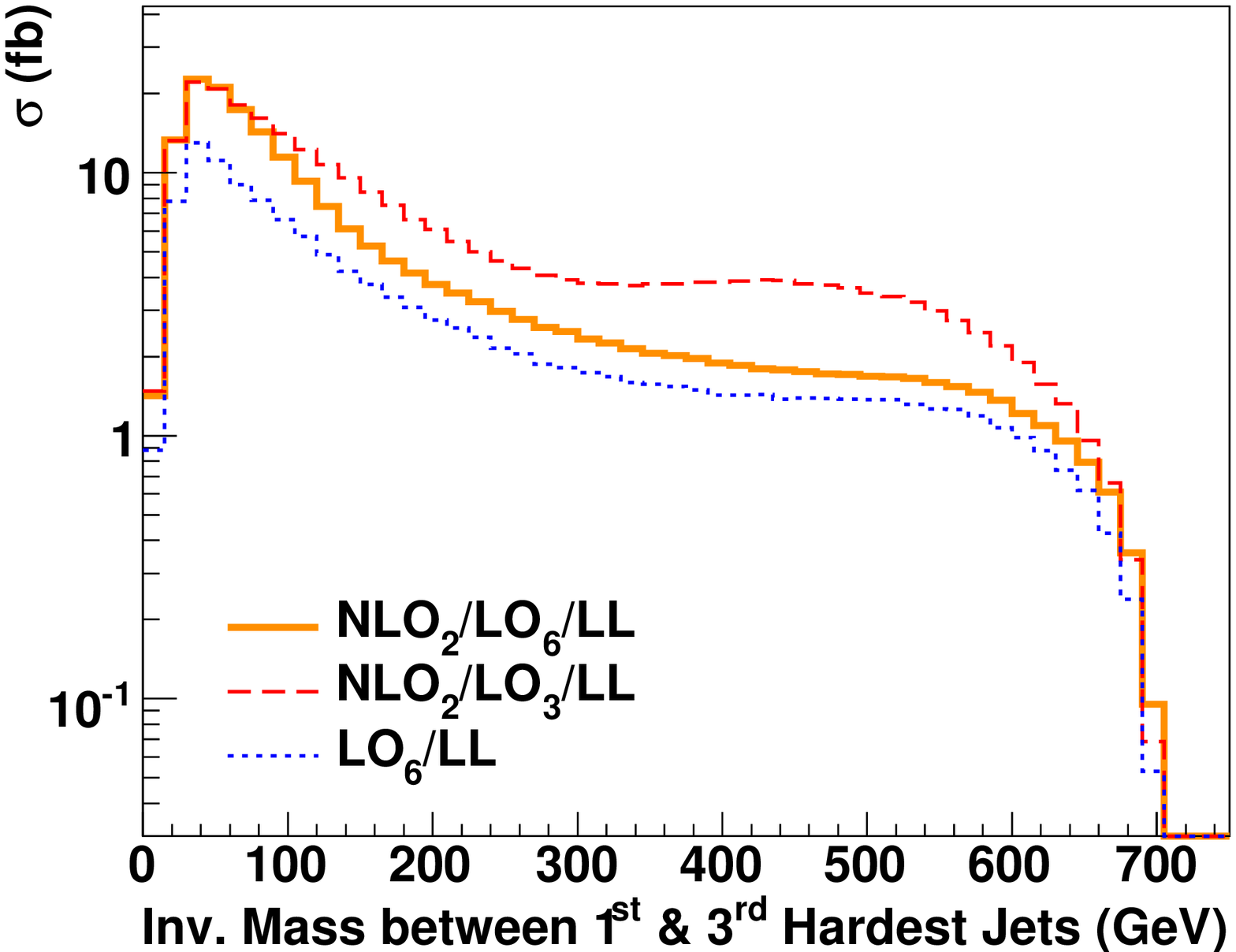}
\caption{Comparison of the \GenEvA\ best sample NLO$_2$/LO$_6$/LL with the PS/ME merging analog LO$_6$/LL and the PS/NLO merging analog NLO$_2$/LO$_3$/LL for two jet-based observables. Left panel: The invariant mass of the 2nd hardest jet, where we can clearly see the destructive interference in the matrix element calculation with $n>3$ partons compared to the results where these additional partons are only generated by the parton shower. The NLO$_2$/LO$_6$/LL sample interpolates between the two comparison distributions, showing that \GenEvA\ can capture important physical effects that cannot be seen by considering the NLO$_2$/LO$_3$/LL or LO$_6$/LL samples alone. Right panel: The invariant mass between the 1st and 3rd hardest jets. Since no particular parton multiplicity dominates at large pairwise invariant masses, the three different samples have slightly different behaviors near the endpoint.}
\label{fig:gencompare}
\end{figure}

On the left panel of \fig{fig:gencompare}, we show the invariant mass of the 2nd hardest jet. Events which are almost 2-jet-like contribute to small values of this observable, while events with well-separated partons contribute to larger values. As we saw in the previous section, adding additional tree-level matrix elements suppresses rates for well-separated partons due to interference effects in the full matrix-element calculations. On the other hand, the NLO information increases the cross section in regions of phase space which contribute mostly to 2-jet-like events. Thus, we expect the NLO$_2$/LO$_3$/LL result to be higher than the LO$_6$/LL sample in the whole kinematic region, which is seen in the figure. Combining these two results, we expect to reproduce the lower rate for well-separated partons, due to the interference from the tree-level matrix elements, while at the same time reproducing the enhanced cross section for almost 2-jet-like events due to the NLO information. This is clearly seen in the NLO$_2$/LO$_6$/LL sample, which thus captures important physical effects that cannot be seen by considering the PS/ME merging or PS/NLO merging results alone.

The right panel of \fig{fig:gencompare} shows the invariant mass between the 1st and 3rd hardest jets. Low values of this pairwise invariant mass correspond to the 2-jet-like region, and again the NLO$_2$/LO$_6$/LL sample inherits the NLO cross section information. As the pairwise invariant mass increases, this observable gets contributions from a variety of different event types, so the NLO$_2$/LO$_6$/LL sample tracks the shape of the LO$_6$/LL result with an overall NLO cross section increase. At very large values of this invariant mass, all three curves give slightly different answers because this extreme kinematic region is sensitive to the exact way in which the jets are clustered, which is strongly affected by the exact ratios of the 4-, 5-, and 6-parton matrix elements.

\begin{figure}[t]
\includegraphics[width=0.5\textwidth]{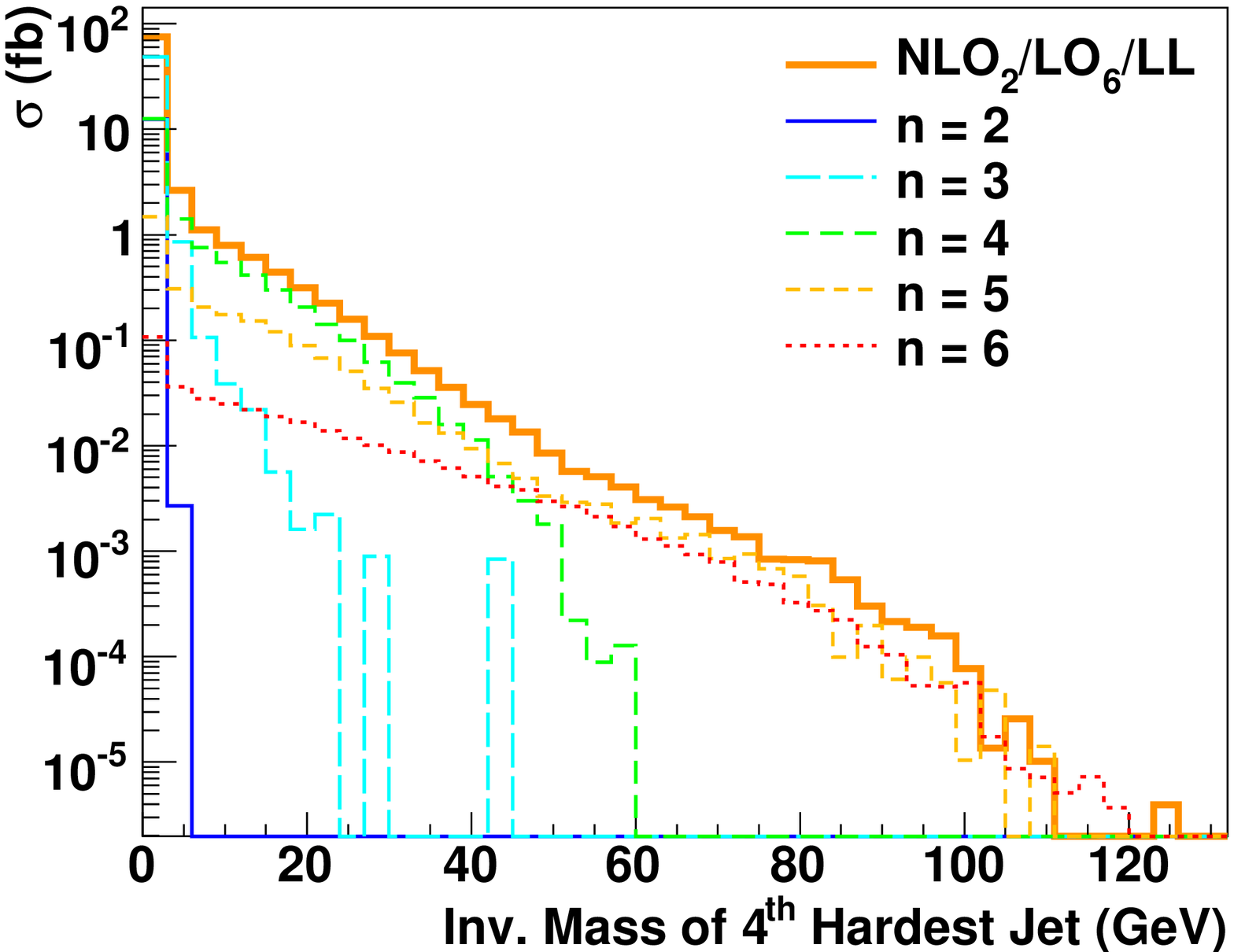}%
\hfill%
\includegraphics[width=0.5\textwidth]{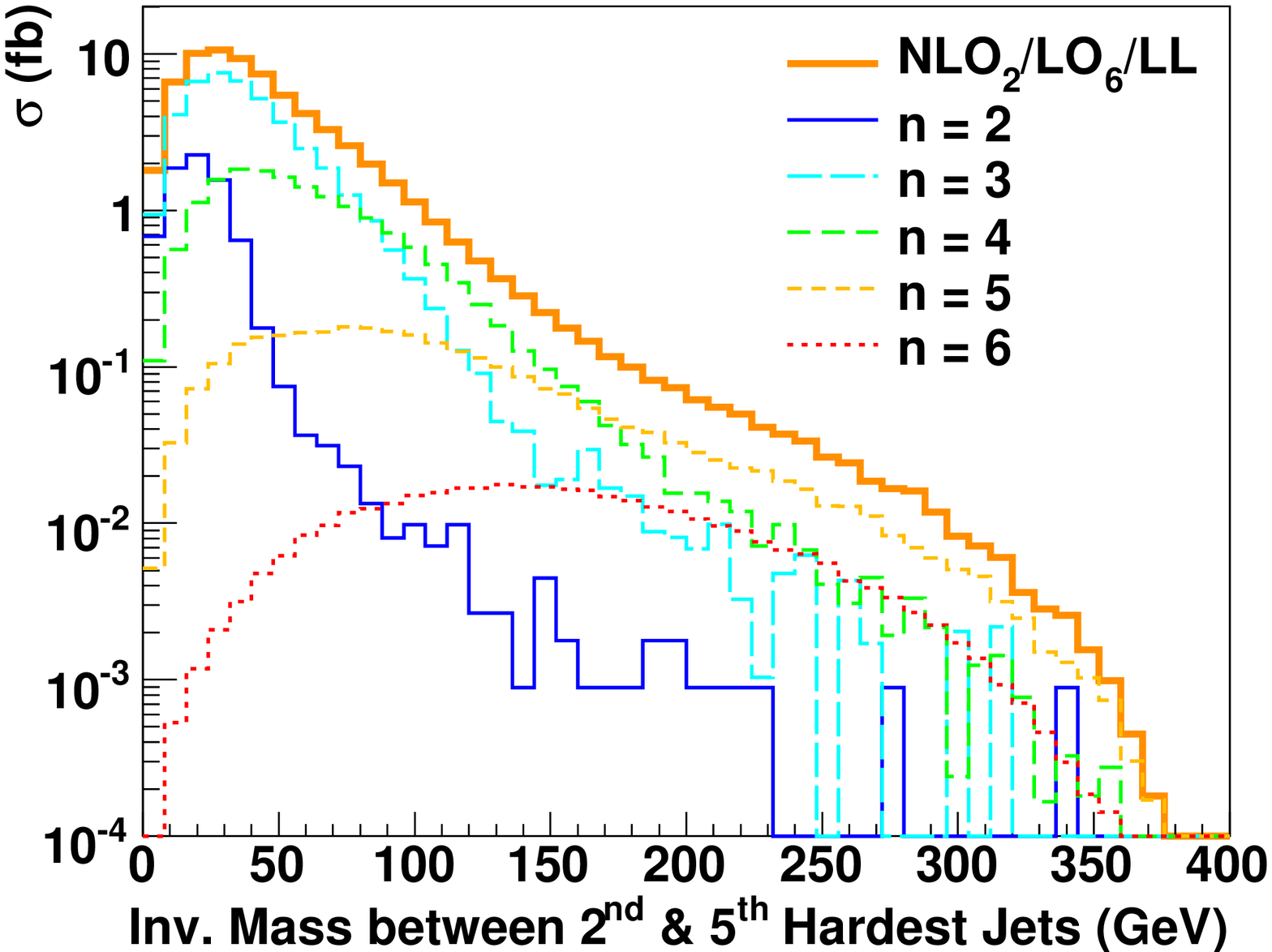}
\caption{The five components of the NLO$_2$/LO$_6$/LL sample for two jet-based observables. The curves for different $n$ show the contributions of the individual $n$-parton matrix elements. Left panel: The invariant mass of the 4th hardest jet. Since \GenEvA\ currently uses virtuality to define the scale $\mu$ that separates different parton-level multiplicities, the invariant masses of jets have sharp cutoffs for the different parton-multiplicity components, with the $n=4$ sample turning off at the matching scale around $50 \GeV$ as expected (the small tail comes from the jet algorithm ``accidentally'' clustering together two different quark flavors). Right panel: The invariant mass between the 2nd and 5th hardest jets. As this pairwise invariant mass increases, the dominant $n$-parton matrix element smoothly changes from $n=3$ to $n=4$ to $n=5$.}
\label{fig:isocompare}
\end{figure}

As seen in the $C$ parameter, the NLO$_2$/LO$_6$/LL sample combines 5 different matrix elements, and it is interesting to see the individual matrix-element contributions for jet-based observables.
On the left panel of \fig{fig:isocompare}, we show the results for the invariant mass of the 4th hardest jet. Since our evolution variable is equal to the virtuality between partons, the parton shower can only generate virtualities below the staring scale of the shower, which is chosen as $\mu = 50 \GeV$. Thus we expect that a 4th jet with invariant mass above 50 GeV can only be generated by a partonic calculation with at least 5 partons in the final state.\footnote{There are $n=4$ events that can have jet masses slightly larger than the matching scale, because the cutoff only applies to singularity-producing partons. One can see the effect of, say, the $e^+ e^- \to u \bar{u} d\bar{d}$ sample in the rare events that have a 4th jet with invariant mass upwards of $60 \GeV$.} This effect is reproduced by \geneva, and while there are sharp cutoffs in the individual contributions of $n=2,3,4$-parton matrix elements, the combined NLO$_2$/LO$_6$/LL result is relatively smooth over the entire range of the invariant mass of the 4th hardest jet. Note that there is a slight kink in the final distribution at the matching scale $50 \GeV$, which gives a sense of the size of the subleading-logarithmic errors one makes in this merging.

For the pairwise invariant mass between the 2nd and 5th hardest jets, shown on the right panel of \fig{fig:isocompare}, the individual partonic contributions do not cut off as sharply as for the previous case. The dominant matrix element smoothly changes from $n=3$ for small invariant mass, to $n=4$ for intermediate masses, to $n=5$ towards the endpoint of the distribution. The total result is again a very smooth function over the entire range of the inter-jet invariant mass, with perhaps a slight kink at $175 \GeV$ from the same $n=4$ to $n=5$ transition that gave a kink in the previous distribution.

\begin{figure}[t]
\includegraphics[width=0.5\textwidth]{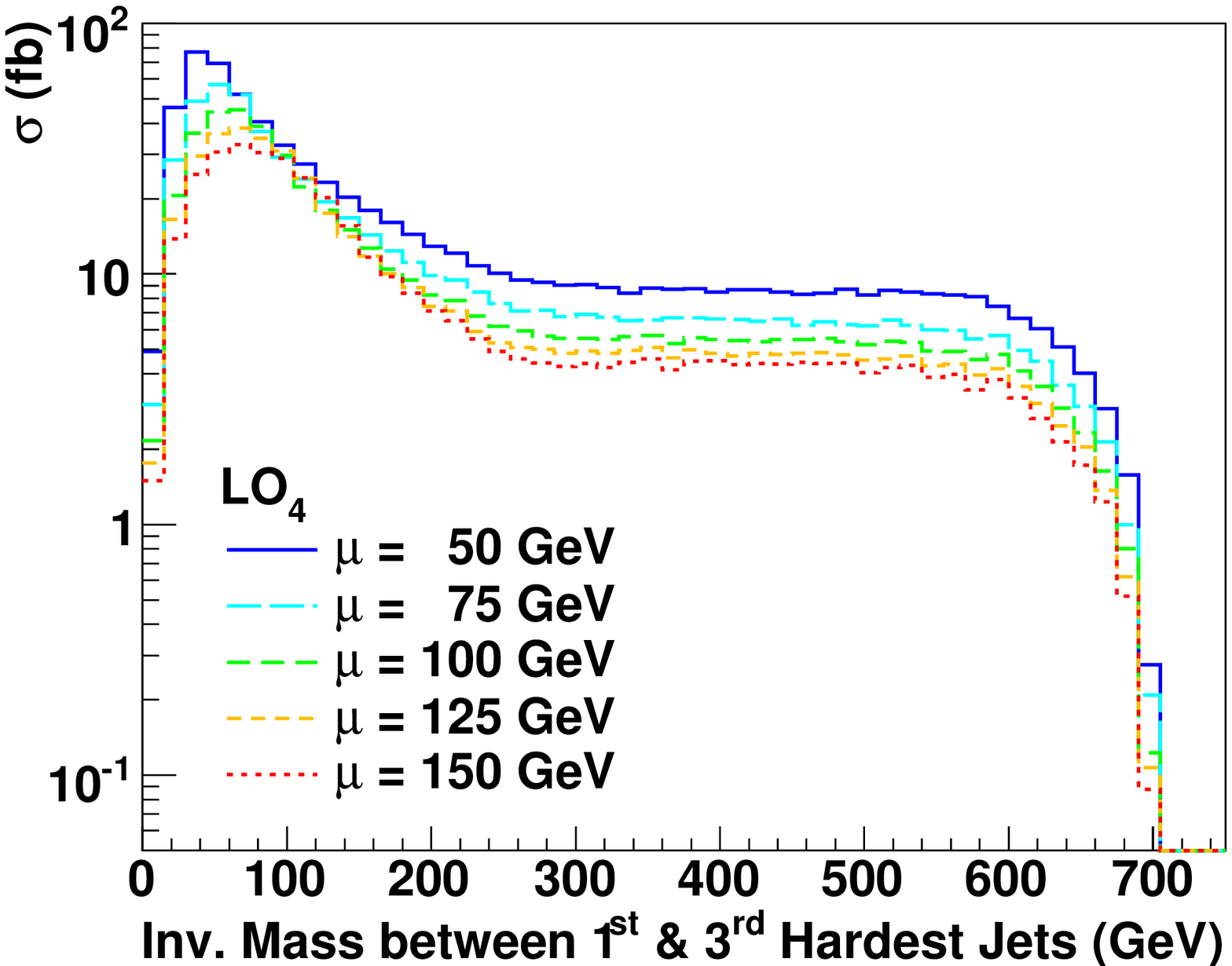}%
\hfill%
\includegraphics[width=0.5\textwidth]{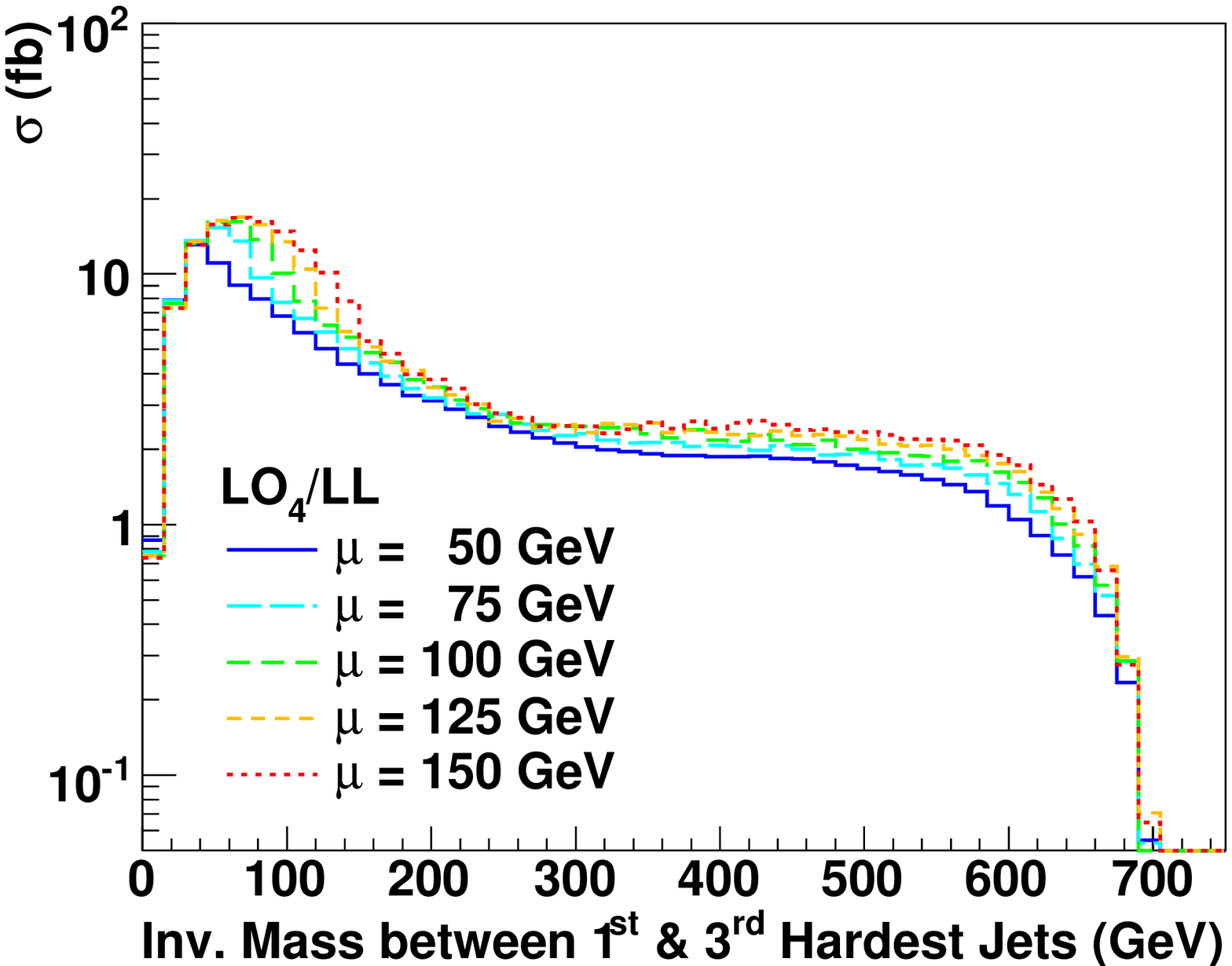}%
\\[2ex]
\includegraphics[width=0.5\textwidth]{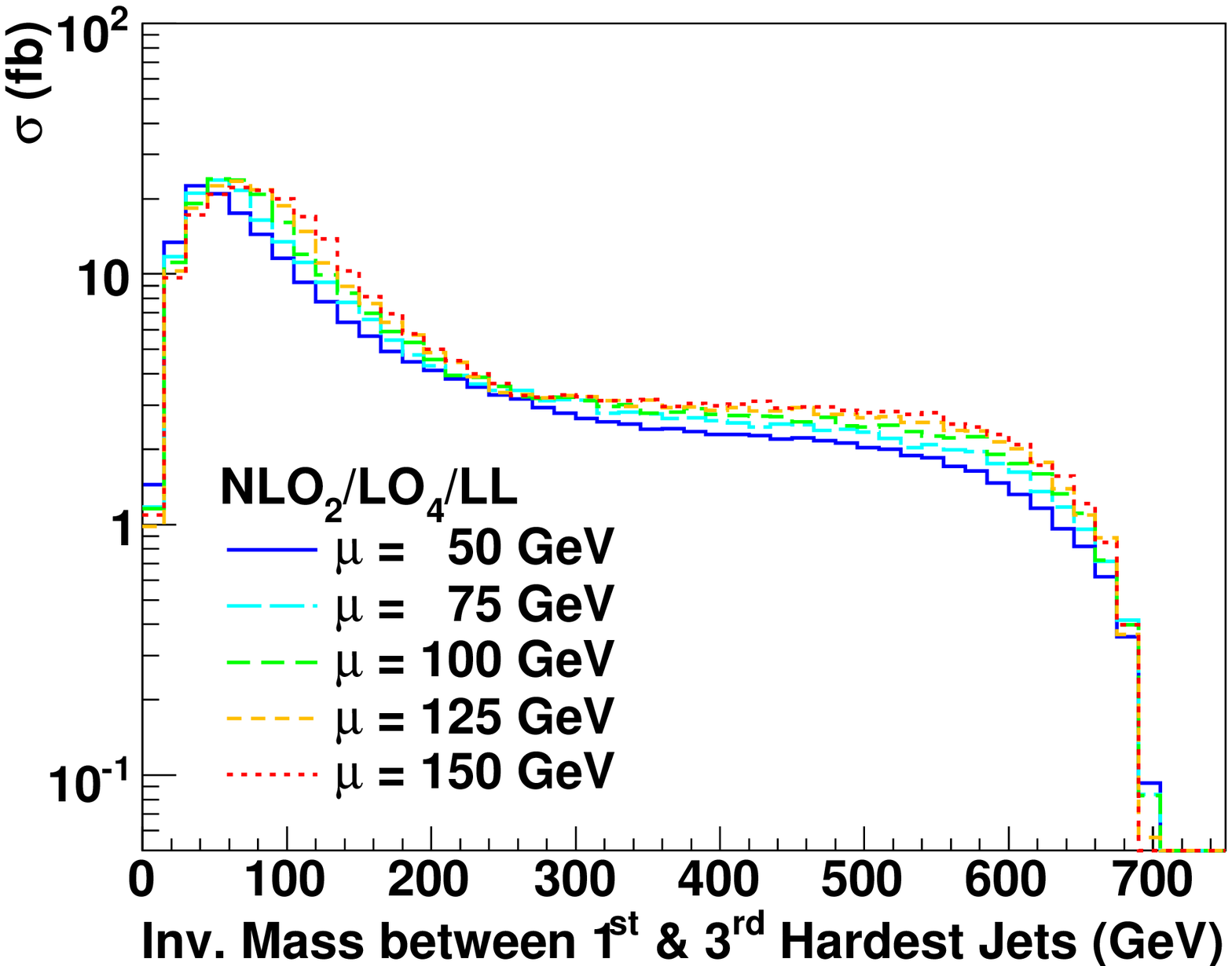}
\caption{Comparison of the $\mu$ dependence of the LO$_4$, LO$_4$/LL, and NLO$_2$/LO$_4$/LL samples. Shown is the distribution of the invariant mass between the 1st and 3rd hardest jets for five different values of the matching scale $\mu$. Just as in \fig{fig:sigma_mudep}, the inclusion of leading-logarithmic and NLO cross-section information reduces the dependence on the unphysical matching scale $\mu$. At large invariant masses, the dominant reduction comes from including leading-logarithmic information in high-multiplicity matrix elements. At small invariant masses, there is an additional reduction in the $\mu$ dependence from the inclusion of NLO cross-section information.}
\label{fig:floatmu}
\end{figure}

As a final application, we study the scale dependence of a differential distribution for $n_\max = 4$. We saw in \fig{fig:sigma_mudep} that the $\mu$ dependence of the total cross section was reduced when going from LO to LO/LL to NLO/LO/LL, and it is interesting to see the extent to which this reduced scale dependence is reflected in differential distributions. In \fig{fig:floatmu}, we show the invariant mass between the 1st and 3rd hardest jets for these three samples. We see that the $\mu$ dependence is lowered going from LO$_4$ sample (top left plot) to the LO$_4$/LL sample (top right plot). We also see how the direction of the scale dependence is reversed, as expected from the scale dependence of the total cross section. In the bottom plot we show the result for the NLO$_2$/LO$_4$/LL sample. Including the NLO information does not change the scale dependence significantly for large values of the invariant mass, since this is the region where the real-emission diagrams are dominant. However, for small values of the invariant mass, \ie\ in the 2-jet region, the scale dependence is reduced, mirroring the reduced scale dependence observed in the total cross section. The residual scale dependence is single logarithmic.

%%%%%%%%%%%%%%%%%%%%%%%%%%%%%%%%%%%%%%%%%%%%%%%%%%%%%%%%%%%%%%%%%%%%%%%%%%%%%%%%
\section{Conclusions}
\label{sec:conclusions}
%%%%%%%%%%%%%%%%%%%%%%%%%%%%%%%%%%%%%%%%%%%%%%%%%%%%%%%%%%%%%%%%%%%%%%%%%%%%%%%%

We have presented a new framework for event generation, \GenEvA, which allows almost any partonic calculation to be interfaced with a phenomenological model such as a parton shower. Because the phenomenological model is assumed to include a description of hadronization, \GenEvA\ offers a method for inclusive partonic information to be used to produce fully exclusive hadronic events.

The main conceptual breakthrough is a definition of phase space with a matching scale. This matching scale cleanly separates partonic calculations performed in QCD from phenomenological models based on QCD. \GenEvA\ avoids phase space double-counting by construction, because once \GenEvA\ is interfaced with a parton shower, every point in perturbative phase space is covered once and only once. By appropriate inclusion of leading-logarithmic information in the partonic calculations, the dominant sensitivity to the unphysical scale which separates the partonic regime from the showering regime is removed.

In this way, \GenEvA\ allows theorists to focus on providing the best possible partonic calculations and not on the algorithmic details of implementing those calculations. While other solutions to the problems of double-counting and shower merging exist in the literature, they are tailored to specific partonic calculations. For example, though there are various methods to merge tree-level calculations with parton showers~\cite{Catani:2001cc, Lonnblad:2001iq, Krauss:2002up, MLM, Mrenna:2003if, Schalicke:2005nv, Lavesson:2005xu, Hoche:2006ph, Alwall:2007fs, Giele:2007di, Lavesson:2007uu, Nagy:2007ty, Nagy:2008ns}, usually entirely different procedures are necessary to merge NLO calculations with parton showers~\cite{Collins:2000gd,Collins:2000qd, Potter:2001ej, Dobbs:2001dq, Frixione:2002ik, Frixione:2003ei, Nason:2004rx, Nason:2006hfa, LatundeDada:2006gx, Frixione:2007vw, Kramer:2003jk, Soper:2003ya, Nagy:2005aa, Kramer:2005hw}. In contrast, \GenEvA\ offers a generic solution to these problems, and the NLO/LO/LL merged sample provides an example of the kinds of improved partonic calculations that can be developed when algorithmic issues are separated from calculational ones.

The name \GenEvA\ is obviously inspired by the site of the upcoming LHC experiment at CERN, and we expect to generalize the \GenEvA\ framework to deal with additional complications present in a hadronic environment. We discuss some of these issues in our companion paper \cite{genevatechnique}, and we argue there that the complications are technical, not conceptual. Much of the legwork has already been done, as modern parton showers implement initial state radiation through backwards evolution, effectively defining the notion of a variable matching scale for hadronic collisions.

From a theoretical point of view, the most interesting developments will be to implement more advanced kinds of partonic calculations. In this work, we only considered loop diagrams involving two final state particles, but there is a growing body of one-loop calculations with large numbers of final states (for a recent review see Ref.~\cite{Bern:2007dw}). Because \GenEvA\ is agnostic as to the method of regulating IR divergences in loop diagrams, the choice about whether to use a slicing method, a subtraction method, or an elegant method to supplement NLO calculations with leading-logarithmic information can be made on the basis of physics considerations alone. Also interesting will be the implementation of calculations~\cite{Bauer:2006mk, Bauer:2006qp, Schwartz:2007ib} based on SCET~\cite{Bauer:2000ew, Bauer:2000yr, Bauer:2001ct, Bauer:2001yt} because it allows for a consistent treatment of subleading logarithms.

From an experimental point of view, \GenEvA\ offers a unique opportunity to assess Monte Carlo systematics, as a single \GenEvA\ event can support multiple different theoretical distributions, allowing theoretical errors to be probed without additional detector simulation time \cite{genevatechnique}. The NLO/LO/LL merged sample also gives a proof-of-concept that multiple different theoretical distributions can coexist within a single Monte Carlo sample, allowing the experiments to use the best theoretical knowledge available for any given point in phase space. With the possibility for sub-GeV measurements of the top quark mass and aggressive use of Monte Carlo to extrapolate the Standard Model up to 14 TeV, we anticipate that the experimental collaborations will benefit from the flexibility and transparency of the \GenEvA\ framework in the LHC era.

%%%%%%%%%%%%%%%%%%%%%%%%%%%%%%%%%%%%%%%%%%%%%%%%%%%%%%%%%%%%%%%%%%%%%%%%%%%%%%%%
\begin{acknowledgments}
We would like to thank Johan Alwall, Lance Dixon, Walter Giele, Beate Heinemann, Zoltan Ligeti, Michelangelo Mangano, Michael Peskin, Matthew Schwartz, Torbj\"{o}rn Sj\"{o}strand, Peter Skands, Iain Stewart, and Dieter Zeppenfeld for many useful discussions. We would also like to thank Johan Alwall, Fabio Maltoni, and Tim Stelzer for patient assistance with \texttt{MadGraph}, and Jeffrey Anderson for help with general computing questions.
This work was supported in part by the Director, Office of
Science, Office of High Energy Physics of the U.S.\ Department of Energy under
the Contract DE-AC02-05CH11231.
CWB acknowledges support from an DOE OJI award and an LDRD grant from LBNL.
JT is supported by a fellowship from the Miller Institute for Basic Research in Science.
\end{acknowledgments}

\bibliographystyle{../physrev4}
\bibliography{../geneva}

\providecommand{\href}[2]{#2}\begin{thebibliography}{10}

\bibitem{genevatechnique}
C.~W. Bauer, F.~J. Tackmann, and J.~Thaler,
\newblock \href{http://www.arXiv.org/abs/arXiv:0801.4028}{arXiv:0801.4028}.
%%CITATION = 0801.4028;%%

\bibitem{Brubaker:2006qt}
Tevatron Electroweak Working Group Collaboration, E.~Brubaker {\em et~al.},
\newblock \href{http://www.arXiv.org/abs/hep-ex/0603039}{hep-ex/0603039}.
%%CITATION = HEP-EX/0603039;%%

\bibitem{:2007ypa}
CDF Collaboration, T.~Aaltonen {\em et~al.},
\newblock Phys. Rev. Lett. {\bf 99}, 151801 (2007),
  [\href{http://www.arXiv.org/abs/arXiv:0707.0085}{arXiv:0707.0085}].
%%CITATION = 0707.0085;%%

\bibitem{Aaltonen:2007ps}
CDF Collaboration, T.~Aaltonen {\em et~al.},
\newblock Phys. Rev. D {\bf 77}, 112001 (2008),
  [\href{http://www.arXiv.org/abs/arXiv:0708.3642}{arXiv:0708.3642}].
%%CITATION = 0708.3642;%%

\bibitem{:1999fq}
ATLAS Collaboration,
\newblock (1999),
\newblock CERN-LHCC-99-14.
%%CITATION = ATLAS-TDR-14;%%

\bibitem{:1999fr}
ATLAS Collaboration,
\newblock (1999),
\newblock CERN-LHCC-99-15.
%%CITATION = ATLAS-TDR-15;%%

\bibitem{Ball:2007zza}
CMS Collaboration, G.~L. Bayatian {\em et~al.},
\newblock J. Phys. G {\bf 34}, 995 (2007).
%%CITATION = JPHGB,G34,995;%%

\bibitem{Lonnblad:1992tz}
L.~L{\"o}nnblad,
\newblock Comput. Phys. Commun. {\bf 71}, 15 (1992).
%%CITATION = CPHCB,71,15;%%

\bibitem{Sjostrand:2000wi}
T.~Sj{\"o}strand {\em et~al.},
\newblock Comput. Phys. Commun. {\bf 135}, 238 (2001),
  [\href{http://www.arXiv.org/abs/hep-ph/0010017}{hep-ph/0010017}].
%%CITATION = HEP-PH/0010017;%%

\bibitem{Sjostrand:2006za}
T.~Sj{\"o}strand, S.~Mrenna, and P.~Skands,
\newblock JHEP {\bf 05}, 026 (2006),
  [\href{http://www.arXiv.org/abs/hep-ph/0603175}{hep-ph/0603175}].
%%CITATION = HEP-PH/0603175;%%

\bibitem{Corcella:2000bw}
G.~Corcella {\em et~al.},
\newblock JHEP {\bf 01}, 010 (2001),
  [\href{http://www.arXiv.org/abs/hep-ph/0011363}{hep-ph/0011363}].
%%CITATION = HEP-PH/0011363;%%

\bibitem{Mangano:2002ea}
M.~L. Mangano, M.~Moretti, F.~Piccinini, R.~Pittau, and A.~D. Polosa,
\newblock JHEP {\bf 07}, 001 (2003),
  [\href{http://www.arXiv.org/abs/hep-ph/0206293}{hep-ph/0206293}].
%%CITATION = HEP-PH/0206293;%%

\bibitem{Maltoni:2002qb}
F.~Maltoni and T.~Stelzer,
\newblock JHEP {\bf 02}, 027 (2003),
  [\href{http://www.arXiv.org/abs/hep-ph/0208156}{hep-ph/0208156}].
%%CITATION = HEP-PH/0208156;%%

\bibitem{Gieseke:2003hm}
S.~Gieseke, A.~Ribon, M.~H. Seymour, P.~Stephens, and B.~Webber,
\newblock JHEP {\bf 02}, 005 (2004),
  [\href{http://www.arXiv.org/abs/hep-ph/0311208}{hep-ph/0311208}].
%%CITATION = HEP-PH/0311208;%%

\bibitem{Gieseke:2006ga}
S.~Gieseke {\em et~al.},
\newblock \href{http://www.arXiv.org/abs/hep-ph/0609306}{hep-ph/0609306}.
%%CITATION = HEP-PH/0609306;%%

\bibitem{Gleisberg:2003xi}
T.~Gleisberg {\em et~al.},
\newblock JHEP {\bf 02}, 056 (2004),
  [\href{http://www.arXiv.org/abs/hep-ph/0311263}{hep-ph/0311263}].
%%CITATION = HEP-PH/0311263;%%

\bibitem{Paige:2003mg}
F.~E. Paige, S.~D. Protopopescu, H.~Baer, and X.~Tata,
\newblock \href{http://www.arXiv.org/abs/hep-ph/0312045}{hep-ph/0312045}.
%%CITATION = HEP-PH/0312045;%%

\bibitem{Boos:2004kh}
CompHEP Collaboration, E.~Boos {\em et~al.},
\newblock Nucl. Instrum. Meth. A {\bf 534}, 250 (2004),
  [\href{http://www.arXiv.org/abs/hep-ph/0403113}{hep-ph/0403113}].
%%CITATION = HEP-PH/0403113;%%

\bibitem{Kilian:2007gr}
W.~Kilian, T.~Ohl, and J.~Reuter,
\newblock \href{http://www.arXiv.org/abs/arXiv:0708.4233}{arXiv:0708.4233}.
%%CITATION = 0708.4233;%%

\bibitem{Cafarella:2007pc}
A.~Cafarella, C.~G. Papadopoulos, and M.~Worek,
\newblock \href{http://www.arXiv.org/abs/arXiv:0710.2427}{arXiv:0710.2427}.
%%CITATION = 0710.2427;%%

\bibitem{Aaltonen:2007dg}
CDF Collaboration, T.~Aaltonen {\em et~al.},
\newblock \href{http://www.arXiv.org/abs/arXiv:0712.1311}{arXiv:0712.1311}.
%%CITATION = 0712.1311;%%

\bibitem{:2007kp}
CDF Collaboration, T.~Aaltonen {\em et~al.},
\newblock \href{http://www.arXiv.org/abs/arXiv:0712.2534}{arXiv:0712.2534}.
%%CITATION = 0712.2534;%%

\bibitem{Catani:2001cc}
S.~Catani, F.~Krauss, R.~Kuhn, and B.~R. Webber,
\newblock JHEP {\bf 11}, 063 (2001),
  [\href{http://www.arXiv.org/abs/hep-ph/0109231}{hep-ph/0109231}].
%%CITATION = HEP-PH/0109231;%%

\bibitem{Lonnblad:2001iq}
L.~L{\"o}nnblad,
\newblock JHEP {\bf 05}, 046 (2002),
  [\href{http://www.arXiv.org/abs/hep-ph/0112284}{hep-ph/0112284}].
%%CITATION = HEP-PH/0112284;%%

\bibitem{Krauss:2002up}
F.~Krauss,
\newblock JHEP {\bf 08}, 015 (2002),
  [\href{http://www.arXiv.org/abs/hep-ph/0205283}{hep-ph/0205283}].
%%CITATION = HEP-PH/0205283;%%

\bibitem{MLM}
M.~L. Mangano,
\newblock ``The so--called MLM prescription for ME/PS matching (2004)'',
\newblock Talk presented at the Fermilab ME/MC Tuning Workshop, October 4,
  2004., 2004.

\bibitem{Mrenna:2003if}
S.~Mrenna and P.~Richardson,
\newblock JHEP {\bf 05}, 040 (2004),
  [\href{http://www.arXiv.org/abs/hep-ph/0312274}{hep-ph/0312274}].
%%CITATION = HEP-PH/0312274;%%

\bibitem{Schalicke:2005nv}
A.~Sch{\"a}licke and F.~Krauss,
\newblock JHEP {\bf 07}, 018 (2005),
  [\href{http://www.arXiv.org/abs/hep-ph/0503281}{hep-ph/0503281}].
%%CITATION = HEP-PH/0503281;%%

\bibitem{Lavesson:2005xu}
N.~Lavesson and L.~L{\"o}nnblad,
\newblock JHEP {\bf 07}, 054 (2005),
  [\href{http://www.arXiv.org/abs/hep-ph/0503293}{hep-ph/0503293}].
%%CITATION = HEP-PH/0503293;%%

\bibitem{Hoche:2006ph}
S.~H{\"o}che {\em et~al.},
\newblock \href{http://www.arXiv.org/abs/hep-ph/0602031}{hep-ph/0602031}.
%%CITATION = HEP-PH/0602031;%%

\bibitem{Alwall:2007fs}
J.~Alwall {\em et~al.},
\newblock Eur. Phys. J. C {\bf 53}, 473 (2008),
  [\href{http://www.arXiv.org/abs/arXiv:0706.2569}{arXiv:0706.2569}].
%%CITATION = 0706.2569;%%

\bibitem{Giele:2007di}
W.~T. Giele, D.~A. Kosower, and P.~Z. Skands,
\newblock Phys. Rev. D {\bf 78}, 014026 (2008),
  [\href{http://www.arXiv.org/abs/arXiv:0707.3652}{arXiv:0707.3652}].
%%CITATION = 0707.3652;%%

\bibitem{Lavesson:2007uu}
N.~Lavesson and L.~L{\"o}nnblad,
\newblock JHEP {\bf 04}, 085 (2008),
  [\href{http://www.arXiv.org/abs/arXiv:0712.2966}{arXiv:0712.2966}].
%%CITATION = 0712.2966;%%

\bibitem{Nagy:2007ty}
Z.~Nagy and D.~E. Soper,
\newblock JHEP {\bf 09}, 114 (2007),
  [\href{http://www.arXiv.org/abs/arXiv:0706.0017}{arXiv:0706.0017}].
%%CITATION = 0706.0017;%%

\bibitem{Nagy:2008ns}
Z.~Nagy and D.~E. Soper,
\newblock JHEP {\bf 03}, 030 (2008),
  [\href{http://www.arXiv.org/abs/arXiv:0801.1917}{arXiv:0801.1917}].
%%CITATION = 0801.1917;%%

\bibitem{MCFM}
R.~K. Ellis and J.~Campbell,
\newblock ``MCFM - Monte Carlo for FeMtobarn processes'',
\newblock \url{http://mcfm.fnal.gov}.

\bibitem{NLOJET}
Z.~Nagy,
\newblock ``NLOJet++'',
\newblock \url{http://nagyz.web.cern.ch/nagyz/Site/NLOJet++.html}.

\bibitem{PHOX}
P.~Aurenche {\em et~al.},
\newblock ``The PHOX Family'',
\newblock \url{http://wwwlapp.in2p3.fr/lapth/PHOX_FAMILY/main.html}.

\bibitem{VBFNLO}
M.~B{\"a}hr {\em et~al.},
\newblock ``VBFNLO - NLO parton level Monte Carlo for Vector Boson Fusion'',
\newblock \url{http://www-itp.particle.uni-karlsruhe.de/~vbfnloweb/}.

\bibitem{Collins:2000gd}
J.~C. Collins and F.~Hautmann,
\newblock JHEP {\bf 03}, 016 (2001),
  [\href{http://www.arXiv.org/abs/hep-ph/0009286}{hep-ph/0009286}].
%%CITATION = HEP-PH/0009286;%%

\bibitem{Collins:2000qd}
J.~C. Collins,
\newblock JHEP {\bf 05}, 004 (2000),
  [\href{http://www.arXiv.org/abs/hep-ph/0001040}{hep-ph/0001040}].
%%CITATION = HEP-PH/0001040;%%

\bibitem{Potter:2001ej}
B.~Potter and T.~Schorner,
\newblock Phys. Lett. B {\bf 517}, 86 (2001),
  [\href{http://www.arXiv.org/abs/hep-ph/0104261}{hep-ph/0104261}].
%%CITATION = HEP-PH/0104261;%%

\bibitem{Dobbs:2001dq}
M.~Dobbs,
\newblock Phys. Rev. D {\bf 65}, 094011 (2002),
  [\href{http://www.arXiv.org/abs/hep-ph/0111234}{hep-ph/0111234}].
%%CITATION = HEP-PH/0111234;%%

\bibitem{Frixione:2002ik}
S.~Frixione and B.~R. Webber,
\newblock JHEP {\bf 06}, 029 (2002),
  [\href{http://www.arXiv.org/abs/hep-ph/0204244}{hep-ph/0204244}].
%%CITATION = HEP-PH/0204244;%%

\bibitem{Frixione:2003ei}
S.~Frixione, P.~Nason, and B.~R. Webber,
\newblock JHEP {\bf 08}, 007 (2003),
  [\href{http://www.arXiv.org/abs/hep-ph/0305252}{hep-ph/0305252}].
%%CITATION = HEP-PH/0305252;%%

\bibitem{Nason:2004rx}
P.~Nason,
\newblock JHEP {\bf 11}, 040 (2004),
  [\href{http://www.arXiv.org/abs/hep-ph/0409146}{hep-ph/0409146}].
%%CITATION = HEP-PH/0409146;%%

\bibitem{Nason:2006hfa}
P.~Nason and G.~Ridolfi,
\newblock JHEP {\bf 08}, 077 (2006),
  [\href{http://www.arXiv.org/abs/hep-ph/0606275}{hep-ph/0606275}].
%%CITATION = HEP-PH/0606275;%%

\bibitem{LatundeDada:2006gx}
O.~Latunde-Dada, S.~Gieseke, and B.~Webber,
\newblock JHEP {\bf 02}, 051 (2007),
  [\href{http://www.arXiv.org/abs/hep-ph/0612281}{hep-ph/0612281}].
%%CITATION = HEP-PH/0612281;%%

\bibitem{Frixione:2007vw}
S.~Frixione, P.~Nason, and C.~Oleari,
\newblock JHEP {\bf 11}, 070 (2007),
  [\href{http://www.arXiv.org/abs/arXiv:0709.2092}{arXiv:0709.2092}].
%%CITATION = 0709.2092;%%

\bibitem{Kramer:2003jk}
M.~Kr{\"a}mer and D.~E. Soper,
\newblock Phys. Rev. D {\bf 69}, 054019 (2004),
  [\href{http://www.arXiv.org/abs/hep-ph/0306222}{hep-ph/0306222}].
%%CITATION = HEP-PH/0306222;%%

\bibitem{Soper:2003ya}
D.~E. Soper,
\newblock Phys. Rev. D {\bf 69}, 054020 (2004),
  [\href{http://www.arXiv.org/abs/hep-ph/0306268}{hep-ph/0306268}].
%%CITATION = HEP-PH/0306268;%%

\bibitem{Nagy:2005aa}
Z.~Nagy and D.~E. Soper,
\newblock JHEP {\bf 10}, 024 (2005),
  [\href{http://www.arXiv.org/abs/hep-ph/0503053}{hep-ph/0503053}].
%%CITATION = HEP-PH/0503053;%%

\bibitem{Kramer:2005hw}
M.~Kr{\"a}mer, S.~Mrenna, and D.~E. Soper,
\newblock Phys. Rev. D {\bf 73}, 014022 (2006),
  [\href{http://www.arXiv.org/abs/hep-ph/0509127}{hep-ph/0509127}].
%%CITATION = HEP-PH/0509127;%%

\bibitem{Frixione:2006gn}
S.~Frixione and B.~R. Webber,
\newblock \href{http://www.arXiv.org/abs/hep-ph/0612272}{hep-ph/0612272}.
%%CITATION = HEP-PH/0612272;%%

\bibitem{Frixione:2007nu}
S.~Frixione, P.~Nason, and G.~Ridolfi,
\newblock \href{http://www.arXiv.org/abs/arXiv:0707.3081}{arXiv:0707.3081}.
%%CITATION = 0707.3081;%%

\bibitem{Bethke:1991wk}
S.~Bethke, Z.~Kunszt, D.~E. Soper, and W.~J. Stirling,
\newblock Nucl. Phys. B {\bf 370}, 310 (1992).
%%CITATION = NUPHA,B370,310;%%

\bibitem{Bauer:2000ew}
C.~W. Bauer, S.~Fleming, and M.~E. Luke,
\newblock Phys. Rev. D {\bf 63}, 014006 (2000),
  [\href{http://www.arXiv.org/abs/hep-ph/0005275}{hep-ph/0005275}].
%%CITATION = HEP-PH/0005275;%%

\bibitem{Bauer:2000yr}
C.~W. Bauer, S.~Fleming, D.~Pirjol, and I.~W. Stewart,
\newblock Phys. Rev. D {\bf 63}, 114020 (2001),
  [\href{http://www.arXiv.org/abs/hep-ph/0011336}{hep-ph/0011336}].
%%CITATION = HEP-PH/0011336;%%

\bibitem{Bauer:2001ct}
C.~W. Bauer and I.~W. Stewart,
\newblock Phys. Lett. B {\bf 516}, 134 (2001),
  [\href{http://www.arXiv.org/abs/hep-ph/0107001}{hep-ph/0107001}].
%%CITATION = HEP-PH/0107001;%%

\bibitem{Bauer:2001yt}
C.~W. Bauer, D.~Pirjol, and I.~W. Stewart,
\newblock Phys. Rev. D {\bf 65}, 054022 (2002),
  [\href{http://www.arXiv.org/abs/hep-ph/0109045}{hep-ph/0109045}].
%%CITATION = HEP-PH/0109045;%%

\bibitem{Bauer:2006qp}
C.~W. Bauer and M.~D. Schwartz,
\newblock Phys. Rev. Lett. {\bf 97}, 142001 (2006),
  [\href{http://www.arXiv.org/abs/hep-ph/0604065}{hep-ph/0604065}].
%%CITATION = HEP-PH/0604065;%%

\bibitem{Bauer:2006mk}
C.~W. Bauer and M.~D. Schwartz,
\newblock Phys. Rev. D {\bf 76}, 074004 (2007),
  [\href{http://www.arXiv.org/abs/hep-ph/0607296}{hep-ph/0607296}].
%%CITATION = HEP-PH/0607296;%%

\bibitem{Bauer:2007ad}
C.~W. Bauer and F.~J. Tackmann,
\newblock Phys. Rev. D {\bf 76}, 114017 (2007),
  [\href{http://www.arXiv.org/abs/arXiv:0705.1719}{arXiv:0705.1719}].
%%CITATION = 0705.1719;%%

\bibitem{Sudakov:1954sw}
V.~V. Sudakov,
\newblock Sov. Phys. JETP {\bf 3}, 65 (1956).
%%CITATION = SPHJA,3,65;%%

\bibitem{Gribov:1972ri}
V.~N. Gribov and L.~N. Lipatov,
\newblock Sov. J. Nucl. Phys. {\bf 15}, 438 (1972).
%%CITATION = SJNCA,15,438;%%

\bibitem{Altarelli:1977zs}
G.~Altarelli and G.~Parisi,
\newblock Nucl. Phys. B {\bf 126}, 298 (1977).
%%CITATION = NUPHA,B126,298;%%

\bibitem{Dokshitzer:1977sg}
Y.~L. Dokshitzer,
\newblock Sov. Phys. JETP {\bf 46}, 641 (1977).
%%CITATION = SPHJA,46,641;%%

\bibitem{Catani:1991hj}
S.~Catani, Y.~L. Dokshitzer, M.~Olsson, G.~Turnock, and B.~R. Webber,
\newblock Phys. Lett. B {\bf 269}, 432 (1991).
%%CITATION = PHLTA,B269,432;%%

\bibitem{Ellis:1993tq}
S.~D. Ellis and D.~E. Soper,
\newblock Phys. Rev. D {\bf 48}, 3160 (1993),
  [\href{http://www.arXiv.org/abs/hep-ph/9305266}{hep-ph/9305266}].
%%CITATION = HEP-PH/9305266;%%

\bibitem{Gustafson:1987rq}
G.~Gustafson and U.~Pettersson,
\newblock Nucl. Phys. B {\bf 306}, 746 (1988).
%%CITATION = NUPHA,B306,746;%%

\bibitem{Catani:1996jh}
S.~Catani and M.~H. Seymour,
\newblock Phys. Lett. B {\bf 378}, 287 (1996),
  [\href{http://www.arXiv.org/abs/hep-ph/9602277}{hep-ph/9602277}].
%%CITATION = HEP-PH/9602277;%%

\bibitem{Catani:1996vz}
S.~Catani and M.~H. Seymour,
\newblock Nucl. Phys. B {\bf 485}, 291 (1997),
  [\href{http://www.arXiv.org/abs/hep-ph/9605323}{hep-ph/9605323}],
\newblock [Erratum-ibid. {\bf B510}, 503 (1998)].
%%CITATION = HEP-PH/9605323;%%

\bibitem{Fabricius:1981sx}
K.~Fabricius, I.~Schmitt, G.~Kramer, and G.~Schierholz,
\newblock Zeit. Phys. C {\bf 11}, 315 (1981).
%%CITATION = ZEPYA,C11,315;%%

\bibitem{Kramer:1986mc}
G.~Kramer and B.~Lampe,
\newblock Fortschr. Phys. {\bf 37}, 161 (1989).
%%CITATION = FPYKA,37,161;%%

\bibitem{Bergmann:1989zy}
L.~J. Bergmann,
\newblock UMI-89-15738.
%%CITATION = UMI-89-15738;%%

\bibitem{Giele:1991vf}
W.~T. Giele and E.~W.~N. Glover,
\newblock Phys. Rev. D {\bf 46}, 1980 (1992).
%%CITATION = PHRVA,D46,1980;%%

\bibitem{Giele:1993dj}
W.~T. Giele, E.~W.~N. Glover, and D.~A. Kosower,
\newblock Nucl. Phys. B {\bf 403}, 633 (1993),
  [\href{http://www.arXiv.org/abs/hep-ph/9302225}{hep-ph/9302225}].
%%CITATION = HEP-PH/9302225;%%

\bibitem{Harris:2001sx}
B.~W. Harris and J.~F. Owens,
\newblock Phys. Rev. D {\bf 65}, 094032 (2002),
  [\href{http://www.arXiv.org/abs/hep-ph/0102128}{hep-ph/0102128}].
%%CITATION = HEP-PH/0102128;%%

\bibitem{Ellis:1991qj}
R.~K. Ellis, W.~J. Stirling, and B.~R. Webber,
\newblock Camb. Monogr. Part. Phys. Nucl. Phys. Cosmol. {\bf 8}, 1 (1996).
%%CITATION = CMPCE,8,1;%%

\bibitem{Murayama:1992gi}
H.~Murayama, I.~Watanabe, and K.~Hagiwara,
\newblock KEK-91-11.
%%CITATION = KEK-91-11;%%

\bibitem{Stelzer:1994ta}
T.~Stelzer and W.~F. Long,
\newblock Comput. Phys. Commun. {\bf 81}, 357 (1994),
  [\href{http://www.arXiv.org/abs/hep-ph/9401258}{hep-ph/9401258}].
%%CITATION = HEP-PH/9401258;%%

\bibitem{Ellis:1980wv}
R.~K. Ellis, D.~A. Ross, and A.~E. Terrano,
\newblock Nucl. Phys. B {\bf 178}, 421 (1981).
%%CITATION = NUPHA,B178,421;%%

\bibitem{Parisi:1978eg}
G.~Parisi,
\newblock Phys. Lett. B {\bf 74}, 65 (1978).
%%CITATION = PHLTA,B74,65;%%

\bibitem{Donoghue:1979vi}
J.~F. Donoghue, F.~E. Low, and S.-Y. Pi,
\newblock Phys. Rev. D {\bf 20}, 2759 (1979).
%%CITATION = PHRVA,D20,2759;%%

\bibitem{Catani:1998sf}
S.~Catani and B.~R. Webber,
\newblock Phys. Lett. B {\bf 427}, 377 (1998),
  [\href{http://www.arXiv.org/abs/hep-ph/9801350}{hep-ph/9801350}].
%%CITATION = HEP-PH/9801350;%%

\bibitem{Cacciari:2006sm}
M.~Cacciari,
\newblock \href{http://www.arXiv.org/abs/hep-ph/0607071}{hep-ph/0607071}.
%%CITATION = HEP-PH/0607071;%%

\bibitem{Bern:2007dw}
Z.~Bern, L.~J. Dixon, and D.~A. Kosower,
\newblock Annals Phys. {\bf 322}, 1587 (2007),
  [\href{http://www.arXiv.org/abs/arXiv:0704.2798}{arXiv:0704.2798}].
%%CITATION = 0704.2798;%%

\bibitem{Schwartz:2007ib}
M.~D. Schwartz,
\newblock Phys. Rev. D {\bf 77}, 014026 (2008),
  [\href{http://www.arXiv.org/abs/arXiv:0709.2709}{arXiv:0709.2709}].
%%CITATION = 0709.2709;%%

\end{thebibliography}

\end{document}